\documentclass[12pt,a4paper]{article}
\usepackage{jheppub}
\usepackage{dsfont}
\usepackage{eucal}
\usepackage{amsmath,amssymb,amscd,amsfonts}

\DeclareMathOperator{\arcsec}{Arcsec}

\newtheorem{theorem}{Theorem}[section]

\newtheorem{lemma}[theorem]{Lemma}

\newtheorem{corollary}[theorem]{Corollary}

\newtheorem{property}[theorem]{Property}
\newtheorem{observation}[theorem]{Observation}
\newtheorem{definition}{Definition}

\newcommand{\RR}{\mathbb{R}}
\newcommand{\CC}{\mathbb{C}}

\newcommand{\TT}{\Gamma}
\newcommand{\HH}{\mathbb{H}^{2}}
\newcommand{\ol}[1]	{\overline{#1}} %for arithmetic mean
\usepackage[toc,page]{appendix}
\usepackage{epsfig}
\usepackage{epstopdf}
\usepackage{latexsym}
\usepackage{graphicx,caption,subcaption}
\usepackage{booktabs}
\usepackage{bbm}
\usepackage{physics}

\makeatletter
\def\@fpheader{\relax}
\makeatother

\def\cK{{\cal K}}
\def\cM{{\cal M}}

\newcommand{\p}	{\partial}

\newcommand{\cC}{\mathcal{C}}

\newcommand{\cH}{\mathcal{H}}
\newcommand{\cI}{\mathcal{I}}

\newcommand{\cS}{\mathcal{S}}

\newcommand\ads {\text{AdS}}
\newcommand\cft {\text{CFT}}
\newcommand{\AdS}	{\ads}

\newcommand{\adsL}  {\mathsf{L}}

\newcommand{\n}     {n_{\text{Page}}}
\newcommand{\e}     {\mathsf{e}}
\newcommand{\s}     {\mathsf{s}}
\newcommand{\ith}   {i^{\text{th}}}
\newcommand{\jth}   {j^{\text{th}}}
\newcommand{\pin}   {p_{\text{in}}}
\newcommand{\pout}  {p_{\text{out}}}
\newcommand{\poutI}  {p_{\text{out}}^I}
\newcommand{\poutII}  {p_{\text{out}}^{II}}

\newcommand{\chiin} {\chi_\text{in}}
\newcommand{\chioutI} {\chi_\text{out}^I}
\newcommand{\chioutII} {\chi_\text{out}^{II}}

\newcommand{\rhoin} {\rho_\text{in}}
\newcommand{\rhooutI} {\rho_\text{out}^I}
\newcommand{\rhooutII} {\rho_\text{out}^{II}}

\newcommand{\betain} {\beta_\text{in}}
\newcommand{\betaoutI} {\beta_\text{out}^I}
\newcommand{\betaoutII} {\beta_\text{out}^{II}}

\title{Topological shadows and complexity of islands in multiboundary wormholes}
\author{Aranya Bhattacharya$^{a}$, Anindya Chanda$^{b}$, Sabyasachi Maulik$^{a}$, Christian Northe$^{c,d}$, Shibaji Roy$^{a}$}
\affiliation{$^a$Theory Division, Saha Institute of Nuclear Physics, HBNI, 1/AF Bidhannagar, Kolkata 700064, India.}

\affiliation{$^b$Department of Mathematics, Florida State University, 1017 Academic Way, Tallahassee, FL 32304, USA.}

\affiliation{$^c$Institut f\"{u}r Theoretische Physik und Astrophysik,\\Julius-Maximilians-Universit\"{a}t W\"{u}rzburg, Am Hubland, 97074 W\"{u}rzburg, Germany}

\affiliation{$^d$W\"{u}rzburg-Dresden Cluster of Excellence ct.qmat,\\Julius-Maximilians-Universit\"{a}t W\"{u}rzburg, Am Hubland, 97074 W\"{u}rzburg, Germany}

\emailAdd{aranya.bhattacharya[at]saha.ac.in, achanda[at]math.fsu.edu} 
\emailAdd{sabyasachi.maulik[at]saha.ac.in, christian.northe[at]physik.uni-wuerzburg.de,shibaji.roy[at]saha.ac.in}

\abstract{
Recently, remarkable progress in recovering the Page curve of an evaporating black hole (BH) in Jackiw-Teitelboim gravity has been achieved through use of Quantum Extremal surfaces (QES).
Multi-boundary Wormhole (MbW) models have been crucial in parallel model building in three dimensions. Motivated by this we here use the latter models to compute the subregion complexity of the Hawking quanta of the evaporating BH in AdS$_{3}$ and obtain the Page curve associated with this information theoretic measure. We use three- and $n$-boundary wormhole constructions to elucidate our computations of volumes below the Hubeny-Rangamani-Takayanagi (HRT) surfaces at different times. Time is represented by the growing length of the throat horizons corresponding to smaller exits of the multi-boundary wormhole and the evaporating bigger exit shrinks with evolving time. We track the change in choice of HRT surfaces with time and plot the volume with time. The smooth transition of Page curve is realized by a discontinuous jump at Page time in volume subregion complexity plots and the usual Page transition is realized as a phase transition due to the inclusion of the island in this context. We discuss mathematical intricacies and physical insights regarding the inclusion of the extra volume at Page time. The analysis is backed by calculations and lessons from kinematic space and tensor networks. 
}

\begin{document}
	
	\maketitle
	\flushbottom
		
	%%%%%%%%%%%%%%%%%
	\section{Introduction}
	%%%%%%%%%%%%%%%%%
	\par In recent times, the AdS/CFT correspondence \cite{Maldacena:1997re} has been used to decode secrets of a quantum theory of gravity through elegant geometrization of concepts from quantum information theory. This quest started by the Ryu-Takayanagi (RT) conjecture \cite{Ryu:2006bv, Ryu:2006ef} and its covariant generalization \cite{Hubeny:2007xt} for computing entanglement between boundary subsystems through bulk calculations. The conjecture was later derived as an instance of generalized entropy for Euclidean gravity solutions in \cite{Lewkowycz:2013nqa}. In its original incarnation, the RT formula seeks to evaluate the entanglement entropy $S_A$ of any subsystem $A$ in the $d$-dimensional dual QFT by computing the area of a codimension-$2$ minimal surface $\gamma$ homologous to $A$ in the bulk space-time following
	\begin{equation*}
		S_A = \frac{\text{Area}\left(\gamma\right)}{4 G_N^{(d+1)}}\,,
	\end{equation*}
	 which remains true as long as we consider pure classical gravity. Corrections due to quantum effects of bulk fields were first discussed in \cite{Faulkner:2013ana} and later explored in \cite{Engelhardt:2014gca, Engelhardt:2019hmr} which introduced the idea of a QES.   %have paved path for newer information theoretic measures over the years, e.g; entanglement of purification \cite{Takayanagi:2017knl, Bhattacharyya:2018sbw}, multipartite purification \cite{Umemoto:2018jpc}, complexity and many more. The Ryu-Takayanagi proposal has also accommodated more and more corrections that have bettered the overall understanding of the interplay between information and geometry.  generalized gravitational entropy\cite{Lewkowycz:2013nqa, Faulkner:2013ana} (motivated by the generalized entropy proposal by Bekenstein), quantum extremal surfaces(QES)\cite{Engelhardt:2014gca,Engelhardt:2019hmr} etc. 
	 \par  Of late, the quantum extremal surface programme has been very successfully utilized to reproduce the Page curve for an evaporating black hole \cite{Page:1993wv} from semi-classical constructions \cite{Almheiri:2019psf, Penington:2019npb, Almheiri:2019hni, Penington:2019kki, Almheiri:2019psy}. %realise the Page curve for black holes in AdS (both evaporating and eternal) within semi-classical domain. This is a very special development since it helps in understanding the combination of black hole and radiation as a unitarily evolving system. It is also a surprise that this has been achieved by staying within semi-classical regime without including any non-perturbative dynamics \cite{Almheiri:2019psf,Penington:2019npb,Almheiri:2019hni}. \footnote{Let us not be misrepresenting in anyway. This does not solve the full information paradox at all. It is still largely believed that to actually understand the paradox, or rather to fully resolve it, one has to have non-perturbative QG dynamics as important input.} 
	  The difficulty in this program was that a systematic description of fine grained (entanglement) entropy was missing which can be applied both to the black hole and the radiation. Hence, the understanding of Page curve remained incomplete and kept running into elusive contradictions. Using the QES, the authors of \cite{Almheiri:2019psf, Penington:2019npb, Almheiri:2019hni, Penington:2019kki, Almheiri:2019psy} were able to show that indeed one can systematically start from a pure state black hole for which, in the process of evaporation, a natural definition for consistent fine grained entropy arises. The curve displayed by this fine grained entropy is the ever-expected Page curve, fully devoid of any contradictions involving fine grained-to-coarse grained shift during the process. In describing such a process successfully, it was found that a bulk region is added to the QES after the Page time and aids in the appearance of the Page curve. These bulk regions are typically known as \textit{islands}\footnote{It is also noteworthy that the Page curves are different for the evaporating and the eternal black holes and so are the islands.}.

	Since the islands came into the picture, they have been greatly investigated and grasping the origin of islands from a more physical perspective is a subject of current research. In this vein, a few classical models have been introduced \cite{Akers:2019nfi,Li:2020ceg,Balasubramanian:2020hfs}, where the picture is purely classical-gravitational and one gets away by working only with HRT surfaces instead of QES. Ideally, in such a situation, we obtain an analogue of an island and a Page curve is also realized. Strictly speaking however, due to the absence of bulk entanglement entropy, this picture is a purely coarse grained approach. Nevertheless, these models have played an important role in understanding the origin of the islands from various perspectives and also realising the analogues of Page curve for other quantum information theoretic measures e.g; reflected entropy \cite{Dutta:2019gen}, entanglement of purification \cite{Takayanagi:2017knl, Bhattacharyya:2018sbw} etc.
	
	These models rely on multi-boundary wormholes in AdS$_{3}$, which are very special objects since they can be constructed as quotients of empty AdS$_3$ by its isometries. Once the fundamental domain is known and one avoids the fixed points to have well defined curvatures at each point of the fundamental domain, the problems become a lot easier to deal with. The radiation quanta themselves are typically modeled as a multipartite (at least bipartite, i.e. three-boundary wormhole ) systems where they are represented by smaller exits of the multi-boundary wormhole. To begin with, the actual black hole is modelled by a bigger exit and if evaporating, it keeps shrinking with time whereas more and more quanta are accumulated in the smaller exits. In these models, the minimal throat horizons at different times play the role of the HRT surface measuring the entanglement between the black hole and the combination of the Hawking quanta. Since the situation is dynamical, at some critical point, the choice of minimal surface changes and an island is included. There have been a few of such models in which the difference is how one stores the emitted quanta in different exits. Different entanglement measures have also been computed within the scope of these models. One among them is the reflected entropy \cite{Faulkner:2013ana, Li:2020ceg} . It measures the entanglement between different parts of a mixed state. For example, one can compute how entangled the different emitted quanta are with each other individually or with the black hole. The Page curve for the reflected entropy differs as well from its entanglement entropy counterpart.
	
\par Another interesting question that this line of study hopes to answer in the long run is the computational complexity associated with the decoding of the information stored within the evaporating black hole and radiation state. The complexity is in general different from entanglement and by definition, it captures the practical hardness of generating the quantum state in Hilbert space through some operations known as gates \cite{Jefferson:2017sdb,Chapman:2018hou,Caceres:2019pgf,Carmi:2016wjl}. Due to the Harlow-Hayden protocol \cite{Harlow:2013tf} and later works by Susskind and collaborators \cite{Brown:2019rox}, there is a general idea in the literature that this kind of state decoding is an exponentially hard task. This is supported by proposals about a state of the art geometric structure known as Python's lunch \cite{Brown:2019rox, Bao:2020hsc} that shows some signs why this is supposed to be such a complicated task. Nevertheless, the gravitational proposals of complexity \cite{Susskind:2014rva, Brown:2015bva, Brown:2015lvg} have not yet been able to find a situation that agrees with this particular suggestion. 
	
	In this paper, motivated by these studies, we study the volumes dual to the throat horizons in the multi-boundary wormhole models sketched above. Primarily put forward by Alishahiha \cite{Alishahiha:2015rta}, the volumes $V(\cS)$ subtended by HRT surfaces $\cS$ are conjectured to represent the so-called \textit{subregion complexity},
	\begin{equation}\label{AlishahihaComplexity}
	 \cC_A=\frac{V(\cS)}{8\pi\adsL G},
	\end{equation}
    where $G$ is Newton's constant and $\adsL$ is the AdS radius. Subregion complexity is argued to measure the difficulty of an algorithm to construct a mixed density matrix. In AdS$_{3}$, this has been studied in details and is understood as a compression algorithm constructed using tensors \cite{Abt:2017pmf}. In the tensor network picture, the number of bonds associated with some cost successfully mimics the behaviour of subregion complexity. Kinematic space provides yet another way of understanding these volumes \cite{Abt:2018ywl}. Usually kinematic space yields a description in which the bulk curves are understood roughly as the number of boundary anchored bulk geodesics crossing that curve \cite{Czech:2015qta} and the volumes as the number of such geodesics along with the chord lengths that each of them contribute to the volume. All of these are mostly understood within AdS$_{3}$/CFT$_{2}$. Since, the multi-boundary wormholes are also best understood in three spacetime dimensions, we use the machineries built in \cite{Abt:2017pmf, Abt:2018ywl} to study the Page curve analogue of subregion complexity in these models. 
	
	%In this paper, motivated by these studies, we study the volumes dual to the throat horizons in the multi-boundary wormhole models sketched above. \cn{modification} The volumes below the HRT surfaces $\cS$ are conjectured to represent the so-called subregion complexity,
	%\begin{equation}
	% \cC_A=\frac{V(\cS)}{8\pi\adsL G}
	%\end{equation}
	%Primarily put forward by Alishahiha \cite{Alishahiha:2015rta}, subregion complexity is argued to measure the difficulty of the algorithm to construct a mixed density matrix. In AdS$_{3}$, this has been studied in details and is understood as a compression algorithm constructed using tensors \cite{Abt:2017pmf}. In the tensor network picture, the number of bonds associated with some cost successfully mimics the behaviour of subregion complexity. Kinematic space provides yet another way of understanding these volumes \cite{Abt:2018ywl}. Usually kinematic space yields a description in which the bulk curves are understood roughly as the number of boundary anchored bulk geodesics crossing that curve \cite{Czech:2015qta} and the volumes as the number of such geodesics along with the chord lengths that each of them contribute to the volume. All of these are mostly understood within AdS$_{3}$/CFT$_{2}$. Since, the multi-boundary wormholes are also best understood in three spacetime dimensions, we use the machineries built in \cite{Abt:2017pmf, Abt:2018ywl} to study the Page curve analogue of subregion complexity in these models.
	
The remainder of the paper is structured as follows. In section \ref{volumes}, we briefly review the subregion volume computations in AdS$_{3}$ and then discuss the calculation of the relevant volumes in the models that we study. In Section \ref{kinematic space}, we describe how to understand the volume plots for Page curve from the point of view of kinematic space and the number of geodesics. In section \ref{tensornets}, we discuss the tensor network methods that can reproduce the peculiarities of the volume plot. Finally, we conclude in section \ref{conclusion}. In all of the sections, our treatment is usually twofold. Since these multi-boundary wormholes are typically objects vastly studied in mathematics, we regularly support our physical arguments by strong mathematical theorems, lemmas and observations. We hope that this will help both the Physics as well as the Mathematics community to build a better understanding of the studies in this direction.  
%%%%%%%%%%%%%%%%%%%%%%%%%
\section{Volumes in AdS$_{3}$ and multi-boundary models:}\label{volumes}
%%%%%%%%%%%%%%%%%%%%%%%%  
In this section, we review the constructions of multi-boundary wormhole geometries in \ref{maths section}. In doing so, we begin with the mathematical framework and follow up in \ref{multimodel} with the physical motivation for their study in the present paper.  
The multiboundary wormholes in $\AdS_{3}$ are geometries with multiple exits connected by a wormhole. All the different exits represent asymptotically $\AdS_{3}$ regions dual to CFT$_{2}$s from a physics perspective. Nevertheless, the construction of these geometries consistently involves a good amount of mathematical understanding, in fact it is a topic of research in mathematics itself. These are commonly known as the pair of pants in hyperbolic plane. In this paper we employ modern mathematical view points on the construction of wormhole geometries, in the hopes of establishing a common ground for mathematicians and physicists alike. Therefore, we will first discuss the construction of these geometries in the vastly known language of mathematics. 

\subsection{Mathematical preliminaries on pairs of pants}\label{maths section}

In this subsection we will briefly review the basic geometries of the hyperbolic plane and constructions of pair of pants and other $n$-hole spheres (commonly known as $n$-boundary wormholes in physics).  Let's begin with the hyperbolic plane,

\subsubsection{A Short Review of The Hyperbolic Plane:}\label{HyperbolicPlane}
In 300 BC, Euclid described the five postulates of geometry which are considered to be the starting assumptions in geometry. The fifth one among these was the most mysterious one. In simple words, the fifth postulate says that `given a line on a plane and a point not on that given line, there exist only one line parallel to the given line and the passing through the point'. In ensuing years mathematicians tried to prove the fifth postulate using the other four or to disprove it, which lead to the birth of a new geometry of plane, namely the hyperbolic geometry, which satisfies all the other four postulates except the fifth one. There are several models of hyperbolic plane, some of the most common models are the upper half-space model, the Poincar\'{e} disk model, the Klein model, the  hyperboloid model and others. here we describe only the upper half-space model. This is a very old and well-developed area in Mathematics and the literature on it is exhaustive. Here we mention only a few references and point to \cite{bonahonlow,katok1992fuchsian,casson1988automorphisms,beardon2012geometry} for more detailed descriptions.\\

\textbf{$\triangle$ The Upper Half-Space Model:} The upper half-space model of hyperbolic plane is described as the set $\HH=\{(x,y)\in\mathbb{R}^2|y>0\}=\{x+iy\in \CC|y>0\}$ endowed with the Riemannian metric $ds^2=\frac{dx^2+dy^2}{y^2}$. 
\par Next we describe the geometric nature of isometries and geodsics on $\HH$. M\"{o}bius transforms on the complex plane $\mathbb{C}$ play a key role in this area. A function $f:\mathbb{C}\rightarrow \mathbb{C}$ is called a M\"{o}bius transformation, if it is of the form $f(z)=\frac{az+b}{cz+d}$ where $a,b,c,d\in \mathbb{C}$ and $ad-bc\neq 0$. In our discussion, we will mainly use M\"{o}bius transformations with $a,b,c,d\in \RR$. 
\begin{observation}
M\"{o}bius transformations are homeomorphisms on $\CC$. For $a,b,c,d\in \RR$, the M\"{o}bius transformation $f(z)=\frac{az+b}{cz+d}$ fixes the X-axis and upper-half plane as sets, i,e. $f(\RR)=\RR$ and $f(\HH)=\HH$.
\end{observation}

\begin{observation}
M\"{o}bius transformations of the from $f:\HH \rightarrow \HH$ preserve the hyperbolic Riemannian metric ds on $\HH$, which means all the M\"{o}bius transformations with real coefficients are isometries on $\HH$. 
\end{observation}
The special linear group $SL(2,\RR)$ acts on the upper-half plane $\HH$ via M\"{o}bius transformations. The action is described as below
$$
\begin{pmatrix}
a & b \\
c & d
\end{pmatrix}
\in SL(2,\RR)\longrightarrow f(z)=\frac{az+b}{cz+d};~z\in \HH.
$$
We note that any matrix $A\in SL(2,\RR)$ and $-I_{2}A$ induces the same map on $\HH$ in the above sense. So we can refine the description above  as an action of the group $PSL(2,\RR)=SL(2,\RR)/\{I_2,-I_2\}$ on $\HH$ via M\"{o}bius transformations. This action also preserves the Riemannian metic of $\HH$, hence $PSL(2,\RR)$ can be viewed as a subgroup of the group of isometries of $\HH$. The following lemma describes the whole group of isometries, $Isom(\HH)$, of the hyperbolic plane.
\begin{theorem}\label{Poincare}
The isometry group of $\HH$, $Isom(\HH)$, consists of the maps of the form $z \rightarrow \frac{az+b}{cz+d}$ and $z\rightarrow \frac{-a\overline{z}+b}{-c\overline{z}+d}$ where $a,b,c,d\in \RR$ and $ad-cb=1$.\\ The maps, $z \rightarrow \frac{az+b}{cz+d}$, are the orientation preserving isometries and the other type is orientation reversing.  
\end{theorem}
In other words, $Isom^+(\HH)$, the group of orientation preserving isometries of $\HH$, is same as the group of M\"{o}bius transformations with real coefficients. 

Next, we will describe different types of elements of $Isom(\HH)$ and their geometric natures. Every M\"{o}bius transformation in $Isom(\HH)$ fixes the X-axis. Also a M\"{o}bius transformation $f(z)=\frac{az+b}{cz+d}$ sends $z=-d/c$ to $\infty$ and $\infty$ is mapped to $a/c$. Hence every M\"{o}bius transformation defines a homeomorphism on the topological space $\HH\, \cup\, X-\text{axis}\, \cup\,\{ \infty\}$. Topologically the set $\HH\, \cup\, X-axis\, \cup\, \{\infty\}$ is homeomorphic to the closed disk $\mathbb{D}^2=\{(x,y)\in \RR^2|x^2+y^2\leq 1\}$. Brouwer's fixed point theorem \cite{munkres2000topology} says that ``every continuous map from $\mathbb{D}^2$ to itself has a fixed point". So every M\"{o}bius transformation in $Isom^+(\HH)$ has at least one fixed point in $\HH\cup X-axis\cup\{\infty\}$. Indeed, consider $f(z)=\frac{az+b}{cz+d}\in Isom^+(\HH)$. To find the fixed points of the map, we solve the equation $z=\frac{az+b}{cz+d}$, which is quadratic in $z$ with at most two distinct solutions. A little calculation shows that if $z_1$ and $z_2$ are the solutions of the equation then,
\begin{equation}\label{fixedPointEquation}
 z_{1,2}=\frac{(a-d)\pm \sqrt{(a+d)^2-4}}{2c}\,.
\end{equation}
We classify the isometries depending on the number and nature of its fixed point(s). 
\begin{definition}\label{HyperbolicElements}
If we relate $f(z)=\frac{az+b}{cz+d}$ with the matrix $A=
\begin{pmatrix}
a & b \\
c & d
\end{pmatrix}$,
 then trace of A, tr$(A)=a+d$, determines the nature of the roots of \eqref{fixedPointEquation}. 

     \begin{enumerate}
        \item if $|$tr$(A)|>$2 The isometry has two real fixed points on the boundary of $\HH$ and $A$ is similar to a matrix of the form 
     	$\begin{pmatrix}
     	\mu & 0 \\
     	0 & \frac{1}{\mu}
     	\end{pmatrix}$. Geometrically, this type of isometries are dilatations. Note that, this dilatation  $z \rightarrow \mu^2 z$ fixes the Y-axis. In general, dilatations fix a hyperbolic line connecting the fixed points and this fixed line is called the axis of the dilatation. These isometries are called $hyperbolic$ elements.
     
     	\item if $|$tr$(A)|=$2, the isometry has only one real fixed point on the boundary of $\HH$ and $A$ is similar to a matrix of the form  
     	$\begin{pmatrix}
     	1 & s \\
     	0 & 1
     	\end{pmatrix}$. Geometrically, the isometry is a translation similar to $z\rightarrow z+s$. These isometries are called $parabolic$ elements. 
     	
     	\item if $|$tr$(A)|<$2 then there are two complex fixed points and as they are the root of the same quadratic polynomial, the roots are conjugate to each other. So only one of them lives in $\HH$. Hence this type of isometry has only one fixed point in the interior $\HH$. The corresponding matrices are similar to
     	\begin{equation*}
     	\begin{pmatrix}
     	cos(t) & -sin(t) \\
     	sin(t) & cos(t)
     	\end{pmatrix}
     	\end{equation*} 
     	In the Poincar\'e Disk model these type of isometries are rotations of the disk, they are called $elliptic$ elements.
     \end{enumerate}

 \end{definition}
  All orientation preserving isometries are finite combinations of these three types. To get an orientation reversing isometry, we need to take an orientation preserving isometry and compose it with reflection w.r.t Y-axis.
  \begin{theorem}
  If $\Phi: \HH \rightarrow \HH$ is an isometry, then $\phi$ is a finite composition of translation, dilatation, rotation and reflection.
  \end{theorem}
  
  Our next goal is to describe the geodesics in the upper-half plane model. A geodesic between two points x and y in $\HH$ is defined to be the curve with the smallest length joining the given points. In general, a curve, even if infinitely long in both directions, is called a \emph{geodesic} if for any two points x and y on the curve, the segment of the curve between x and y is the path with the shortest length joining them. \par
  Suppose $\gamma: t\rightarrow \left(\gamma_{1}\left(t\right),\gamma_{2}\left(t\right)\right),~t\in[a,b]$ is a curve in $\HH$. Then the hyperbolic length of $\gamma$, denoted as $\mathcal{L}_{hyp}(
  \gamma)$, is calculated through
  \begin{equation}\label{geodesicLength}
   \mathcal{L}_{hyp}(\gamma)=\int_{a}^{b}\frac{\sqrt{\gamma_1^{'}(t)^{2}+\gamma_2^{'}(t)^{2}}}{\gamma_{2}(t)}dt
  \end{equation}
  \begin{lemma}
  Suppose p and q are two points on a line in $\HH$ perpendicular to the X-axis. Then the vertical line joining p and q is the hyperbolic geodesic between them.
  \end{lemma}
  \textbf{proof:} If L is a line perpendicular to X-axis, then we can map L to the Y-axis via a translation and we have already seen that translations on $\HH$ preserve distance between two points. So it is sufficient to prove the lemma for two points $p$ and $q$ on the Y-axis. \\
  Suppose $\alpha:t\rightarrow(\alpha_{1}(t),\alpha_{2}(t)), t\in [a,b]$ is a curve joining $p=(0,y_1)$ and $q=(0,y_2)$ (assume $y_2>y_1$). Then 
  $$\mathcal{L}_{hyp}(\alpha)=\int_{a}^{b}\frac{\sqrt{\alpha_{1}'(t)^{2}+\alpha_{2}'(t)^{2}}}{\alpha_2(t)}dt\geq\int_{a}^{b}\frac{\sqrt{\alpha_2^{'}(t)^2}}{\alpha_{2}(t)}dt=ln\left(\frac{y_2}{y_1}\right)$$
  So, $ln\left(y_{2}/y_{1}\right)$ is the lower bound of the length for any curve joining $(0,y_{1})$ and $(0,y_{2})$. That means any curve joining $(0,y_{1})$ and $(0,y_{2})$ with exact length $ln\left(y_{2}/y_{1}\right)$ is the geodesic connecting them. We consider the vertical line $\gamma:t\rightarrow (0,t), t\in[y_{1},y_{2}]$. The hyperbolic length of that line is $\int_{y_1}^{y_2}\frac{dt}{t}=ln\left(y_{2}/y_{1}\right)$. 
  Hence the vertical line segment joining $p$ and $q$ is the geodesic between them.$\square$ 
  
  \par Before we describe all other types of geodesics in $\HH$, we mention a couple of important geometric properties of M\"{o}bius transformations. As isometries of $\HH$ are M\"{o}bius transformations, these properties help us to understand their geometries.
  \begin{property}\label{prop2.6}
  Every M\"{o}bius transformation maps line and circles to line and circles. M\"{o}bius transformations also preserve angles. 
  \end{property}
  \begin{property}\label{prop2.7}
  For any two points p and q in $\HH$ not on a vertical line, there exists a M\"{o}bius transformation $\Phi$ with real coefficients such that $\Phi(p)$ and $\Phi(q)$ lie on a vertical line in $\HH$. 
  \end{property}
  The next lemma describes all types of geodesic in $\HH$.
  \begin{lemma}
  The collection of geodesics in $\HH$ consists of all vertical lines and the half-circles in $\HH$ with centres on the X-axis. Between any two points p and q in $\HH$ there exists a unique geodesic segment either of the previous types connecting them. 
  \end{lemma}
\par For a proof of lemma 2.8 we refer \cite{bonahonlow}. We note that if we fix two points p and q on $\HH$ then we can connect them by either a vertical straight line or a semi-circle perpendicular to $X-axis$, so we can connect any two points on $\HH$ by a hyperbolic geodesic. 

  \subsubsection{Hyperbolic Surfaces and their Construction}
  \label{sec:HyperbolicSurfaces}
  A topological surface $\Sigma$ is called a hyperbolic surface with geodesic boundary if there exists a collection of pairs $\{(U_{\alpha},\Phi_{\alpha})\}$ where $\{U_{\alpha}\}$ is an open cover on $\Sigma$ and each $\Phi_{\alpha}$ is a homeomorphism from $U_{\alpha}$ to an open set in the half plane of $\HH$, $\{(x,y)|x\geq 0; y>0\}$. Moreover, the maps $\Phi_{\beta}^{-1}\circ \Phi_{\alpha}$ should be restrictions of isometries on $\HH$. Here, we will first very precisely discuss how to construct hyperbolic surfaces and then we will focus on constructions of pair of pants and general spheres with $n$-holes.   
  
  \par Suppose $\Gamma$ is a subgroup of $Isom(\HH)$. Then we can define a group action of $\TT$ on $\HH$ as $(\gamma,x)\rightarrow \gamma(x)$ for all $\gamma\in \Gamma$ and $x\in \HH$. Before we proceed further, let's recapitulate a few important definitions related to group actions.
  \begin{definition}
  For the group action of $\TT$ on $\HH$
  \begin{itemize}
      \item the $\TT$-orbit of a point $x\in\HH$ is defined as the set $\{\gamma(x)|\gamma\in \TT\}$ and denoted as $\TT(x)$.
      \item the group action is \textit{properly discontinuous} if for every compact set $K\subset\HH$, the set $\{\gamma\in \TT|\gamma(K)\cap K\neq \O\}$ is finite.
      \item the action is free if for all $\gamma\in \TT$ and $x\in\TT$, $\gamma(x)\neq x$.
  \end{itemize}
  \end{definition}
  We construct the quotient space of the action, denoted by $\HH/\TT$, as the topological space $\HH/\TT=\frac{\HH}{ x \sim \gamma(x)};\forall \gamma\in \TT$. 
  \begin{lemma}
  If the action of $\TT$ on $\HH$ via isometries is free and properly discontinuous then the projection map $\HH\rightarrow \HH/\TT$ is a covering space map and $\HH/\TT$ is a hyperbolic surface. 
  \end{lemma}
  Hence using the lemma 2.9 we can construct hyperbolic surfaces by choosing appropriate $\TT$ and considering its quotient space $\HH/\TT$. Our choice of $\TT$ can be more precise and we need to use only discrete subgroups of $Isom^{+}(\HH)$, called the \textit{Fuchsian} groups. Before we describe the geometric way to construct any particular hyperbolic surface, we  define the \textit{fundamental domain} of the action of $\TT$ as below:
  \begin{definition}
  If  $\TT$ is a group of isometries acting on $\HH$, then the fundamental domain of the action is a closed subset $D$ of $\HH$, such that 
  \begin{itemize}
      \item the interior of $D$, $int(D)$, is non-empty,
      \item for all $\gamma\neq Id$, $\gamma(int(D))\cap int(D)=\O$,
      \item the $\TT$-translates of $D$ cover the whole $\HH$, that is $\bigcup_{\gamma \in \TT}\gamma(D)=\HH$. In other words, $D$ has a representative for all $\Gamma$-orbits. 
    
  \end{itemize}
  \end{definition}
  The interesting geometric fact is that if the action of $\TT$ is free and properly-discontinuous then the fundamental domain is a hyperbolic convex region whose boundary is a combination of geodesic segments, geodesic rays, full geodesics and segments of the $X$-axis. In fact, all hyperbolic surfaces can be constructed by starting with an appropriate hyperbolic convex domain as described and attaching its edges pairwise via isometries of $\HH$. In that case, the subgroup $\TT$ generating the surface $\Sigma=\HH/\TT$ is the group generated by elements of $Isom(\HH)$ attaching the edges.
   
  %\begin{theorem}[Poincar\'{e} Polygon Theorem]
  %Let $P\subset\HH$ be a compact polygon whose sides are identified in pairs by isometries of $\HH$, and let $\TT$ be the group generated by the isometries. Suppose that, for each $\TT$-orbit of vertices of P, the internal angles at the vertices in that orbit add up to $2\pi$. Then $\TT$ is a discrete group acting freely and properly discontinuously on $\HH$; moreover, P is a fundamental domain for the action.
 % \end{theorem}
   We are now ready to describe the construction of \textit{pair of pants} and $n$-hole spheres, lets denote them as $\Sigma_n$. 
  \par
\begin{figure}[t]
      \includegraphics[scale=0.5]{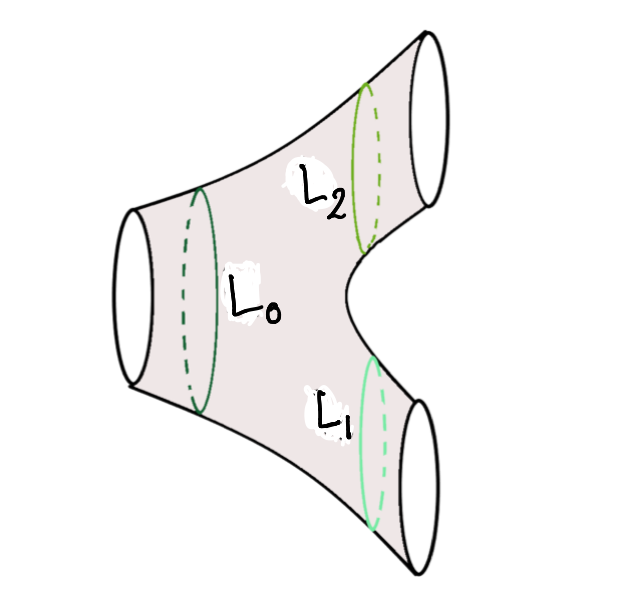}
      \includegraphics[scale=0.45]{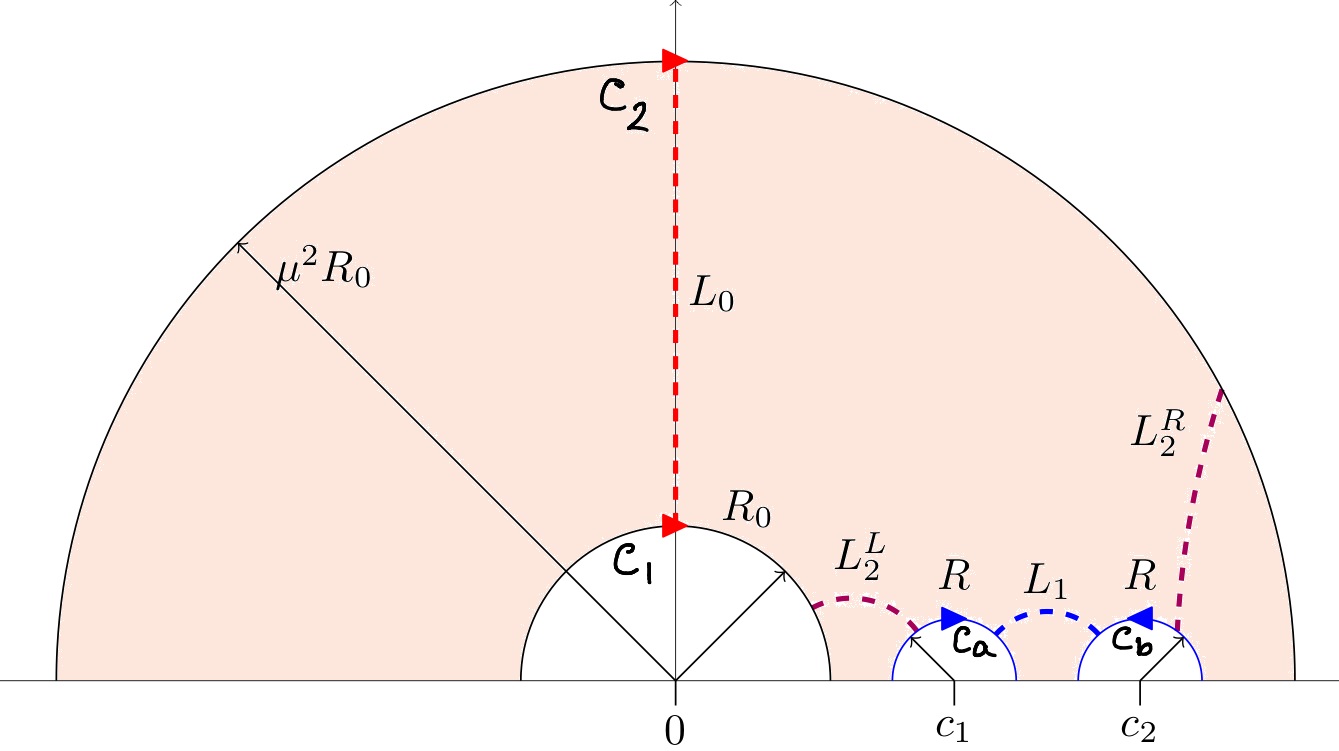}
      \caption{Pair of Pants and Polygonal Representation. In the figure, concentric geodesic edges are denoted by $C_{1}$ and $C_{2}$, whereas $C_{a}$ and $C_{b}$ stand for geodesics that are not concentric. $c_{1}$ and $c_{2}$ denote the centres of the non-concentric geodesics on the horizontal plane.}
      \label{fig:pants_construction}
  \end{figure}

$\bullet$ \textbf{Pair of pants:} A \textit{pair of pants} or $\Sigma_3$ is a hyperbolic surface which is homeomorphic to a sphere with three closed disks removed. We construct it from the convex domain on the hyperbolic plane $\HH$ depicted in figure \ref{fig:pants_construction}. 
  
  First consider the geodesic edges $C_1$ and $C_2$, they lie on two concentric half circles perpendicular to $X-axis$ with radii $R_0$ and $\mu^{2}R_0$ where $\mu>1$. Suppose the circles are centred at $(0,0)$ and given by the parametrizations $\{C_1:t\rightarrow R_{0}e^{\pi t}\}$ and $\{C_2:t\rightarrow \mu^{2}R_{0}e^{\pi t}\}$ with $0\leq t \leq 1$. Then we glue geodesics with the orientation preserving isometry $z\rightarrow \mu^{2}z$. The corresponding matrix representation of the M\"{o}bius transformation is $\gamma_1=
  \begin{pmatrix}
  \mu & 0 \\
  0 & 1/\mu \\
  \end{pmatrix}$. Next we glue the geodesic edges $C_a$ and $C_b$ in a reverse orientation, as shown in figure \ref{fig:pants_construction}. For simplicity we assume both of the circles has the same radius $R$. Suppose $C_a$ is the semi-circle $\{C_a: t\rightarrow -Re^{-\pi t}+\alpha \}$ and $C_b$ is  $\{C_b : t\rightarrow Re^{\pi t}+\beta\}$ where $0<t<1$ and $R_{0}<\alpha -R<R+\alpha <\beta-R < R+\beta< \mu^{2}R_{0}$. To describe the isometry which identifies $C_b$ with $C_a$ with a reverse orientation, we start with $C_b$. First apply the translation $z\rightarrow z-\beta$. This transformation maps $C_b$ to the unit semi-circle centered at the origin. Next, apply the inversion map $z\rightarrow -\frac{1}{z}$ which fixes the unit circle but reverses its orientation and maps the inside of the circle to its outside. Finally apply the transformation $z\rightarrow z+\alpha$, which maps the unit semi-circle to the circle $C_a$. The corresponding matrix representation $\gamma_2$ of the M\"{o}bius transformation which identifies $C_b$ to $C_a$ with a reverse orientation is given by 
  \[\gamma_{2}=
  \begin{pmatrix}
  1 & \alpha\\
  0 & 1
  \end{pmatrix}
  \begin{pmatrix}
  0 & 1 \\
  -1 & 0
  \end{pmatrix}
  \begin{pmatrix}
  1 & -\beta\\
  0 & 1
  \end{pmatrix}
 \] 
  The \textit{Fuchsian group} $\TT$ is generated by $\gamma_1$ and $\gamma_2$ and $\HH/\TT$ in this construction is homeomorphic to a sphere with 3-holes and the convex region in figure \ref{fig:pants_construction} is the fundamental domain of that action. The \textit{throat horizons} of $\Sigma_3$ are denoted by $L_{0}, L_{1}$ and ${L_2}$, where $L_2$ is constructed as the union of $L_{2}^{L}$ and $L_{2}^{R}$ on the fundamental domain. On $\HH$ the curves $L_{0}, L_{1}, L_{2}^{L}$ and $L_{2}^{R}$ lie on the perpendicular geodesics which represents the shortest distances between respective pair of geodesics they intersect. For example, $L_0$ is the shortest geodesic segment connecting $C_1$ and $C_2$.
  
  If we want to construct a pair of pants with boundaries and with geodesic throat horizons, we can consider the quotient space of $\TT$-action on $\HH_{\epsilon}$ for a very small $\epsilon$, where $\HH_\epsilon$ is the component of $\HH$ bounded below by a simple, not necessarily straight line $L$ such that $L$ lies in the $\epsilon$-neighborhood of $X-axis$ representing the same curves in each $\gamma$-image of the fundamental domain and if $x\in L$ then $\gamma(x)\in L$ for all $\gamma \in \TT$. $\HH_\epsilon$ includes $L$. The elements of $\TT$ may not be bijective anymore on $\HH_\epsilon$, but the quotient space construction of $\frac{\HH_{\epsilon}}{x\sim\gamma(x)}$ still works and the line $L$ corresponds to the topological boundary (may not be a geodesic) of the pair of pants.
  \begin{figure}[t]
      \centering
      \includegraphics[scale=0.2]{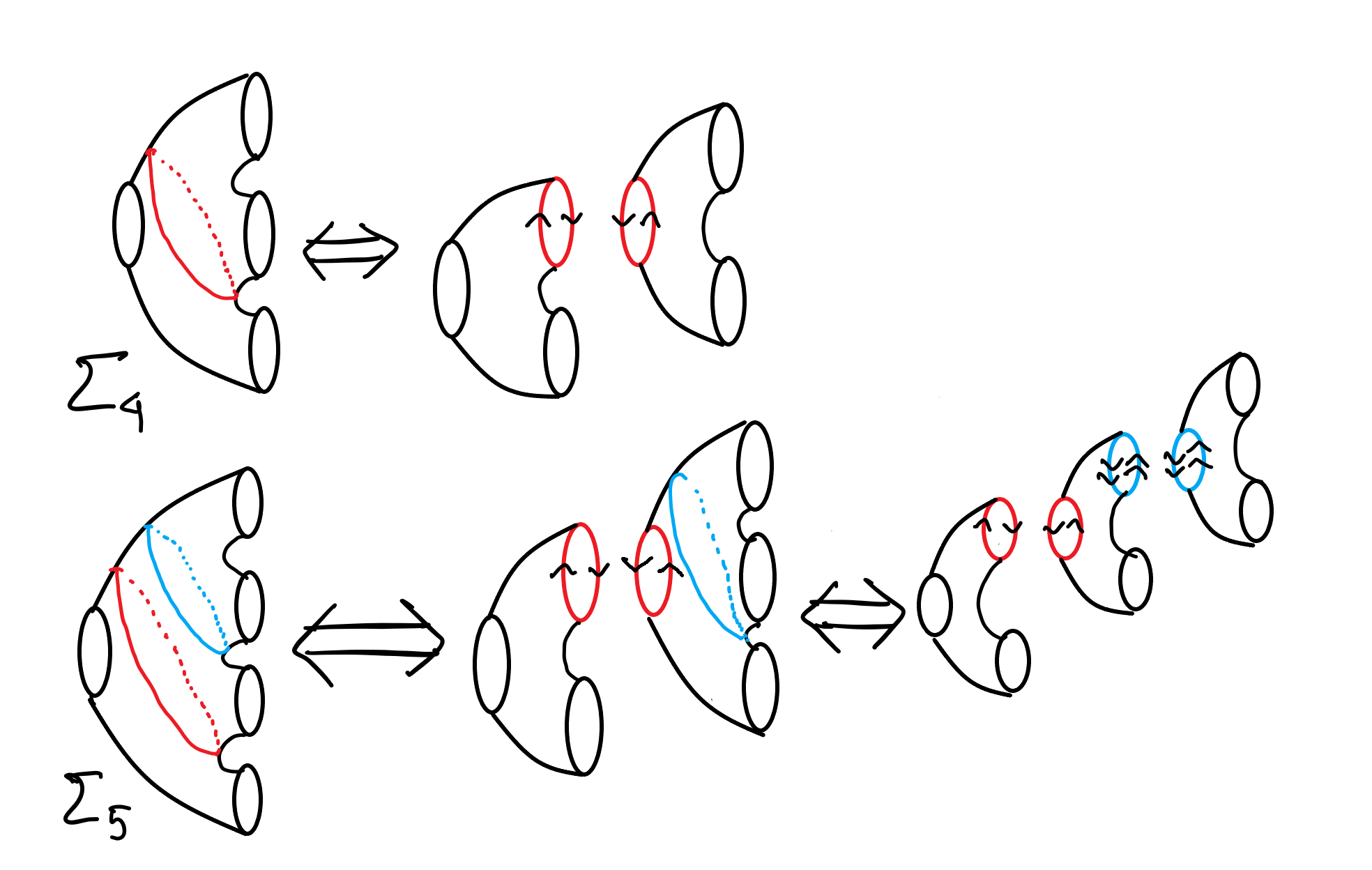}
      \caption{Describing higher number of boundaries.}
      \label{fig:fourfive}
  \end{figure}
  
 % \par Before we describe the construction of general n-boundary sphere $\Gamma_n$ for $n>3$, we will visualize the polygonal fundamental domain to construct the pair of pants in a different way. Note that we can transform the polygon in figure 1 to the polygon in figure 2 via an element of $Isom(\HH)$, say $ \Phi$. So the new set of isometries to identify the edges will be $\{\Phi \gamma_{1}\Phi^{-1}, \Phi\gamma_{2}\Phi^{-1}\}$. \\
 % \begin{figure}
 %     \centering
  %    \includegraphics[scale=0.23]{3BU.png}
   %   \caption{Caption}
      
  %\end{figure}
  
  $\bullet$\textbf{ $n$-hole Sphere $\Sigma_n$}: Now we construct general spheres with $n$ holes, $\Sigma_n$, via the procedure described above for a single pair of pants. Before we describe the construction, here is an important result from surface geometry,
  \begin{lemma}[\cite{schultens2014introduction}]\label{Pant Decomposition of Surfaces}
  For any compact surface $S$ there exists a collection of pairwise disjoint simple closed curves $\{c_1,c_2,...,c_n\}$ such that each component of $S-\{c_1,c_2,...,c_n\}$ is a pair of pants with boundaries.
  \end{lemma}
    
  \begin{figure}[t]
  	\begin{subfigure}{0.36\textwidth}
  		\centering
  		\includegraphics[width=\textwidth]{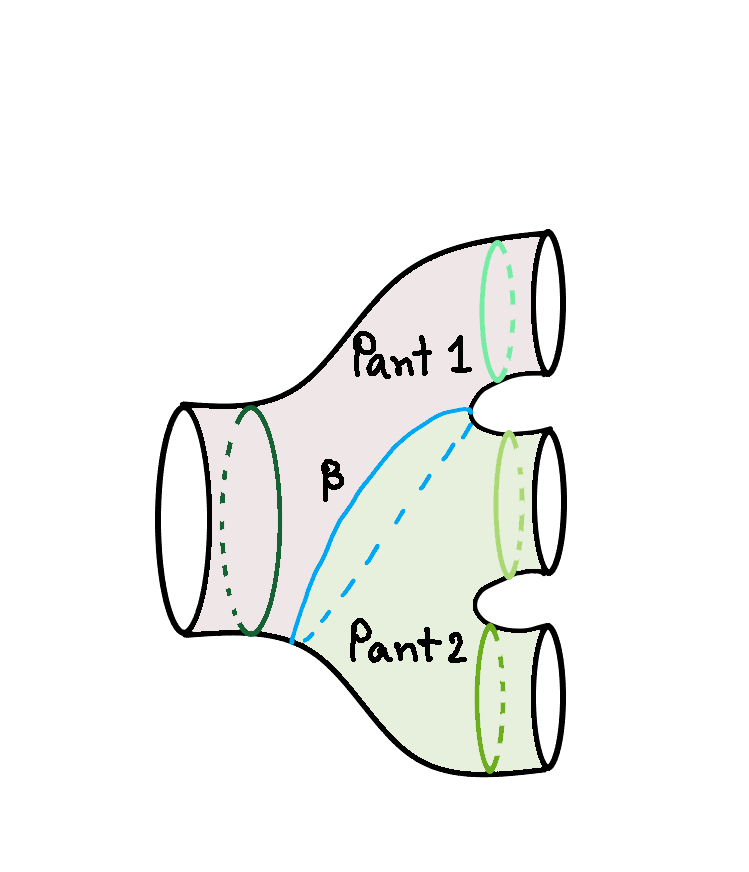}
  	\end{subfigure}
  	\hfill
  	\begin{subfigure}{0.63\textwidth}
  		\centering
  		\includegraphics[width=\textwidth]{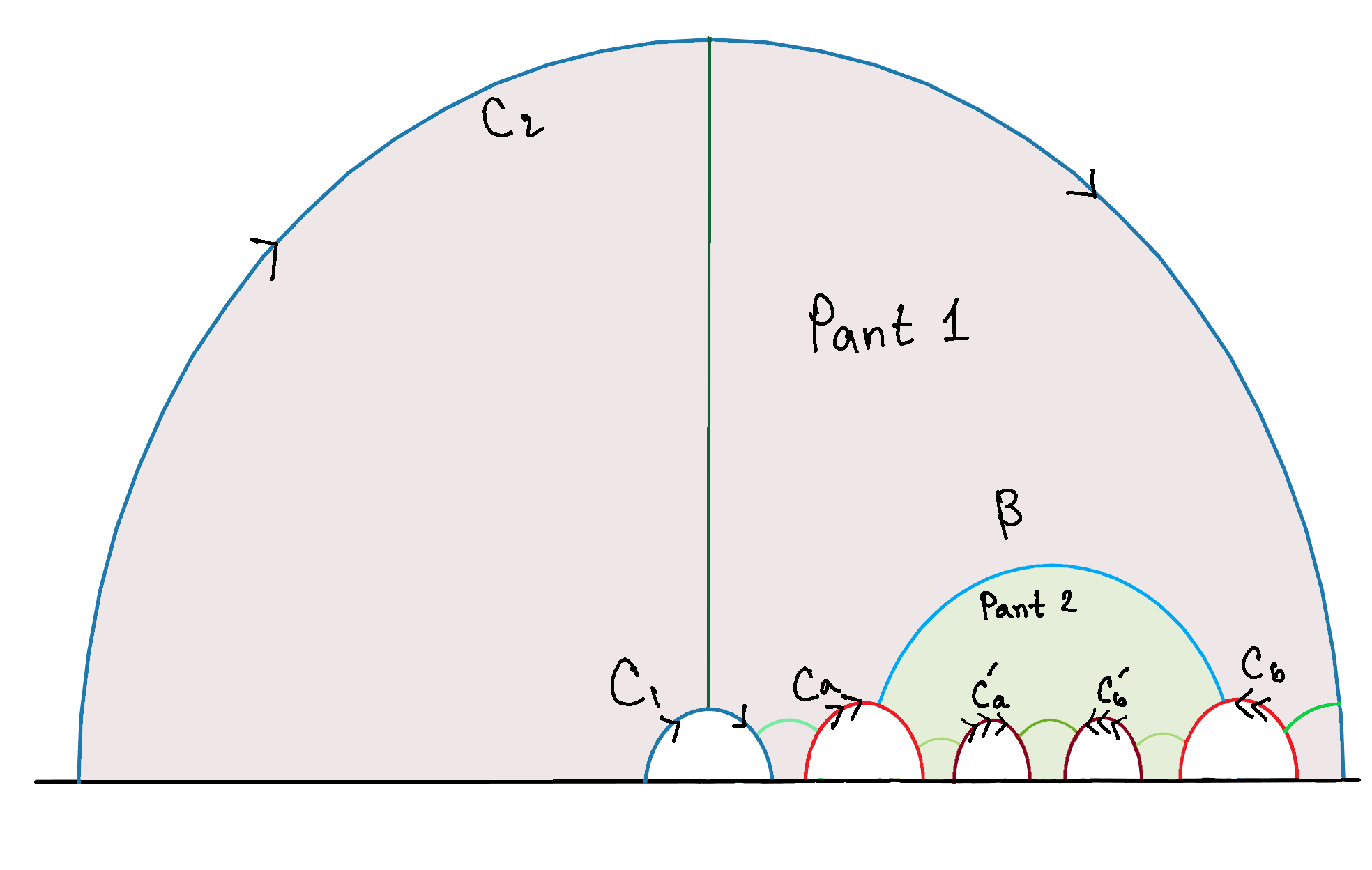}
  	\end{subfigure}
      %\centering
      %\includegraphics[scale=0.3]{3-bddPD.png}
      %\includegraphics[scale=.2]{2-pantU.png}
      \caption{Four boundary case in upper half space model.}
      \label{fig:fourbdy}
  \end{figure}

  In figure \ref{fig:fourfive} we show the pair of pants decomposition of $\Sigma_4$ and $\Sigma_5$ with boundaries and the same process works for any higher $n$. Though the sphere with four holes is not a compact surface, we can draw an intuition for the choice of the fundamental domain for $\Sigma_4$ from the pant decomposition of spheres with four boundaries. In figure \ref{fig:fourbdy} we consider the blue curve $\beta$ on $\Sigma_4$ and note $\beta$ cuts $\Sigma_4$ in two copies of $\Sigma_2$. That information indicates that we can construct the convex fundamental domain for $\Sigma_4$ on $\HH$ by taking two copies of fundamental region of $\Sigma_2$ and attaching them along the curve $\beta$, the resulting domain on $\HH$ is given in figure \ref{fig:fourbdy}. To get $\Sigma_4$ we identify the circles $C_1$ and $C_2$ by preserving their orientations and glue the other pair of circles $\{C_a, C_b\}$ and $\{C_{a}^{'},C_{b}^{'}\}$ by reversing their orientation as shown in figure \ref{fig:fourbdy}. 
    
  To get any general $\Sigma_n$, we will start with a convex domain for $\Sigma_3$ and iterate the above described process $(n-3)$ times, hence we will have a convex domain with $(4n-4)$-sides for $\Sigma_n$ and we need to attach $(n-1)$ pairs of edges, the corresponding Fuchsian group is generated by $(n-1)$ elements. In all cases, if we want our surfaces with boundaries then we can take the quotient space of the respective Fuchsian group actions on $\HH_\epsilon$ for $\epsilon$ very small and positive.
  
  %\par To find the corresponding volumes, we resort to the construction of multiboundary wormholes as quotient of AdS described in section \ref{sec:HyperbolicSurfaces}. The idea is well-known and has been discussed in e.g. \cite{Brill:1995jv, Aminneborg:1997pz, Skenderis:2009ju, Balasubramanian:2014hda, Caceres:2019giy}. We will largely follow \cite{Caceres:2019giy} which gave a simple algorithm to obtain any wormhole with arbitrary genus and number of boundaries.
\subsubsection*{Translation to physics}
\begin{figure}[t]
	\centering
	\includegraphics[scale=0.20]{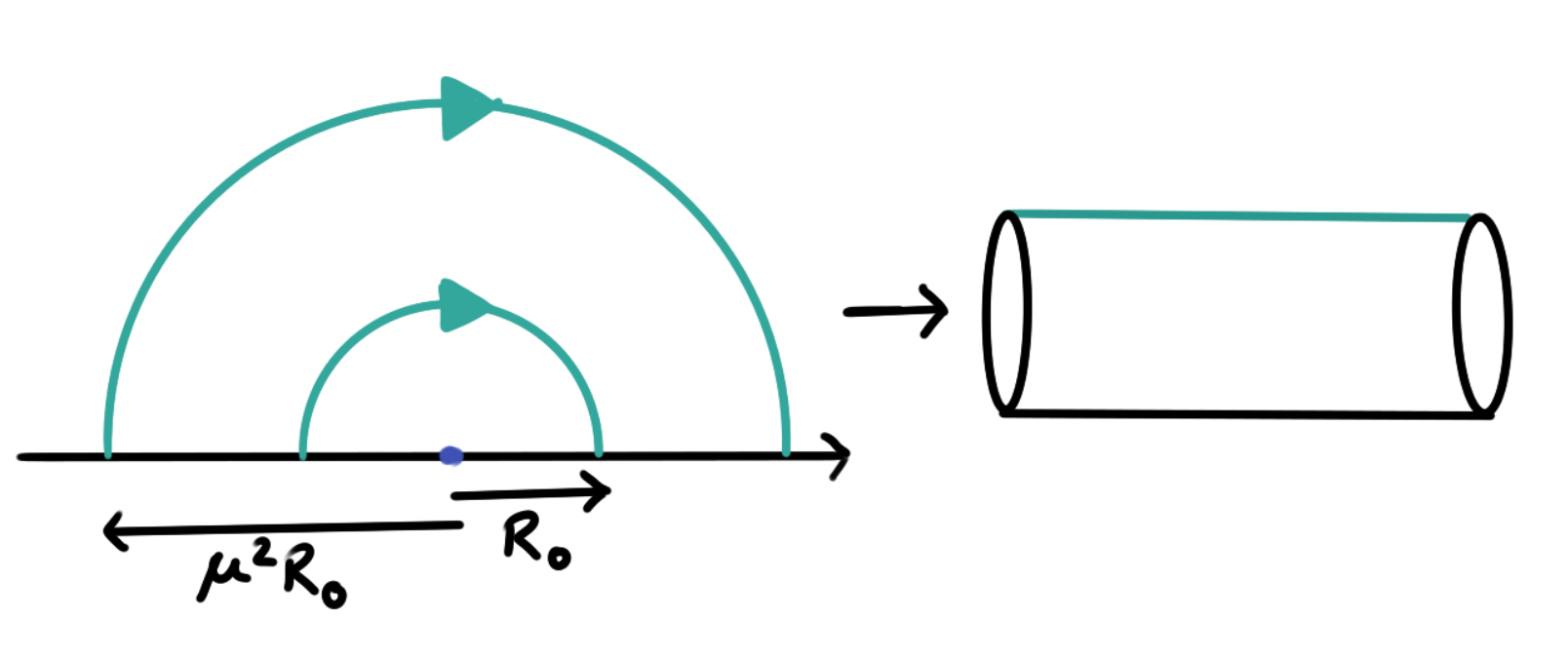}
	\caption{The $t = 0$ slice of the two-sided BTZ obtained by quotienting the upper half plane by dilatation.}
	\label{figs:rs2side}
\end{figure}

The procedure described above is well known in physics \cite{Brill:1995jv, Aminneborg:1997pz, Skenderis:2009ju, Balasubramanian:2014hda, Caceres:2019giy}, albeit usually pronounced with different words. The standard example is the quotient of the $t=0$ slice of $\AdS_3$, which coincides with $\HH$, by a discrete dilatation. It identifies two concentric semicircles in $\HH$ and gives rise to a BTZ black hole \cite{Banados:1992wn, Banados:1992gq}. Its topology is that of a cylinder as shown in figure \ref{figs:rs2side}.
%Very naively this can be thought of as identifying geodesics in the upper half plane, which are semicirles. The simplest quotient of $\AdS_3$ is the BTZ black hole \cite{Banados:1992wn, Banados:1992gq}. It is obtained from $\AdS_3$ by quotienting with dilatation which implies identifying two concentric semicircles in the upper half plane.

%More complicated geometries can be constructed by combinations of isometries.
A three-boundary wormhole is obtained from the two-boundary surface by further identification of a pair of geodesics. They must have opposite orientations and lie \textit{on the same side}%
\footnote{We point out that one can also quotient geodesics lying on distinct sides of the first throat horizon. This quotient gives rise to a geometry with a single boundary only.}
of the throat horizon corresponding to the first quotient, which is $L_0$ in figure \ref{fig:pants_construction}.
%See figure \ref{figs:pinchfold}.
The algorithm to add more boundaries is straightforward: for each new boundary one identifies an additional pair of geodesics with opposite orientations lying on one side of a throat horizon. This is equivalent to the nesting depicted in \ref{fig:fourfive}.
%This construction is mentioned in section \ref{HyperbolicPlane} starting from lemma \ref{Pant Decomposition of Surfaces}.
An $n$-boundary wormhole thus requires identification of $\left(2 n - 2 \right)$ such semicircles. In the following we explain how this construction is employed to model black hole evaporation.

% \begin{figure}[t]
% 	\centering
% 	\includegraphics[scale=0.80]{pinchfold.pdf}
% 	\caption{The three-boundary and one-boundary, one-genus geometries as quotients.} 
% 	\label{figs:pinchfold}
% \end{figure}

\subsection{Introducing the MbW models of black hole evaporation}\label{multimodel}

Having constructed multi-boundary wormholes in hyperbolic geometry, let us discuss the precise models we are interested in. We concentrate on two models which effectively capture some of the central ideas associated with the island program. In both models, we start from a three-boundary wormhole. One of its exits is much larger than the other two, which have coinciding size. The bigger exit is the analogue of the evaporating black hole whereas the smaller ones model the radiation quanta being emitted from the BH. The two models we consider are distinguished by the way the geometry changes with time as more and more quanta get stored in the radiation geometry whereas the BH keeps getting smaller.\\
\begin{figure}[t]
\centering
\includegraphics[scale=0.50]{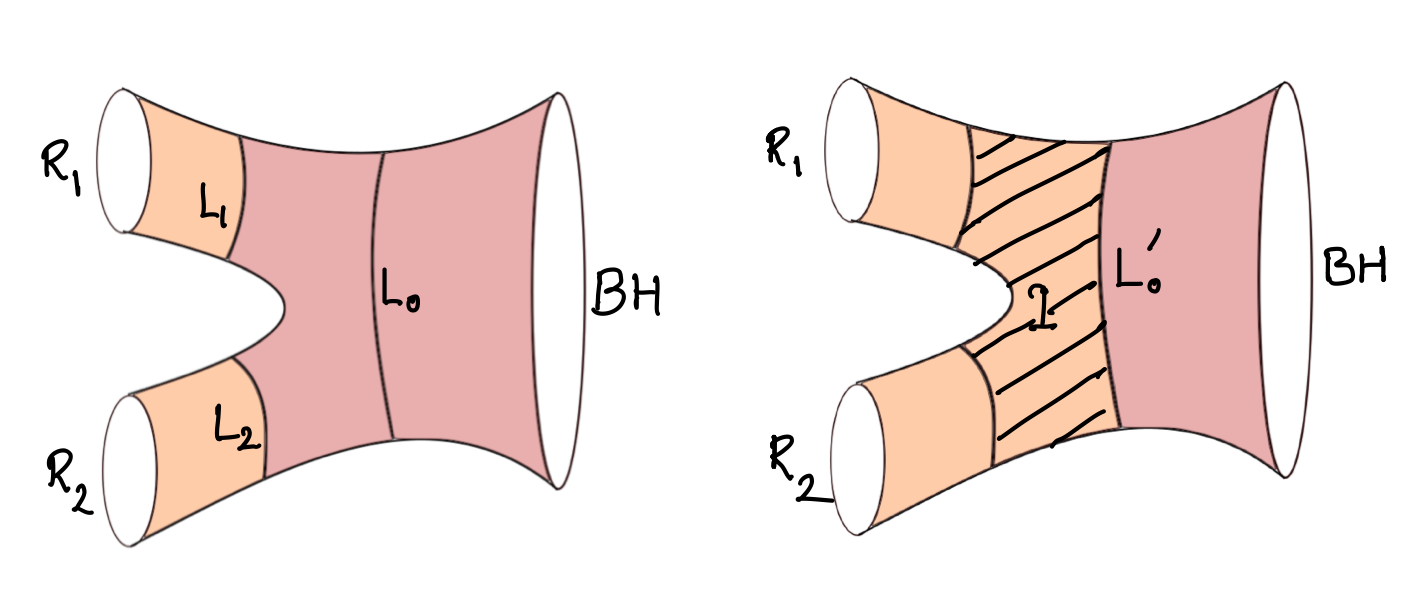}
\caption{Change of preferred HRT in $3$ boundary model. Left: Before the Page time the HRT surface separates the $R_i$ from the remainder of the pair of pants. Right: After the Page time $L_0$ has shrunk to $L_0'$ and the HRT surface has jumped to include the island $I$.}
\label{fig: threebdy}
\end{figure}

\begin{figure}[t]
	\centering
	\includegraphics[scale=0.60]{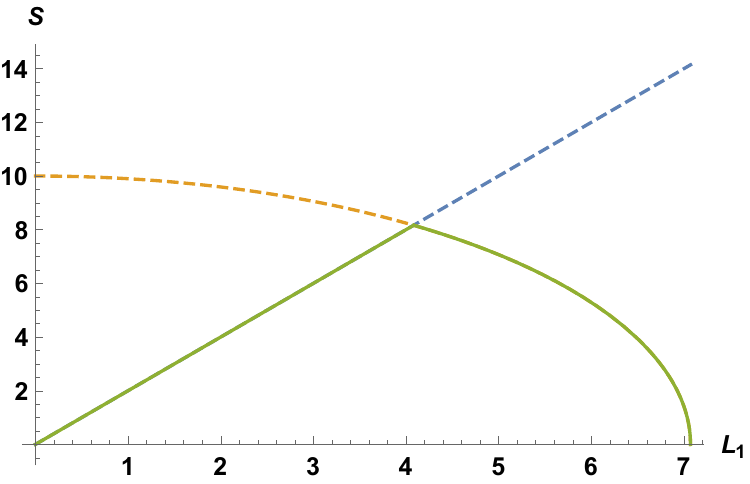}
	\includegraphics[scale=0.45]{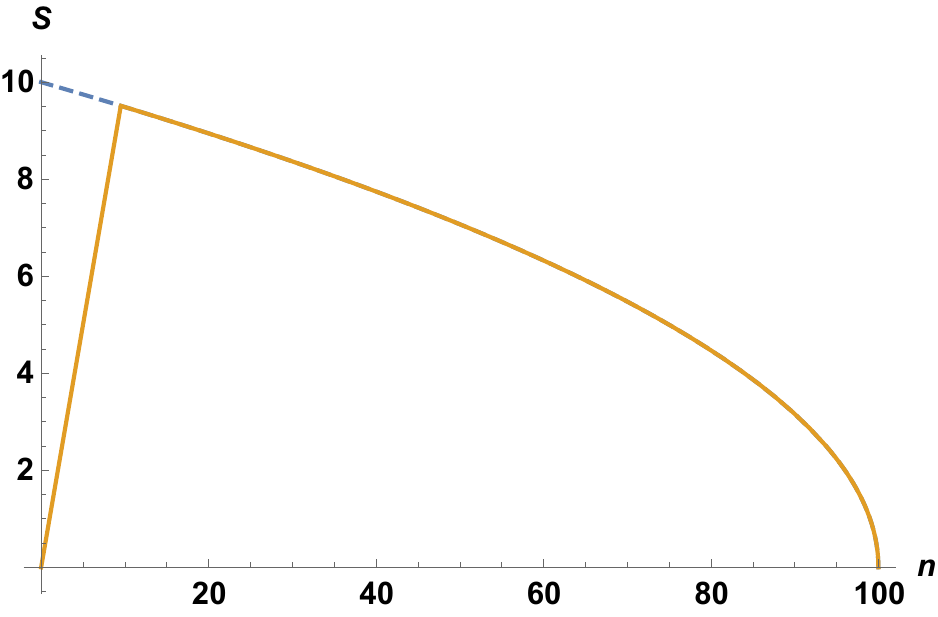}
	\caption{Page curves corresponding to the left: three and right: n boundary models.}
	\label{fig: Pagecurves}
\end{figure}

\textbf{a) Three Boundary Model:}  In the first model, we evolve the exit sizes of a three-boundary wormhole as the system moves forward in time. Therefore, in this model the size of the bigger exit (BH) decreases with time while the smaller exits increase. We insist that the sizes of the smaller exits remain the same as time evolves. Hence, both the smaller exits increase at the same rate. We track the minimal throat horizon lengths corresponding to the union of smaller exits (Hawking quanta) with time . There is a shift in the choice of minimal geodesic at certain timescale, the Page time, after which the connected minimal throat horizon (corresponding to the bigger exit) is the favored choice as opposed to the disconnected unions (throat horizons of the smaller exits)\footnote{It is to be noted that both these choices are homologous to the BH as well as union of Hawking quanta for all times. Hence they are the candidate HRT surfaces.}. This change of preference gives rise to the Page curve in this model. The situation is shown through the pair of pant geometry in figure \ref{fig: threebdy}. After the Page transition the region I is added to the entanglement wedge of the Hawking quanta. This is the representative island in this model. The corresponding Page curve is shown in left hand side of figure \ref{fig: Pagecurves}. Note that the topology of this model never changes; it remains a three-boundary wormhole at all times.\\

\begin{figure}[t]
	\centering
	\includegraphics[scale=0.50]{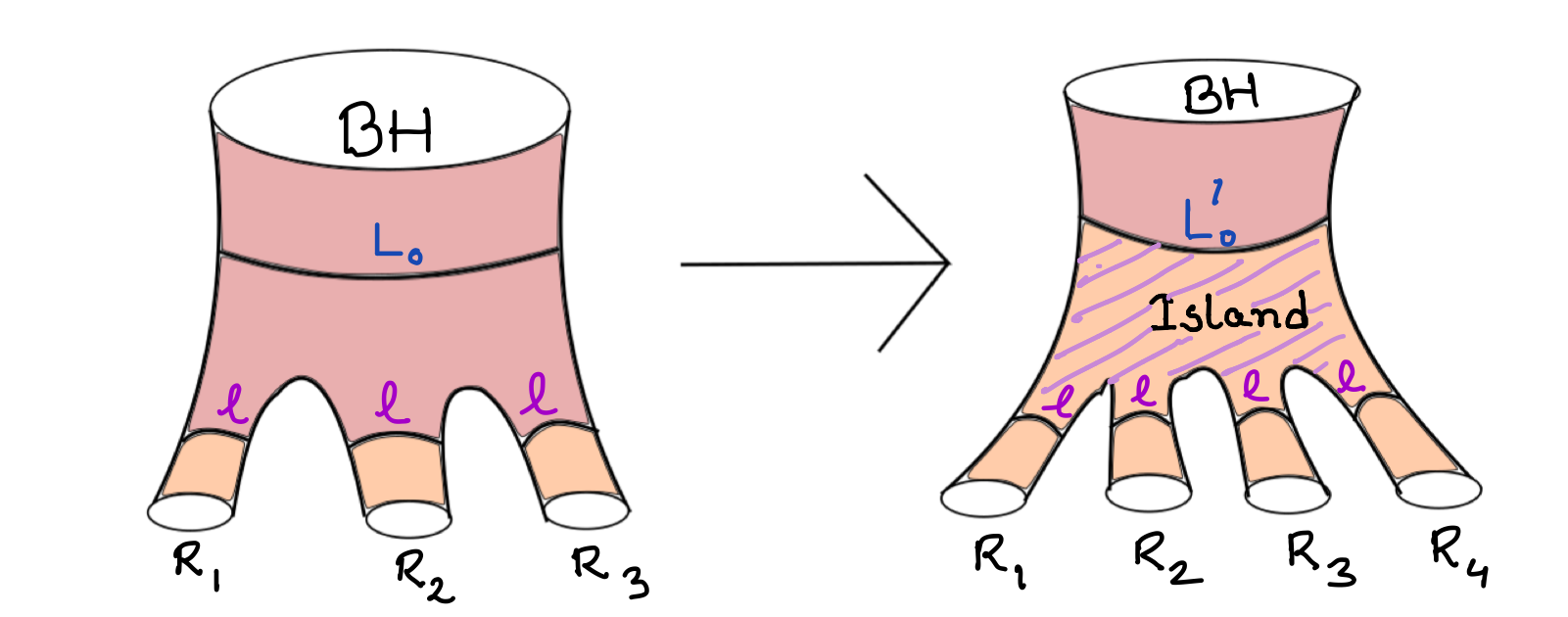}
	\caption{Change of preferred HRT in $n$-boundary model.}
	\label{fig: nbdy}
\end{figure}

\textbf{b) $n$-Boundary Model:}  In the second model, instead of increasing the size of the smaller exits, we increase the \textit{number} of smaller exits. Hence, in this model, topology changes with each time step and the no. of exits $n$ represents this time. Although it is hard to realize dynamically from Einstein's equations, it is perfectly reasonable as discrete snapshots at different times during the radiation. All the different topologies are time reflection symmetric. The bigger exit, similar to the three-boundary model, keeps decreasing and again a transition of HRT surface corresponding to the union of the Hawking quanta (union of the ($n-1$) smaller exits in this case) takes place at certain point of time ($n_{page}$), this is shown in figure \ref{fig: nbdy}. The corresponding Page curve is shown in right hand side of figure \ref{fig: Pagecurves}. 

\subsection{Volumes in AdS$_{3}$}
\label{sec:volumesAdS3}
In this paper we content ourselves with constant time slices of $\AdS_3$ space-time. The HRT formula \cite{Ryu:2006bv, Ryu:2006ef, Hubeny:2007xt} suggests that the entanglement entropy of any region $\mathit{A}$ on the boundary of $\AdS_3$ is equivalent to the length of the bulk geodesic $\gamma_{RT}(A)$ anchored at $\p A$; one also needs to introduce a cutoff surface $\gamma_{\epsilon}$ near the boundary for regularization. Our primary interest is in the volume of the co-dimension-$1$ surface $\Sigma$ with boundary $\partial\Sigma = \gamma_{RT}(A)\,\cup\,A_{\epsilon}$, where $A_\epsilon$ is the segment of the cutoff surface $\gamma_\epsilon$, which hovers over $A$. This volume appears in the original definition of holographic complexity in \eqref{AlishahihaComplexity}. In this work we employ, however, an alternate definition of subregion complexity put forward in \cite{Abt:2017pmf}.
\begin{definition}
 Let $\Sigma\subset\HH$ be a hyperbolic surface with boundary $\partial\Sigma = \gamma_{RT}(A)\,\cup\,A_{\epsilon}$ for boundary interval $A$. Its topological subregion complexity is defined through
\begin{equation}\label{cmplxtydefn}
\mathcal{C}\left(A\right) \equiv -\frac{1}{2}\int_{\Sigma}R\,d\sigma\,,
\end{equation}
where $R$ is the scalar curvature of the bulk space-time.
\end{definition}
In the cases of interest in this paper $R$ is a constant so that the topological subregion complexity \eqref{cmplxtydefn} and the original proposal \eqref{AlishahihaComplexity} differ only in normalization. One benefit of using topological complexity is that it is naturally dimensionless as desired for complexities.
However, the main advantage of \eqref{cmplxtydefn} lies in the fact that it determines the complexity completely by topological data through use of the Gauss-Bonnet theorem
\begin{theorem}
 Let $\Sigma$ be an orientable, compact, two-dimensional Riemannian manifold with piecewise smooth boundary $\p\Sigma$ and scalar curvature $R$. Denote by $k_g$ the geodesic curvature of the curve carved out by $\p\Sigma$. Then
 \begin{equation}\label{cmplxty}
-\frac{1}{2}\int_{\Sigma}R\,d\sigma = \int_{\partial \Sigma} k_g\,ds + \sum_{i=1}^r\alpha_i - 2\pi\chi\left(\Sigma\right),
\end{equation}
where $\chi\left(\Sigma\right)$ is the Euler characteristic of $\Sigma$. $r$ is the number of corners in $\p\Sigma$ and $\alpha_i$ are the corner angles at which the piecewise smooth segments of $\p\Sigma$ intersect. 
\end{theorem}
The \textit{geodesic curvature} $k_g$ measures how much $\partial\Sigma$, or any other curve under scrutiny, deviates from a geodesic. If we anchor $\Sigma$ at a boundary interval $A$, then the left hand side is of course the topological complexity $\mathcal{C} \left(A\right)$. Moreover, in this case the corner angles $\alpha_i$ are always $\pi/2$ \cite{Rangamani_2017}, since geodesics $\gamma_{RT}$ intersect the cutoff surface perpendicularly.

\par Let us illustrate the formula with standard examples. In the simplest case the subsystem $\mathit{A}$ is a single connected interval $\mathit{A} \in \left[x_1,x_2\right]$. The boundary has two corners, those at which $\gamma_{RT}$ and $A_{\epsilon}$ intersect, each of which contributes $\pi/2$, yielding
%As explained in \cite{Abt:2017pmf}, the integral in \eqref{cmplxty} receives contribution from $\gamma_{\epsilon}$ and the two corner angles where $\gamma_{RT}$ and $\A_{\epsilon}$ intersect. Since $\gamma_{RT}$ is a bulk geodesic, it intersects the boundary perpendicularly \cite{Ryu:2006ef} and hence its net contribution from the integral amounts to
\begin{equation*}
\int_{\partial \Sigma} k_g\,ds +\sum_{i=1}^2\alpha_i = \frac{x_2 - x_1}{\epsilon} + 2\times\frac{\pi}{2}\,.
\end{equation*}
The Euler characterisic of $\Sigma$ is $1$ as it is topologically equivalent to a disk, thus we obtain
\begin{equation}
\mathcal{C}\left(A\right) = \frac{x_2 - x_1}{\epsilon} - \pi\, .
\end{equation}
As another example let us consider two disjoint sub-regions $\mathit{A} = \mathit{A_1}\cup\mathit{A_2}$, where $\mathit{A_1}=\left[x_1,x_2\right] \text{~and~} \mathit{A_2}=\left[x_3,x_4\right], \left(x_1<x_2<x_3<x_4\right).$ There are two candidate HRT surfaces for this configuration. In phase I the complexity is simply the sum of that for each subregion, i.e. 
\begin{equation}
\mathcal{C}_I = \frac{x_2-x_1}{\epsilon}+\frac{x_4-x_3}{\epsilon}-2\pi
\end{equation} 
In phase II where $\Sigma$ is a connected surface, only $\chi\left(\Sigma\right)$ is different and hence
\begin{equation}
\mathcal{C}_{II} = \frac{x_2-x_1}{\epsilon}+\frac{x_4-x_3}{\epsilon}+4\times\frac{\pi}{2}-2\pi = \frac{x_2-x_1}{\epsilon}+\frac{x_4-x_3}{\epsilon}
\end{equation}
Thus subregion complexity exhibits a discontinuous jump at the transition. It is easy to generalize this result for arbitrary number of intervals and has been shown in \cite{Abt:2017pmf}, which also considers non-zero temperature.
\begin{figure}[t]
	\begin{subfigure}{0.2\textwidth}
		\centering
		\includegraphics[scale=0.15]{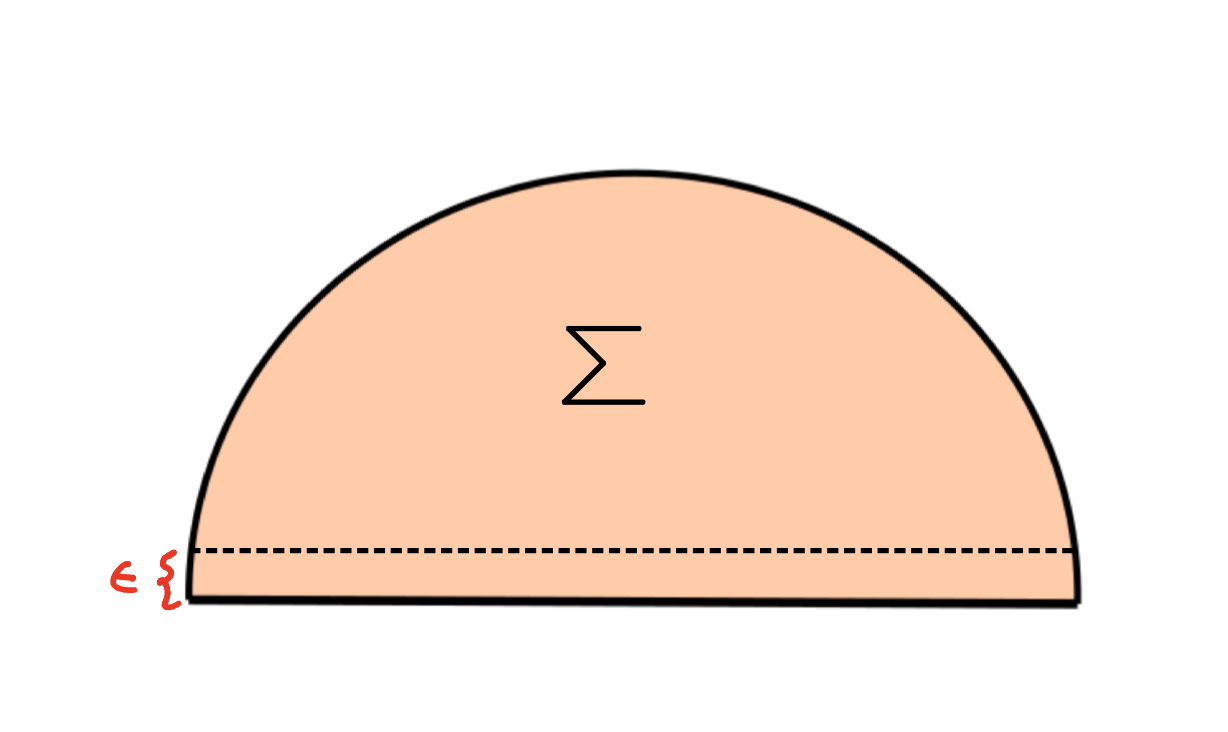}
		\caption{}
	\end{subfigure}
	\hfill
	\begin{subfigure}{0.30\textwidth}
		\centering
		\includegraphics[scale=0.15]{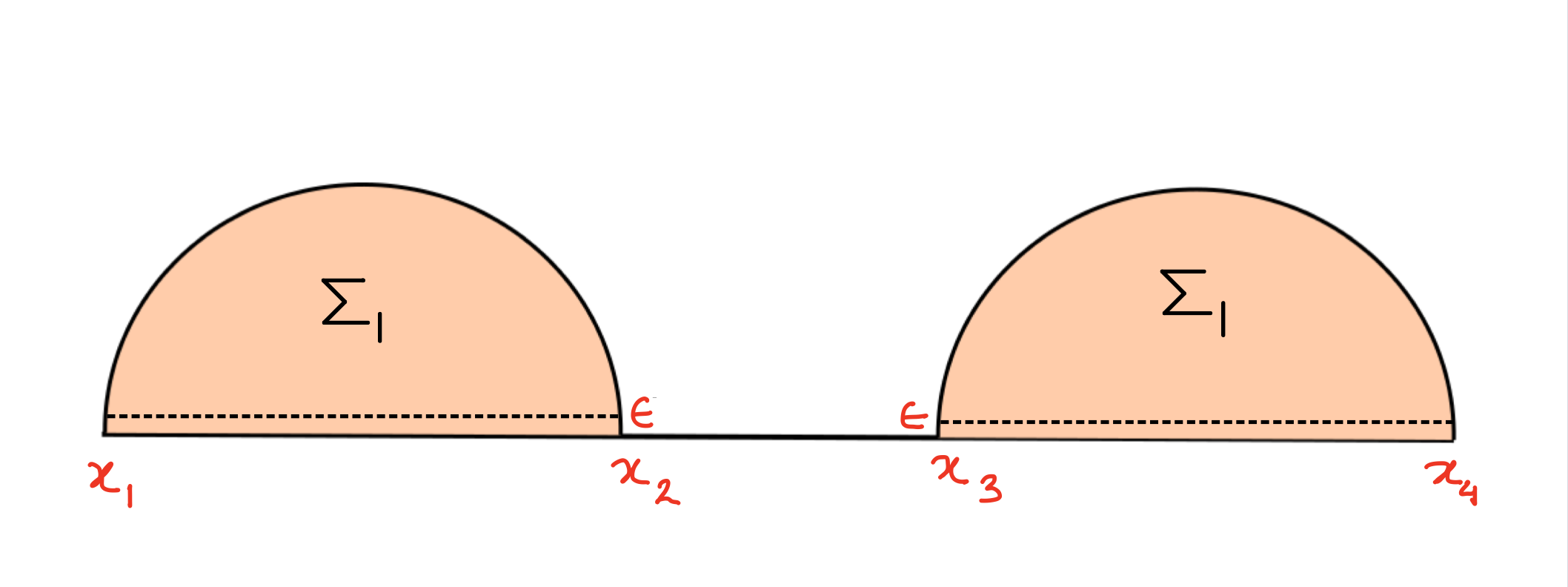}
		\caption{}
	\end{subfigure}
	\hfill
	\begin{subfigure}{0.30\textwidth}
		\centering
		\includegraphics[scale=0.15]{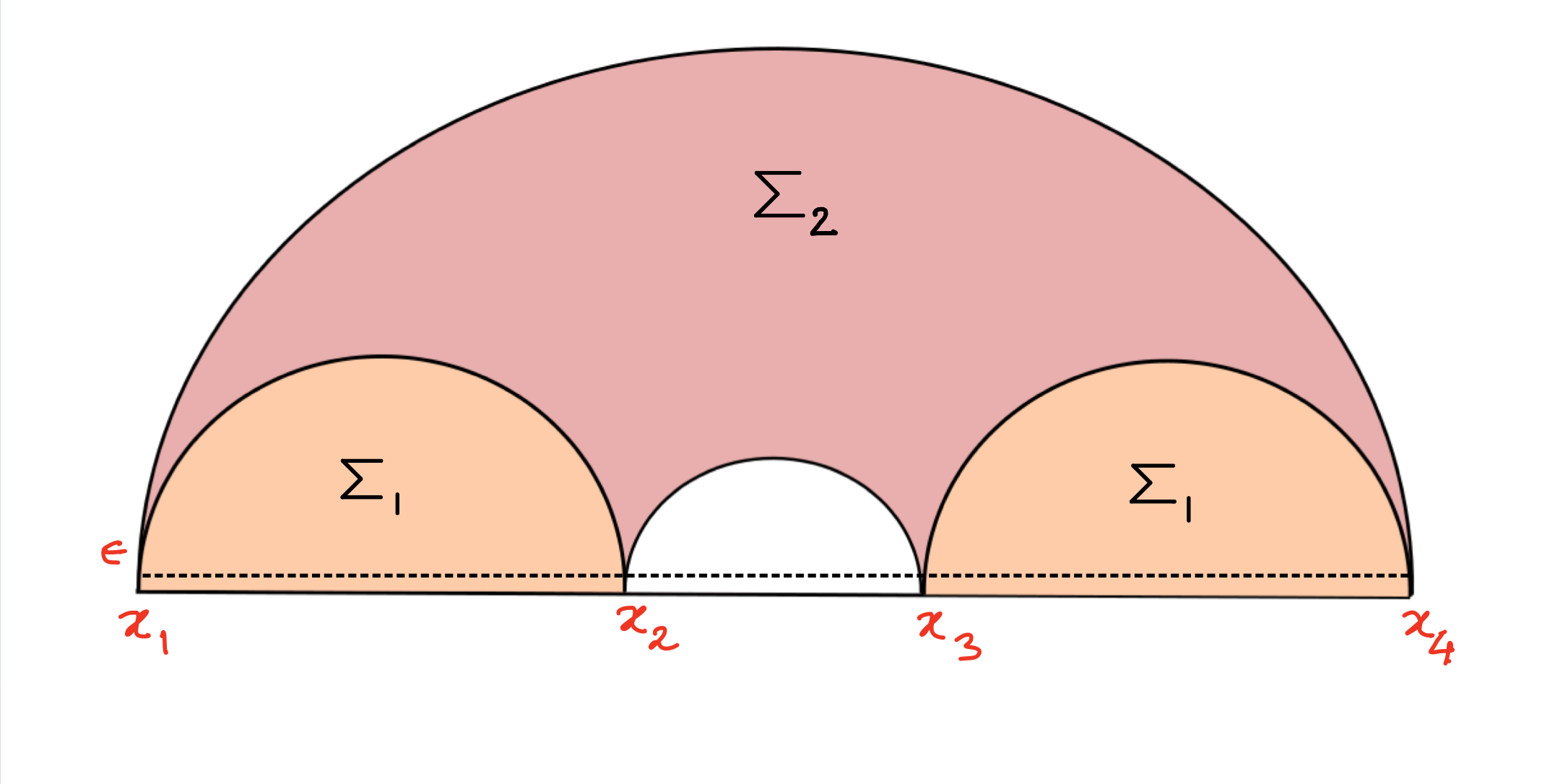}
		\caption{}
	\end{subfigure}
	\caption{HRT surfaces and entanglement wedges in $\AdS_3$ for one (a) and two intervals. The latter has two phases, (b) Phase I and (c) Phase II.}
	%\hfill
\end{figure}

\subsubsection{Volumes in Multiboundary Wormholes: }\label{multvol} Let us now consider the multiboundary wormhole model. We are interested in the evolution of subregion complexity associated with the Hawking radiation during the evaporation. In the toy model of \cite{Akers:2019nfi} the evaporation is described by an initial large black hole regurgitating smaller black holes, which represent the Hawking quanta. For simplicity, all such black holes are considered to be placed in their own separate asymptotically-$\AdS$ space-time. As the evaporation proceeds, the asymptotically-$\AdS$ regions are connected by a wormhole with an increasing number of exits. The subregion we are concerned with is the union of all the smaller exits at one instant of time.
\par As described in \cite{Akers:2019nfi} and \ref{multimodel} above, there are two competing HRT surfaces for the sub-region of our choice, viz. $\cup_{j=2}^{n} \ell_{j}$ and $L_0$. The corresponding entanglement wedges have been illustrated in figure \ref{fig: nbdy}. At the Page transition, the entanglement wedge changes which results in a constant shift of complexity.

\textbf{a) Three boundary model:}
As explained before, we fix the two smaller boundaries by identifying the pair of geodesics that are not concentric. We assume these two semicircles to be of the same radius in our consideration, as shown in \ref{fig:pants_construction}. In addition, we also assume that the corresponding throat horizons are of the same length. This assumption constrains the choice of parameters in the fundamental domain in a particular way as mentioned in \cite{Balasubramanian:2020hfs, Li:2020ceg}. The relation is between center of the non-concentric semicircles. Among these two, let the center of the semicircle near $x=0$ be $c_1$ and the other one be $c_2$. Note that these are the semicircles removed to create a three-boundary wormhole from the two-boundary case. Recall that for the latter, one has to identify two concentric semicircles in the UHP. These two concentric semicircles have their center at $x=0$ and their radii $R_{0}$ and $\mu^{2}R_{0}$ respectively, where $\mu>1$. For the three-boundary evaporating model, we have $\mu\geq1$, which saturates at the end of the evaporation process. The relation constrained by the fact that the two throat horizons are of equal length is $c_{2}=\mu c_{1}$. Let us label the radii of these two semicircles as $R_{1}$ and $R_{2}$. In this paper, we work with the particular choice $R_{1}=R_{2}=R$. \footnote{In \cite{Li:2020ceg}, the authors assumed $R_{2}=\mu R_{1}$. But in that case, one ends up with negative volumes for the smaller exits, which is unsatisfying physically.} We also make the following choice for $c_1$, and thus also for $c_2$, motivated by \cite{Li:2020ceg},
\begin{equation}\label{centchoice}
c_{1}= \frac{\mu+1}{2} R_{0},\,\qquad c_{2}=\mu c_{1}.
\end{equation}
Our parameter choices secure positivity of the volumes of the smaller exits for all times, as desired. We should nevertheless keep in mind that among these two equal throat horizons, one is connected whereas the other is disconnected according to the construction, see figure \ref{fig:pants_construction}. Let us call the connected one $L_{1}$ and the disconnected one $L_{2}=L_{2}^{L}+L_{2}^{R}=L_{1}$, where the superscripts stand for left and right. It is easy to see that once $R_{0}$ is specified and we assume that with time $L_{1}$ and $L_{2}$ increase, while the primarily bigger vertical throat horizon $L_{0}$ keeps decreasing via $L_{0}^{\prime}=\sqrt{L_{0}^{2}-2L_{1}^{2}}$, the only time dependence left to be solved for a consistent construction is the time dependence of $R$. In this case, we replace time by the increasing length $L_{1}$ (or equivalently $L_{2}$) and plot the volumes with increasing $L_{1}$. There are two solutions of $R=R(L_{1})$. Ideally $R$ should also depend upon $\mu$. But since $L_{0}^{\prime}$ can be written either simply in terms of $L_{1}$ or equivalently in terms of $\mu$, there is a relation between these two, $\mu=e^{\frac{\sqrt{L_{0}^{2}-2L_{1}^{2}}}{2}}$, with $L_{0}$ chosen to be a constant (the starting length of the vertical throat horizon).

The expressions for $L_{1}$ and $L_{2}$ are the following once the equality constraint, hyperbolicity condition \footnote{This is mentioned in section \ref{HyperbolicPlane},\ref{HyperbolicElements}.} and the equation \eqref{centchoice} are used 
\begin{multline}
    L_{1}= \log \left[\cot \left[\frac{1}{2} \arcsec\left(\frac{\mu ^2-1}{\sqrt{\left(\mu ^2-1\right)^2-16 R^2}}\right)\right]\right]\\ -\log \left[\tan \left[\frac{1}{2} \arcsec\left(\frac{\mu ^2-1}{\sqrt{\left(\mu ^2-1\right)^2-16 R^2}}\right)\right]\right]
\end{multline}
\begin{multline}
    L_{2}= \log \left[\cot \left[\frac{1}{2} \arcsec\left(\frac{\mu  \left(\mu ^2-1\right) \sqrt{\mu ^2 \left(\left(\mu ^2-1\right)^2-16 R^2\right)}}{\mu ^6-2 \mu ^4+\mu ^2-8 \left(\mu ^2+1\right) R^2}\right)\right]\right]\\ -\log \left[\tan \left[\frac{1}{2} \arccos\left(\frac{\mu ^5+\mu -2 \mu ^3 \left(4 R^2+1\right)-8 \mu  R^2}{\left(\mu ^2-1\right) \sqrt{\mu ^2 \left(\left(\mu ^2-1\right)^2-16 R^2\right)}}\right)\right]\right]
\end{multline}

Given the above expressions of $L_{1}$ and $L_{2}$, we solve for $R$ asking for a linear growth of $L_{1}$ so that we can use it as an analogue of time.\footnote{It is important to note that solving this $R$ for given parameter choices is just for exactness and calculation of volume. In general, for any constant or functional dependence of $R$, $L_{1}$ and $L_{2}$, although they apparently look different, scale in the exactly similar way with $\mu$.} There are two solutions, both of which feature positive volumes for any instance of time, as required by consistency, in particular of the fundamental domain. It is also easy to check that for both of the solutions, $L_{1}$ and $L_{2}$ are indeed equal to each other. %The volumes being positive definite for both the smaller exits is satisfying from the perspective of the well behaved fundamental domain. 

%\begin{figure}[h!]
	%\centering
	%\includegraphics[scale=1.0]{3bdybasic.pdf}
	%\caption{The $t = 0$ slice of the three-boundary wormhole with parameters pointed.}
	%\label{figs:3bdybasic}
%\end{figure}

\par Just to be precise, let us mention the volumes of the smaller exits at any particular instant in terms of the parameters of fundamental domain. 

\begin{equation}
    V_{1}= \frac{\left(c_{1}-R-R_{0}\right)+\left(\mu^{2} R_{0}- c_{2}-R\right)}{\epsilon},
\end{equation}
and
\begin{equation}
    V_{2}= \frac{\left(c_{2}-c_{1}-2R\right)}{\epsilon}
\end{equation}
where $\epsilon$ is again a UV cutoff. The total volume is simply $V=V_{1}+V_{2}$. At the Page time, when the minimal surface corresponding to the union of the smaller exits changes from $L_{1}+L_{2}$ to (the decreased) $L_{0}^{\prime}$, a volume is added to the previous volume $V$. We will come back to this point in the next subsection and where we present plots of the volumes corresponding to the two solutions of $R=R(L_{1})$.

\textbf{b) $n$-boundary model:} This is a good time to explain how we wish to perceive black hole evaporation á la \cite{Akers:2019nfi} from the quotient perspective with more details about the explicit construction. Recall figure \ref{fig: threebdy}. We start with three exits and at each time-step include two more geodesics with opposite orientations, which upon identification provide a new boundary. For simplicity, we consider all semicircles to have the same radius at any moment. The radius is thus a function of the number of exits, which is an analogue of discretized time.
\par The moduli space of an $n$-boundary wormhole contains $n$ physical parameters that characterize the system. These are the periodic geodesics between two identified semicircles in our quotient picture. Consider the $3$-boundary construction in figure \ref{fig:pants_construction}. The dashed lines denote the geodesics which after performing proper identification become closed and the metric outside the causal development of these closed curves is the BTZ metric \cite{Skenderis:2009ju}. Thus the periodic geodesics can be identified as black hole horizons and in fact constitute the candidate HRT surfaces in the evaporation model. In figure \ref{fig:pants_construction} we have denoted the identification of each geodesic with the corresponding BH horizon for the 3-boundary wormhole.
\par The sub-region complexity is essentially determined by the volume under the horizons. Before Page time, it is the volume under $\cup_{j=2}^{n} \ell_{j}$ while after Page time it is that under $L_{0}^{\prime}$ as marked in figure \ref{fig: nbdy}. The explicit formulae for the volumes are given below. Here we only point out that they depend on the radii of the semicircles and the length of the horizons. The horizon lengths are in general difficult to compute, the authors of \cite{Caceres:2019giy} provide two of the three lengths for the $3$-boundary wormhole
\begin{align}
L_0 &= \adsL \log \left(\mu^2\right), \label{hlength1}\\
L_1 &= 2\adsL \, \,\text{arcsinh} \left[\sqrt{\left(\frac{d}{R}\right)^2-1}\right], \label{hlength2}
\end{align}
but an analytic answer for $L_2 = L_2^L \cup L_2^R$ remains elusive. Here $\adsL$ is the $\AdS$ radius and $d$ is the distance between the centers of the orientation reversed semicircles, other parameters are explained in figure \ref{fig:pants_construction}. After identification, $L_0$ becomes the horizon of the parent black hole. Throughout the calculation, we shall follow the footsteps of \cite{Akers:2019nfi} and assume all smaller horizons have equal length $L_1$.
\par In our model, we demand that all smaller semicircles have identical radii, $R$, at any moment in time. Since we accommodate an increasing number of semicircles, hence also boundaries, in the same region as time progresses, $R$ cannot remain constant. Also, starting from the three-boundary wormhole, as we increase number of boundaries, the distance (say $d_{1}$) between the centers of the semicircles are managed in a way to make sure that all the other disconnected throat horizons, except for the one that is attached to the concentric semicircles, have the same length as the connected one between the first set of  orientation reversed semicircles. Therefore, in our model, we make sure that out of the $\left(n-1\right)$ smaller exits, $\left(n-2\right)$ have the same horizon length and only the remaining one is assumed to have a constrained equality. 
There is no way to fix the time dependence of $R$ explicitly. We can however assert that it must satisfy the constraint
\begin{equation}\label{evaproto}
R\left(n\right) < \frac{\left(\mu^2\left(n\right)-1\right)}{4\left(n - 1\right)}R_0\,,
\end{equation}
where $n$ denotes the number of smaller exits. This constraint makes sure that the adequate number of semicircles are accommodated within the interval $\left(\mu^2\left(n\right)-1\right)R_0$.
\par Choosing a good function, one that satisfies \eqref{evaproto}, we can determine the volume and complexity through the help of \eqref{hlength1}. As we will see, it exhibits a finite discontinuity at the Page transition. The source of this discontinuity is purely topological, which we explain in section \ref{gbhp}. Thereafter we give explicit formulae for the volumes and show complete evolution of complexity during entire evaporation process.

\subsubsection{Gauss-Bonnet \& hyperbolic polygons}\label{gbhp} As explained in subsection \ref{sec:volumesAdS3}, the Gauss-Bonnet theorem plays a central role in the calculations of volumes in AdS$_{3}$. Here, we discuss another consequence of the Gauss-Bonnet theorem \eqref{cmplxty}, which regards the computation of the area of hyperbolic triangles. 

\begin{corollary}
Consider a 2d hyperbolic surface. Let it be tessellated by triangles with angles $\left(\alpha, \beta, \gamma\right)=\left(\frac{2\pi}{p},\frac{2\pi}{q}, \frac{2\pi}{r}\right)$. Then the Gauss-Bonnet theorem along with the triangle group imply the following relation% for 2d hyperbolic space
\begin{equation}
    \frac{\pi}{p}+\frac{\pi}{q} +\frac{\pi}{r} < \pi.
\end{equation}
The area of the hyperbolic triangle therefore becomes $|(\pi - \alpha -\beta-\gamma)|\adsL^{2}$, where $\adsL$ stands for an intrinsic length scale, which is the AdS radius.
\end{corollary}
We choose $\adsL=1$ for the remainder of this section. Next, we aim for the computation of the volumes%
\footnote{Strictly speaking, our volumes are of course areas, but we stick with conventional terminology of higher dimensional geometries.} 
of the different kinds of causal shadow regions that crossed our way when contemplating multi-boundary wormholes.
%The next thing we are going to look at is the compute the areas (or in terms of our previously used language, volume) of the different kinds of causal shadow regions that we came across in our multi-boundary models.
As explained previously, these regions correspond to the analogue of islands in our models. In the following, we describe a simple way to compute such volumes in two-dimensional hyperbolic space. In the following, we will only make use of the above-mentioned area of hyperbolic triangle to compute area of any hyperbolic polygon in two dimensional hyperbolic space.

\textbf{A general look into causal shadows:} Let us first point out to the reader that the causal shadow volumes that are added to the entanglement wedge of the radiation subsystem after the Page time, both in case of the three-boundary as well as the $n$-boundary model, are hyperbolic polygons in general.
 
\begin{observation}
For the three-boundary case, the region is a hyperbolic octagon, where as for the $n$-boundary scenario, the region is a hyperbolic $4\left(n_{Page}-1\right)$-gon. $n_{Page}$ stands for the $n$-value at which the Page transition, or in case of volume, the wheel-eyeglass phase transition\cite{Peach:2017npp}, occurs. 
\end{observation}

Therefore, the first thing to understand is that in case of $n$-boundary model, the structure of the causal shadow depends upon the Page time. Now, let us understand the volumes of general hyperbolic polygons in terms of hyperbolic triangles. Firstly, we discuss the three-boundary causal shadow and then generalize it to general number of boundaries.

\begin{figure}[t]
	\begin{subfigure}{0.4\textwidth}
		\centering
		\includegraphics[scale=0.30]{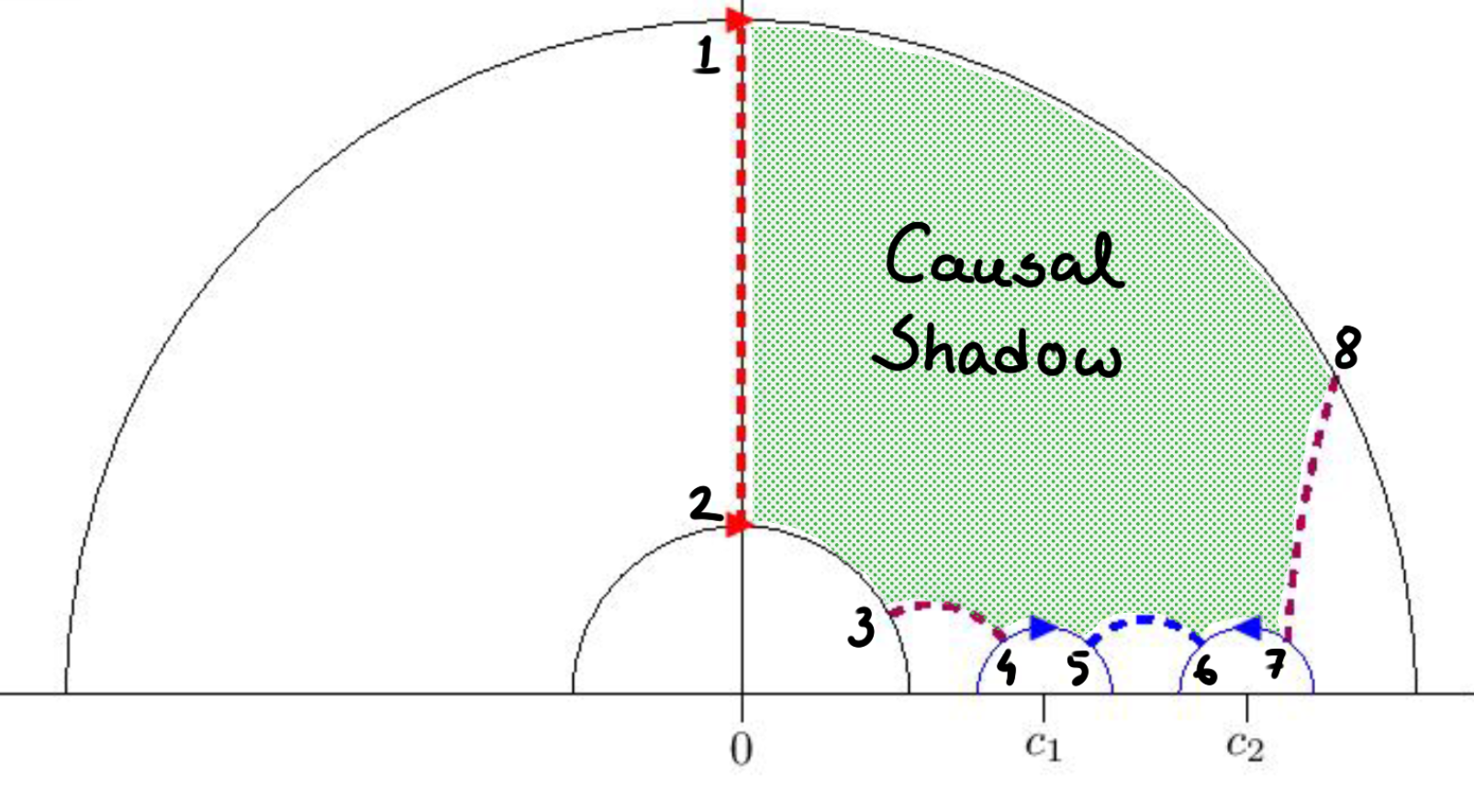}
		\caption{}
		\label{fig: octagon1}
	\end{subfigure}
	\hfill
	\begin{subfigure}{0.40\textwidth}
		\centering
		\includegraphics[scale=0.30]{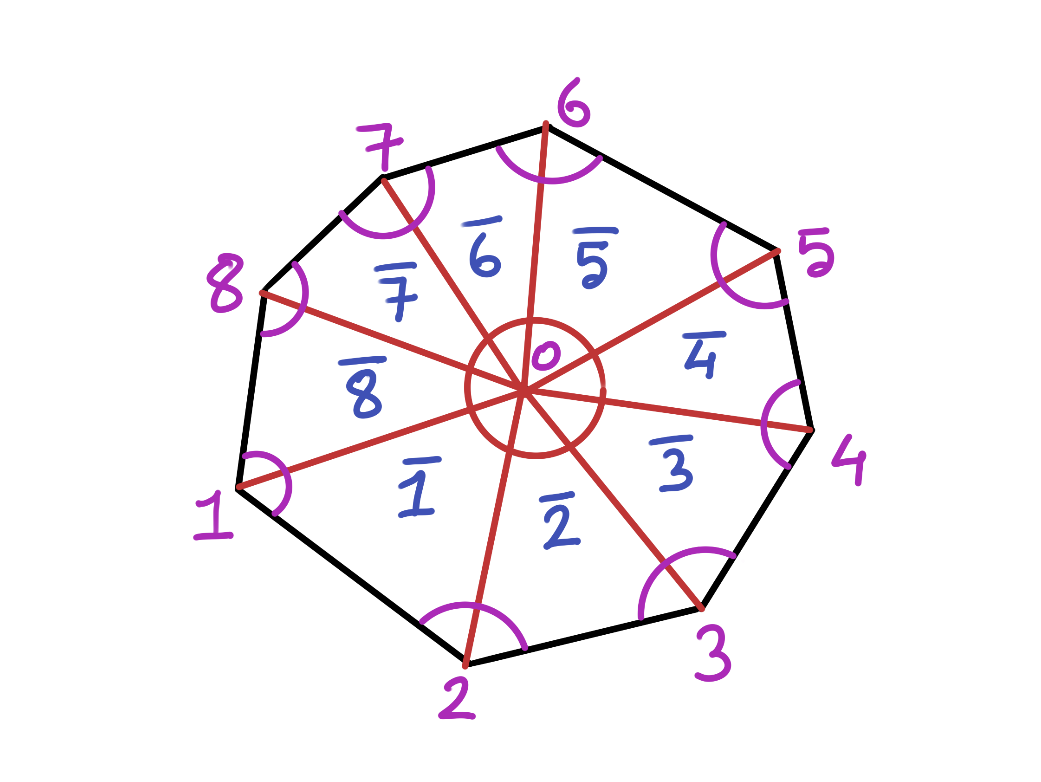}
		\caption{}
		\label{fig: octagon2}
	\end{subfigure}
	\caption{Hyperbolic octagon and Causal Shadow in three-boundary wormhole model.}
	%\hfill
\end{figure}

\textbf{Hyperbolic octagon:}
First, we discuss the three-boundary case. In this case, as mentioned in \ref{multvol}, the minimal surface change gives an additional contribution to the volume of the radiation subsystem. Now, from the Figure \ref{fig: octagon1}, we can see that this is the causal shadow region. For the three-boundary case, as has been marked in the figure, there are eight vertices constructing a hyperbolic polygon. In general it can have any volume depending on the nature of the edges of the polygon. However in our case, we easily see that at each vertex at least one of its edges is always a geodesic (throat horizon) in the fundamental domain of the three-boundary wormhole. Now, any bulk curve or geodesic in the fundamental domain is bound to hit the boundary of the domain with a corner angle $\frac{\pi}{2}$.\footnote{Another way of understanding these bulk geodesics and the corner angles is as entanglement wedge cross sections as pointed out in \cite{Bao:2018fso, Bhattacharya:2020ymw} and as proved in \cite{Nguyen:2017yqw} using Klein coordinates.}

Knowing the corner angles, we can use the formula for the area of the hyperbolic triangle in computing the area of the hyperbolic octagon by dividing it into eight triangles as shown in Figure \ref{fig: octagon2}.

\begin{figure}[t]
	\begin{subfigure}{0.48\textwidth}
		\centering
		\includegraphics[width=\textwidth]{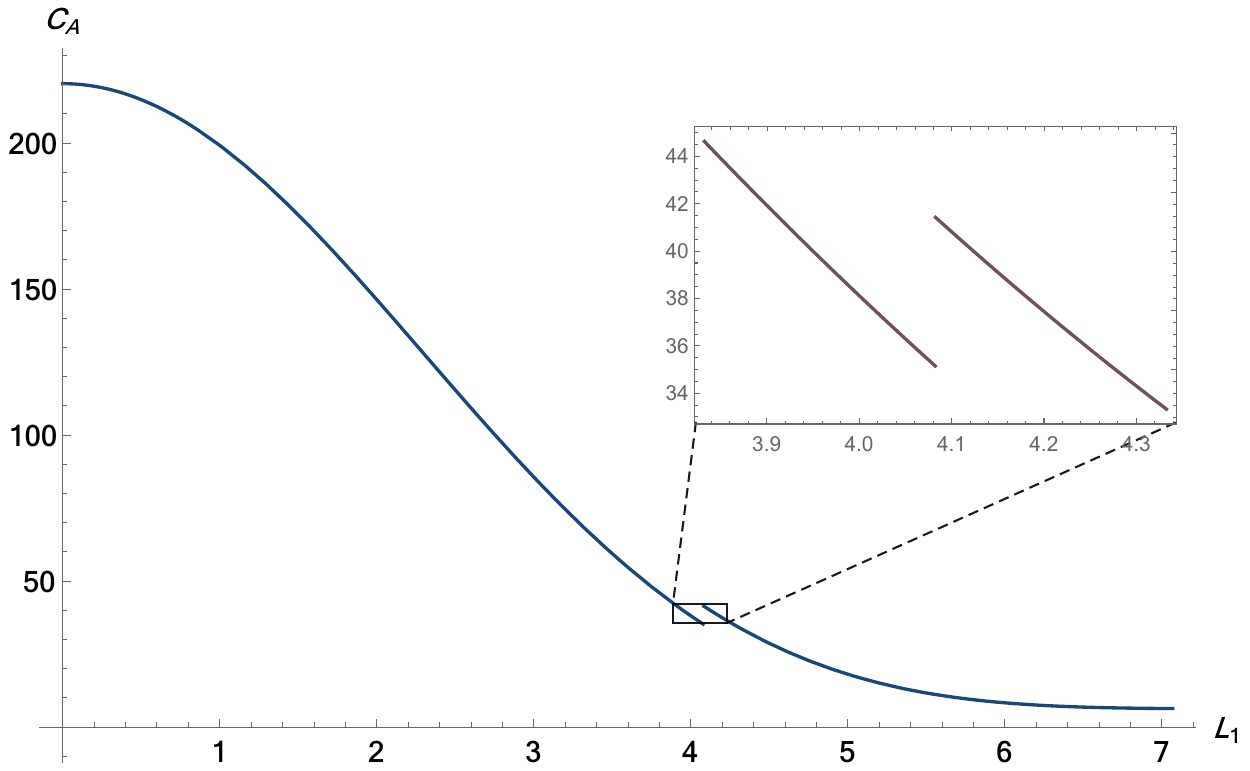}
		\caption{}
	\end{subfigure}
	\hfill
	\begin{subfigure}{0.48\textwidth}
		\centering
		\includegraphics[width=\textwidth]{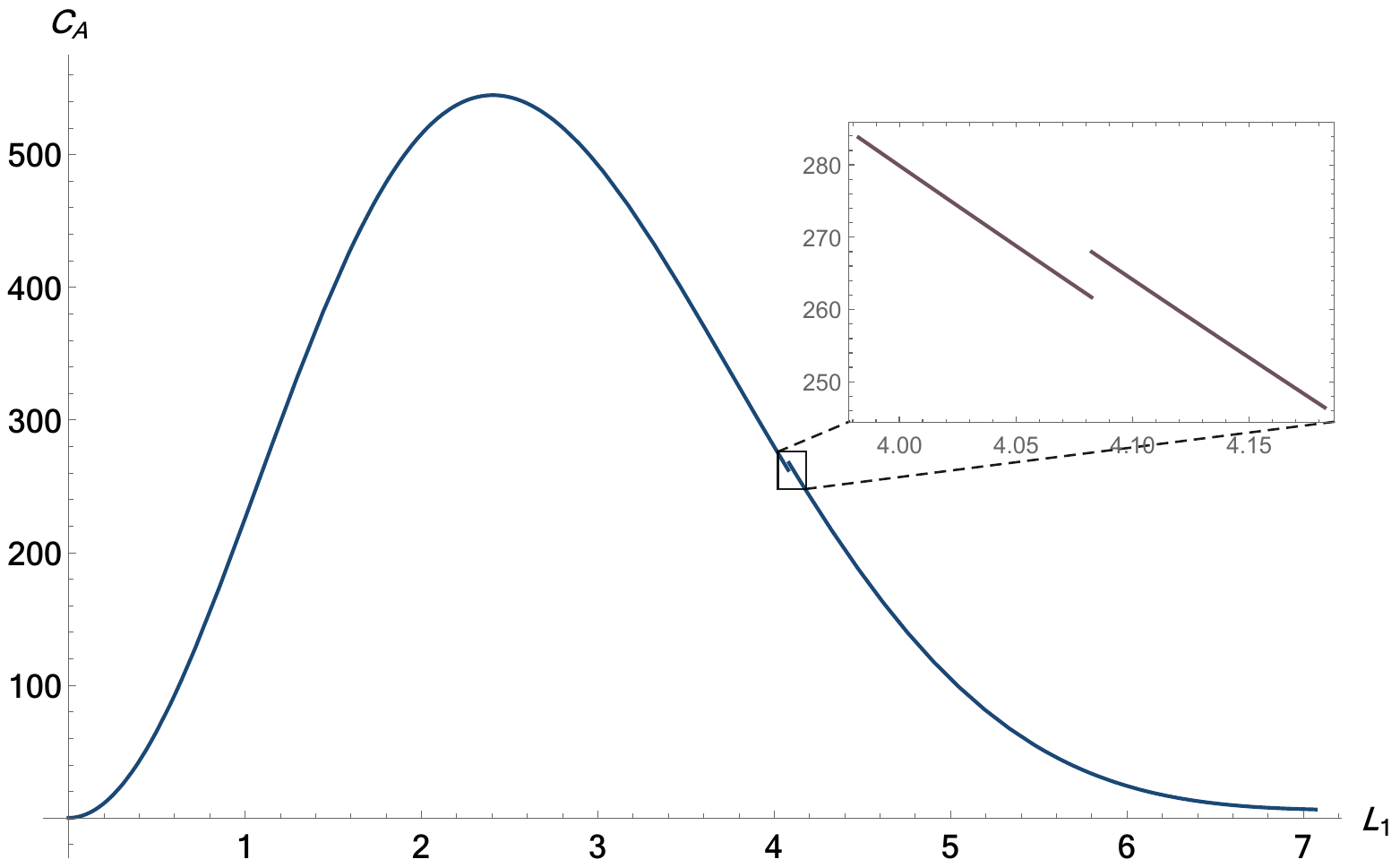}
		\caption{}
	\end{subfigure}
	\caption{Complexity plots of $3$ boundary island model for two choices of R}
	\label{fig: 3bdycomp}
\end{figure}

\begin{observation}
The vertices of the octagon are marked by the numbers $i=1,2,...,8$ and the eight triangles that we divide this octagon into have a common vertex $0$. The sum of all angles joined at the center $0$, we call these $\sphericalangle i0j$ with $i,j=1,2,...,8$, is of course $2\pi$. This allows for a simple derivation of the octagon's volume,
 
 \begin{multline}
     \text{Area of the octagon} (\Delta V_{(3)})= \sum_{i,j(i\neq j)} \Delta (i0j)= \sum_{\overline{I}} \Delta (\overline{I}) \, \, \,, \, (\overline{I}=1,2,..,8)\\ = 8\pi- \sum_{i,j(i\neq j)}{ \sphericalangle i0j} - \sum{\text{Corner angles}}= 8\pi -2\pi- (8 \times \frac{\pi}{2})= 2\pi.
 \end{multline}
\end{observation} 
 Hence, the area of the hyperbolic octagon is constant in our case and the volume experiences a jump of $2\pi$ at the Page transition (wheel-eyeglass phase transition). In Figure \ref{fig: 3bdycomp}, we have shown the volume vs time plots for the two solutions of $R$ (time dependent radius of the non-concentric pair of semicircles) as mentioned in \ref{multvol}.
 
\textbf{Hyperbolic $m$-gon:}
Now we generalize our previous computation for any general $m$-gon of the given kind, i.e; the corner angles being $\frac{\pi}{2}$. In this case as it turns out again, we can divide it into $m$ hyperbolic triangles and the area simply becomes,
\begin{equation}
    \text{Area of $m$-gon} = m \pi -2\pi-m\frac{\pi}{2} = \pi(\frac{m}{2}-2).
\end{equation}
 Now for a given $n$-boundary wormhole, we find that the value of $m$ becomes $m=4(n-1)$. Therefore, for the $n$-boundary wormhole, the volume that is added at the Page transition becomes,
 \begin{equation}\label{jump}
     \text{Jump in volume:}\,\Delta V_{(n)}= \left[2\left(n_{Page}- 1\right)-2\right]\pi = \left(2n_{Page}-4\right)\pi.
 \end{equation}

 \begin{figure}[t]
 	\begin{subfigure}{0.48\textwidth}
 		\centering
 		\includegraphics[width=\textwidth]{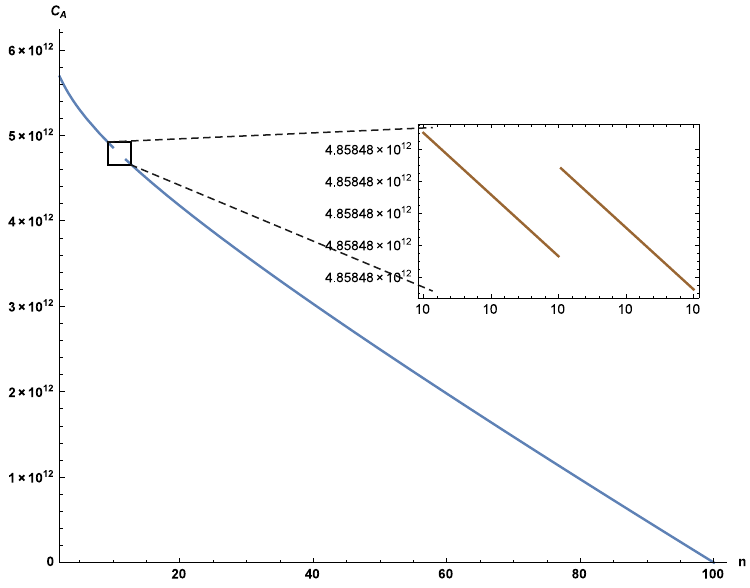}
 		\caption{$R\left(n\right) = R_a\left(n\right)$}
 	\end{subfigure}
	\hfill
	\begin{subfigure}{0.48\textwidth}
		\centering
		\includegraphics[height = 5.7 cm, width=\textwidth]{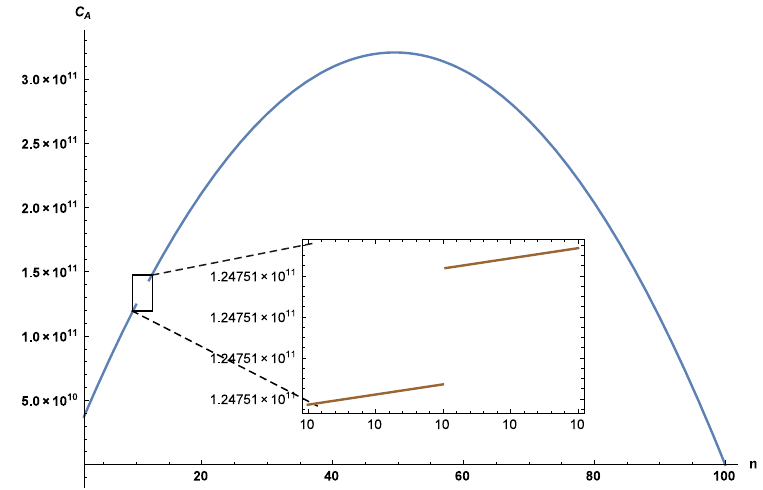}
		\caption{$R\left(n\right) = R_b\left(n\right)$}
	\end{subfigure}
	\caption{Complexity plots of $n$-boundary island model for two choices of R}
	\label{fig: nbdycomp}
\end{figure}
Hence, we find that for the $n$-boundary model, the jump in volume depends on the Page time whereas for the three-boundary model, it does not. For the three-boundary case, there is no topology change in the process of evaporation and therefore, the previous result, $2\pi$, is recovered by setting $n_{Page}=3$.
\par The full evolution of holographic complexity during the evaporation is illustrated in figure \ref{fig: nbdycomp} for two different choices of $R(n)$, both of which obey \eqref{evaproto}. While there exists a large pool of choices for $R(n)$ producing distinct plots, we emphasize that all of them have similar qualitative features as either of our two choices for $R(n)$,
\begin{align}\label{Rchoices}
	R_a\left(n\right) &= \frac{100-n}{(2n + 2)^{1.05}} \nonumber \\
	R_b\left(n\right) &= \left(100-n\right)\times 10^{-3}
\end{align}
Indeed, our choices seem to have been conjured out of the blue. They are not completely ad-hoc, though. While certainly not unique; these are two of the many choices which ensure that at any moment all circles can be sufficiently accommodated in the fundamental domain as well as generate physically meaningful plots of the volume. In absence of any true analytical time-dependence of $R$, these are our best guides to fix a meaningful evolution of complexity. Further, we assumed that the disconnected throat horizons were equally divided into two halves with each of them having length $\frac{L_1}{2}$, where $L_1$ is the length of the sole connected horizon and we have already assumed all horizons to have the same length. Under these assumptions and with the help of equations \eqref{hlength1} and \eqref{hlength2}, we can express the volume associated with each smaller horizon as
\begin{equation}
	V_{initial} = \frac{4(n-1) \left(\cosh \left(\frac{\ell}{4}\right)-1\right)  R\left(n\right)}{\epsilon}+\frac{2 \left(\cosh \left(\frac{\ell}{2}\right)-1\right) R(n)}{\epsilon}
\end{equation}
This is the volume that goes into the complexity before Page time, after the Page transition there's a constant addition \eqref{jump} to the volume. The figures clearly display these required features.

%%%%%%%%%%%%%%%%%%%%%%%%%%%%%%%%%%%%%%%%%%%%%%%%%%%%%%%%%%%%%%%%%%%%%%%%%%%%%%%%
\section{Islands and Kinematic Space:}\label{kinematic space}
\label{sec: kinematicSpace}
%%%%%%%%%%%%%%%%%%%%%%%%%%%%%%%%%%%%%%%%%%%%%%%%%%%%%%%%%%%%%%%%%%%%%%%%%%%%%%%%
In this section we employ kinematic space to obtain an understanding of the quantum information involved in the volume of an island. We explain how island volumes are computed in kinematic space and combine this with our prior results to derive integral identities of trigonometric functions. We begin with an introduction of kinematic space geared towards wormholes in subsection \ref{sec: kinSpaceIntro}. In subsection \ref{sec:kinspaceVolumes} we explain how kinematic space captures volumes. In subsection \ref{sec: kinSpaceCausalShadow} we apply these tools to a causal shadow in $\ads_3$ and derive a first integral identity. Finally, in subsection \ref{sec: kinSpaceIslands} we apply kinematic space islands and derive another integral identity.

%%%%%%%%%%%%%%%%%%%%%%%%%%%%%%%%%%%%%%%%%%%%%%%%%%%%%%%%%%%%%%%%%%%%%%%%%%%%%%%%
\subsection{A crash course on kinematic space}
\label{sec: kinSpaceIntro}
%%%%%%%%%%%%%%%%%%%%%%%%%%%%%%%%%%%%%%%%%%%%%%%%%%%%%%%%%%%%%%%%%%%%%%%%%%%%%%%%
Kinematic space is an intermediate geometry between the gravity side and the CFT side of the gauge/gravity correspondence. Its power resides in its aptitude to translate geometric properties of the bulk theory into information theoretic objects in the boundary theory \cite{Czech:2015qta, Czech:2016xec}. 

We work with static, asymptotically $\ads_3$ spacetimes $\cM$, which satisfy
\begin{equation}\label{eq: AdS}
	ds^2
	\sim
	-\frac{r^2}{\adsL^2}dt^2+\adsL^2\frac{dr^2}{r^2}+r^2d\phi^2
	\qquad
	\text{as}
	\quad
	r\to \infty\,,
\end{equation}
where $\phi\sim\phi+2\pi$ is an angular variable and $\adsL$ is the AdS radius. Note that in this section we work with global $\ads_3$ rather than the Poincar\`e patch. This will not pose a problem however, as our intent is to extract statements on the connectivity of wormhole geometries, and these do not depend on which patch we quotient to obtain a wormhole geometry.   

For a fixed value of time $t=const$, the kinematic space $\cK$ associated with \eqref{eq: AdS} is the space of all boundary anchored, oriented geodesics. In pure $\ads_3$ any tuple of boundary points $(u,v)$ is associated uniquely%
\footnote{In quotient geometries this need not be the case.}
with one geodesic and hence with a point in $\cK$. It is convenient to introduce another set of coordinates on $\cK$,
\begin{equation}\label{KinSpaceCoordinates}
    \theta = \frac{1}{2}(v+u)\in[0,2\pi),
    \qquad
    \alpha = \frac{1}{2}(v-u)\in[0,\pi]
%	u = \theta - \alpha\,,
%	\quad
%	v = \theta + \alpha\,.
\end{equation}
The intuition for these coordinates is as follows. The tuple $(u,v)\in\cK$, with $v>u$, naturally delimits a boundary interval $[u,v]$, i.e. a CFT subregion. The center of this subregion is given by $\theta$ and $\alpha$ is the interval's opening angle. A point $(\theta,\alpha)\in\cK$ and $(\theta+\pi,\pi-\alpha)$ encode the same geodesic, albeit with reversed orientation. 

Entanglement entropy $S(u,v)$ in a CFT, being dependent on a boundary interval, naturally becomes a function on $\cK$. It plays a prominent role, as it induces a metric and a volume form on $\cK$ \cite{Czech:2015qta}
\begin{subequations}\label{eq: kinSpaceGeometry}
\begin{align}
	\label{eq: ds kin. sp.}
	ds^2_\cK &= \partial_u\partial_vS\,du\,dv
	%= \frac{1}{2}(\partial^2_\theta-\partial^2_\alpha)S\,(-d\alpha^2+d\theta^2)
	=-\frac{1}{2}(\partial^2_\alpha S)\,(-d\alpha^2+d\theta^2)
	\,,
	\\
	\label{eq: omega}
	\omega
	&= \partial_u\partial_v S\,du\wedge dv
	%= \frac{1}{2}(\partial^2_\theta-\partial^2_\alpha)S\,d\theta\wedge d\alpha
	= -\frac{1}{2}(\partial^2_\alpha S)\,d\theta\wedge d\alpha
	\,.
\end{align}
\end{subequations}
In the second equality on each line we imposed rotational symmetry $S(\theta,\alpha)=S(\alpha)$.
\begin{definition}
 The two-form $\omega$ is called the Crofton form. It is a volume form on $\cK$.
\end{definition}
\begin{observation}
 The Crofton form \eqref{eq: omega} is a measure on $\cK$ and invariant under the isometries of the hyperbolic plane \cite{Czech:2015qta}. The Crofton form is a line density, i.e. it associates a measure to each geodesic, similar to how $\sqrt{|g|}\,d^dx$ associates a measure to each point on a manifold.
\end{observation}
We stress that even though we started out with a fixed time slice of $\ads_3$, which is a Euclidean manifold, its associated kinematic space is Lorentzian and its light-cone coordinates are given by $u$ and $v$. In fact $\cK$ is de-Sitter spacetime \cite{Czech:2016xec}.

Being Lorentzian, $\cK$ naturally carries a causal structure. Any interval $[u_1,v_1]$ lies in the past of $[u_2,v_2]$ if $[u_1,v_1]\subset[u_2,v_2]$. This implies that the point $(u_1,v_1)\in\cK$ lies in the backward lightcone of $(u_2,v_2)\in\cK$. We anticipate that any pair of geodesics which we identify in order to obtain a wormhole, needs to be time-like related \cite{ZhangChen}. A geodesic $(\theta,\alpha)\in\cK$ and its orientation reversal $(\pi+\theta,\pi-\alpha)$ are space-like related. Note that the CFT spacetime is also represented in $\cK$, since for any $(\theta,\alpha)$ the limit $\alpha\rightarrow0$ shrinks the geodesic to the point $\theta=u=v$ on the boundary of $\ads$. 

It should be clear that kinematic space $\cK$ can be constructed for any CFT as space of subregions $[u,v]$ without invoking holography. In this case we may still think of an auxiliary $\ads$ spacetime in which each $[u,v]$ is associated with a geodesic $(\theta,\alpha)$. If however, the CFT is holographic, then this auxiliary $\ads$ is promoted to the actual geometry of the gravity dual. Of course, for holographic CFTs, $S$ also measures the length $\ell(u,v)=S(u,v)/4G_N$ \cite{Ryu:2006bv} of a boundary anchored geodesic. Because $S(u,v)$ essentially determines $\cK$, cf. \eqref{eq: kinSpaceGeometry}, we establish that $\cK$ acts as intermediary geometry between CFT and $\ads$ as claimed at the beginning of this section. %translates information theoretic properties of the CFT into the gravity dual.

\subsection*{Lengths in AdS as integrals in $\cK$}
\begin{figure}[t]
\begin{center}
\includegraphics[scale=0.4]{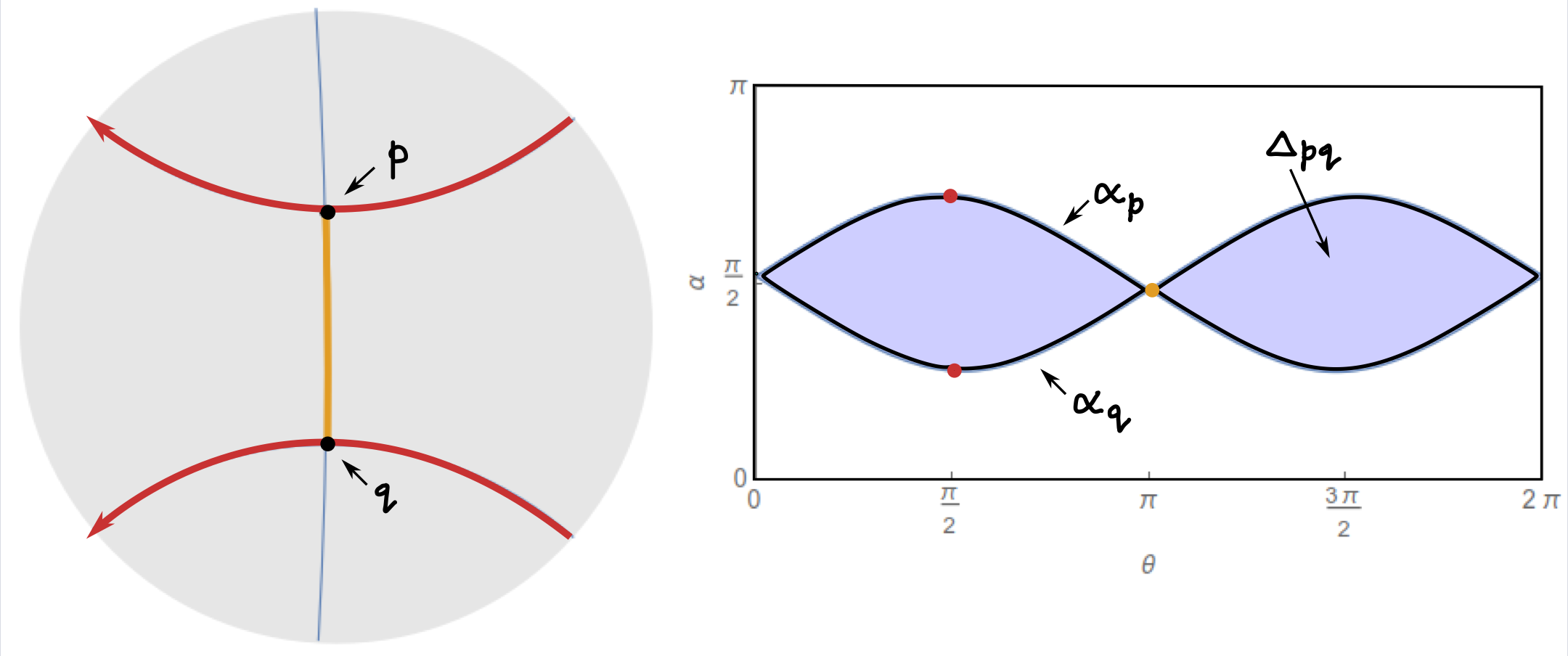} 
\end{center}
\caption{Left: A BTZ black hole generated in global $\ads_3$ through identification the two red geodesics. The event horizon is drawn in orange. Right: $\cK$ of global $\ads_3$. The thick black graphs are point curves $\alpha_p(\theta)$, $\alpha_q(\theta)$ and they enclose the region $\Delta_{pq}$. It captures all geodesics crossing the event horizon.}
\label{fig: EventHorizonGeod}
\end{figure}

As a first application we discuss how to compute lengths of curves in $\ads$ using kinematic space. Just as a (boundary anchored, spacelike) geodesic in $\ads$ corresponds to a single point in $\cK$, a point $p\in\ads$ is associated with a spacelike geodesic in $\cK$. 
\begin{definition}
 Let $\alpha_p(\theta)$ be the curve in $\cK$ which collects all boundary anchored spacelike geodesics running through $p\in\ads$. It is called a point curve.
\end{definition}
\begin{observation}
 Point curves for pure $\ads$ are spacelike geodesics in $\cK$ \cite{Czech:2014ppa}.
\end{observation}
\begin{observation}
In a cylindrical coordinate system the metric of a constant time slice in $\ads_3$, the Poincar\'e disk $\mathbb{D}_2$, assumes the form $ds^2=d\rho^2+\sinh^2\rho\, d\chi^2$ with $\chi\sim\chi+2\pi$. Then the point curve of a point $p=(\rho_0,\chi_0)\in\mathbb{D}_2$ is parametrized through
\begin{equation}\label{pointCurve}
 \alpha_p(\theta)=\arccos\left(\tanh\rho_0\cos(\chi_0-\theta)\right).
\end{equation}
\end{observation}
An example is given in figure \ref{fig: EventHorizonGeod}. We have picked out two points $p,q\in\ads$ and their corresponding point curves in $\cK$ are the black curves delimiting the blue region. Both points are intersected by the orange geodesic (which for reasons to be discussed momentarily is drawn only partly in orange and has blue tails) and thus both of their point curves run through the orange geodesic's point in $\cK$. On the contrary, each point is intersected just by one of the red geodesics and so each point curve contains only one red dot in $\cK$.

%--------------------
%--------------------
Say we want to compute the distance $d(p,q)$ between $p$ and $q$. This distance is given by the orange geodesic. The orange part is indeed only the geodesic distance between $p,q$ and the blue tails complete it to a boundary geodesic, thus naturally associating with it the orange point in $\cK$. The distance is then computed by integrating the Crofton form.

\begin{theorem}
  Let $\Delta_{pq}\in\cK$ be the region enclosed by two point curves $\alpha_p(\theta),\,\alpha_q(\theta)$, which encode two points $p,\,q\in\ads$ in kinematic space. Then the distance $d(p,q)$ in a constant time slice of AdS is computed as a volume integral over the Crofton form
  \begin{equation}
        \label{eq: geodesic distance}
        \frac{d(p,q)}{4G_N}
        =
        \frac{1}{4}\int_{\Delta_{pq}}\omega\,,
    \end{equation}
\end{theorem}

The region $\Delta_{pq}$ encapsules all geodesics which intersect the orange line and the integral assigns a weight $\omega$ to them; an example is found in figure \ref{fig: EventHorizonGeod}. This explains why $\omega$ may be thought of as a line density. It was proven in \cite{Czech:2015qta} that \eqref{eq: geodesic distance} is indeed equivalent to \eqref{geodesicLength}.

Now we are in a position to give a first statement on black holes in conjunction with kinematic space. In order to produce a BTZ black hole, we need to quotient $\ads_3$ by a hyperbolic element of the Fuchsian group. This translates to the statement that the two geodesics which we identify need to be timelike%
\footnote{In the Poincar\`e patch this incorporates the orientation flip of the identified geodesics.}
related in $\cK$ \cite{ZhangChen}. 
\begin{observation}
 Hyperbolic elements of the Fuchsian group, as in definition \ref{HyperbolicElements}, induce time-like transformations on $\cK$.
\end{observation}
In figure \ref{fig: EventHorizonGeod} we can thus identify the two red geodesics, since one lies in the future of the other. The orange line, being the geodesic distance between the two red geodesics, is identified with the event horizon. Thus for this example, equation \eqref{eq: geodesic distance} computes the length of the event horizon since the labels on $\Delta_{pq}$ are taken to be the endpoints of the orange line. For future purposes note that if we shrink the orange line, the region $\Delta_{pq}\in\cK$ shaded in blue shrinks accordingly.

%%%%%%%%%%%%%%%%%%%%%%%%%%%%%%%%%%%%%%%%%%%%%%%%%%%%%%%%%%%%%%%%%%%%%%%%%%%%%%%%
\subsection{Volumes of AdS subspaces as integrals in $\cK$}
\label{sec:kinspaceVolumes}
%%%%%%%%%%%%%%%%%%%%%%%%%%%%%%%%%%%%%%%%%%%%%%%%%%%%%%%%%%%%%%%%%%%%%%%%%%%%%%%%
%The integral \eqref{eq: geodesic distance} is also able to compute lengths of non-geodesic curves, in fact that was the original motivation for that equation \cite{Czech}. In this case some geodesics 
The next natural step is to ask how to compute volumes. An extended discussion can be found in \cite{Abt:2017pmf, Abt:2018ywl}; here we settle for an explanation in terms of examples geared towards wormholes. %\cn{Place this the following further down?} Note that even though formally accessible, computing volumes through kinematic space is technically involved. Thus we will not attempt to execute the resulting integrals since, granted the discussion above, we already know the answer and the kinematic space expressions are guaranteed to yield the same result. Kinematic space holds one major advantage though, as it grants insight into how the information in the CFT arranges itself to contribute to a particular volume. This is in fact the main reason to include this discussion in this work. 

\begin{definition}
 Let $Q$ be a hyperbolic surface on the constant time slice of $\ads_3$. It is naturally associated with a region $\cK_Q\subseteq\cK$, which collects all boundary anchored geodesics $\gamma=(\theta,\alpha)$ of $\ads$ with non-vanishing intersection with $Q$,
 \begin{equation}
  \cK_Q=\{\gamma\in\cK|\gamma\cap Q\neq\O\}\, .
 \end{equation}
\end{definition}

\begin{figure}[t]
\begin{center}
\includegraphics[scale=0.45]{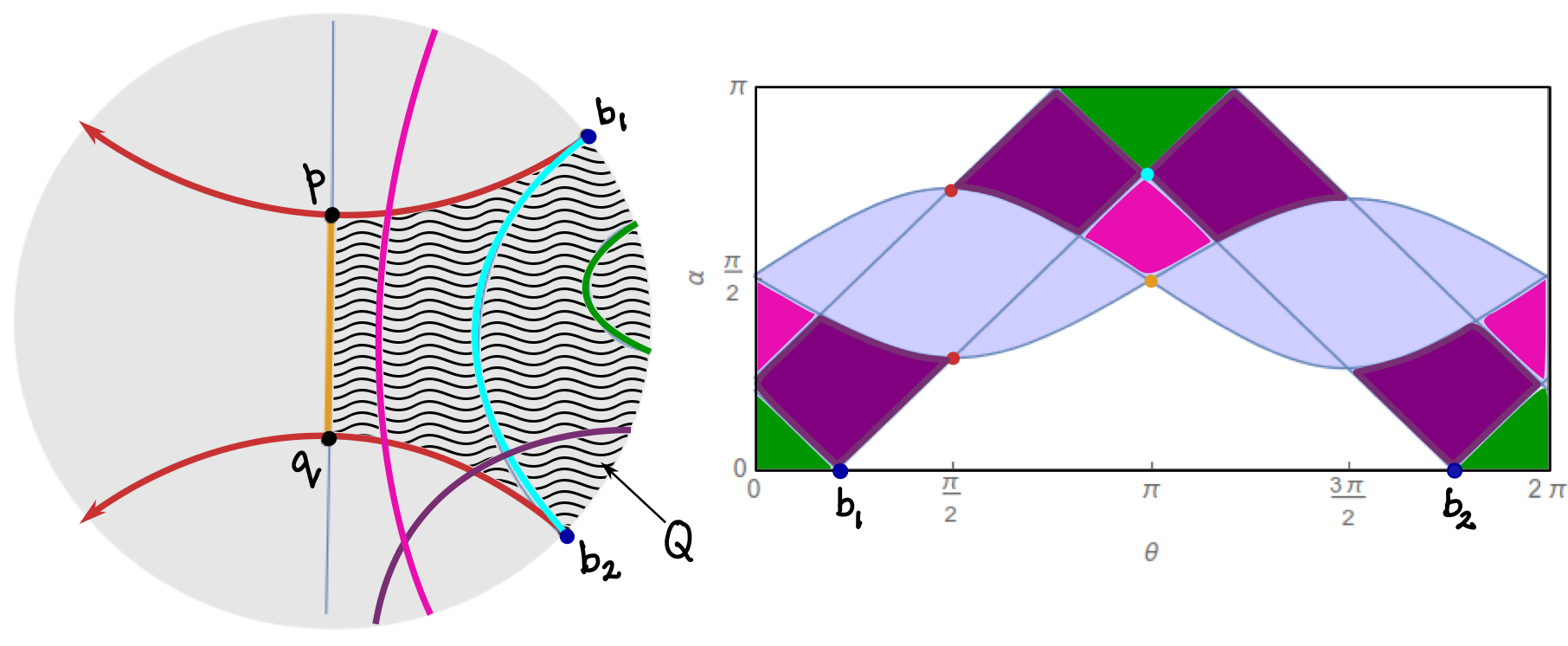} 
\end{center}
\caption{Left: $Q$ is the region filled in with wavy lines, i.e. the right side of event horizon. Geodesics are distinguished by the locus of their endpoints on the boundary of $\ads_3$. Color coding is described in the text. Each geodesic contributes only with its chord $\lambda_Q$, i.e. its intersection with $Q$, here the wiggly region. Right: The colored region is $\cK_Q\subset\cK$. Its sectors are color-coded according to the harbored type of geodesic.}
\label{fig: BTZoneSideVolume}
\end{figure}

As an example consider $Q$ to be the hyperbolic surface to the right of the event horizon in figure \ref{fig: BTZoneSideVolume}. Observe that the geodesics inside $\Delta_{pq}$ intersect $Q$. However, there are more such candidates. One way to fix the remaining geodesics is as follows. 

Because the red geodesics define our BTZ black hole, they obviously also encode the CFT boundary regions to either side of the event horizon. The endpoints of the red geodesics on the right, we call them $b_i\in\p\ads$ with $i=1,2$, which after identification are the same point, are key to determining the remaining geodesics intersecting $Q$.
Finding the point curves for the $b_i$ is simple, because points on $\p\ads$ always have lightlike point curves in $\cK$.
%Point curves of points on the boundary of $\ads$ are always lightlike in $\cK$. This makes finding the point curves corresponding to the $b_i$  simple.
Thus, starting from the red points in $\cK$ we follow lightlike paths so that on the lower boundary of $\cK$, which we recall is identified with $\p\ads$, we hit $b_1$ and $b_2$ as shown in figure \ref{fig: BTZoneSideVolume}. We see that the intersection of these lightlike paths hands us the geodesic corresponding to the CFT boundary interval delimited by the $b_i$; in the picture this geodesic is highlighted in sky blue. 

%-------------------------------------------------------
%-------------------------------------------------------

We find three types of geodesics, which contribute to the volume of $Q$ but do not intersect the event horizon.
\begin{itemize}
 \item \textbf{Green geodesics} are all those that lie fully contained below the sky blue geodesic. They correspond to all subregions of the CFT.
 \item \textbf{Violet geodesics} are all those that have one end in the CFT spacetime and the other in one of the boundary intervals corresponding to one of the two red geodesics.
 \item \textbf{Pink geodesics} are all those geodesics that have one end on the boundary intervals  corresponding one red geodesic and the remaining end on the boundary interval of the other red geodesic.
\end{itemize}
The white regions in $\cK$ are all those geodesics that do not contribute to the volume of $Q$. In order to answer how each $(\theta,\alpha)=\gamma\in\cK_Q$ contribtues to the volume of $Q$, we need another ingredient.
\begin{definition}
 Let $\gamma\in\cK_Q$. We define the chord $\lambda_Q(\gamma)$ to be the length of the intersection of $\gamma$ with $Q$. Given the collection of points $\pin\in\p Q$ where $\gamma$ enters $Q$ and the collection of points $\pout\in\p Q$ where $\gamma$ exits $Q$, the chord $\lambda_Q(\gamma)$ is computed via \eqref{eq: geodesic distance}.
\end{definition}

Every $\gamma\in\cK_Q$ contributes with its chord $\lambda_Q(\gamma)$ to the volume  of $Q$; geodesics $\gamma\notin\cK_Q$ do not contribute.
\begin{theorem}
Let $Q$ be a hyperbolic surface on a constant time slice of $\ads_3$. Its volume is computed in kinematic space $\cK$ through
 \begin{equation}\label{volumeTheorem}
 V(Q)=\frac{2G_N}{\pi}\int_{\cK_Q}\lambda_Q\,\omega.
\end{equation}
\end{theorem}
This formula simply assigns a weight to each chord of $Q$ and sums them up. It was proven in \cite{Abt:2018ywl} that the volume of $Q$ is indeed obtained in this manner.

We now sketch how this formula is applied to compute the volume highlighted in figure \ref{fig: BTZoneSideVolume}. The integration domain $\cK_Q$ is the combination of the colored regions in the figure,
\begin{equation}
 V(\text{RHS})=\frac{2G_N}{\pi}\left(\int_{\text{green}}+\int_{\text{violet}}+\int_{\text{blue}}+\int_{\text{pink}}\right)\lambda_Q\,\omega.
\end{equation}
We have already seen how the colored regions are determined. Now we check that the integrands, more precisely the chords $\lambda_Q$, can in principle be computed using CFT data only if we have access to the CFT regions to either side of the event horizon. For this, we investigate the integral \eqref{eq: geodesic distance}. What we need to check is that the point curves, which provide the integration domain of the chord, are determined through the CFT. 

The green geodesics are the simplest as they describe subregions lying contained within the CFT spacetime. Therefore these geodesics are contained fully in the volume of interest. Their point curves are lightlike and emanate from the CFT boundary points which delimit said CFT subregion. If we pick one representative green geodesic, $\gamma_g=(\theta_g,\alpha_g)$, then its chord length will evaluate to the entanglement entropy, $\lambda_\text{green}=S(\alpha_g)$, of the corresponding CFT subregion.

The other geodesics lie only partially within the volume we wish to compute. Say we pick a geodesic $\gamma_v=(\theta_v,\alpha_v)$ in the violet region, i.e. $\gamma_v$ crosses a red geodesic. One of its point curves corresponds to the boundary point contained within the CFT spacetime; therefore it is lightlike and obviously accessible with CFT data. What about the point curve $\alpha_*(\theta)$ corresponding to the intersection point $p_*=(\rho_*,\chi_*)$ with the red geodesic? This point lies in the bulk and so we have to reconstruct it. That this is possible is evident once we recall that point curves are geodesics, meaning straight lines, in $\cK$, and given the prescription \eqref{pointCurve}, we only need to know two points of $\alpha_*(\theta)$ to fix the $(\rho_*,\chi_*)$.
We know that the point curve passes through $\gamma_v$ giving $\alpha_*(\theta_v)=\alpha_v$; it also passes through the red geodesic $\gamma_r=(\theta_r,\alpha_r)$ giving $\alpha_*(\theta_r)=\alpha_r$. Taken together these constraints give rise to
\begin{equation}\label{pointCurveConstraint}
 \frac{\cos(\theta_v-\chi_*)}{\cos(\theta_r-\chi_*)}=\frac{\cos(\alpha_v)}{\cos(\alpha_r)}\, ,
\end{equation}
which determines $\chi_*=\chi_*(\theta_r,\alpha_r, \theta_v,\alpha_v)$ and $\alpha_*(\theta_r)=\alpha_r$ subsequently fixes  $\rho_*=\rho_*(\theta_r,\alpha_r, \theta_v,\alpha_v)$. In general the functional dependence will be rather involved. Nevertheless, for the computation of the chord length \eqref{eq: geodesic distance} this poses no major obstacle with these point curves. However, for the volume we need to integrate this newly found chord against the Crofton form over the violet region in figure \ref{fig: BTZoneSideVolume} which is parametrized by $(\theta_v,\alpha_v)$. This type of integral that is very hard to control, even for pure $\ads_3$.

This procedure is repeated in the same fashion for the pink geodesics $\gamma_p=(\theta_p,\alpha_p)$. The only difference is that non of its point curves are lightlike and have to be computed through \eqref{pointCurveConstraint} (with all instances of subscript $v$ replaced by subscript $p$). Similarly for the blue geodesics, which has one lightlike point curve and its other one is computed through \eqref{pointCurveConstraint} with $(\theta_v,\alpha_v)\rightarrow(\theta_b,\alpha_b)$ and $(\theta_r,\alpha_r)\rightarrow(\theta_o,\alpha_o)$, where subscript $o$ stands for orange and labels the event horizon geodesic, see figure \ref{fig: BTZoneSideVolume}.

Momentarily, we will illustrate this procedure exemplarily for the causal shadow appearing for two disconnected CFT boundary regions in pure $\ads$. Because we have access to the correct answer for the volume of the present example already through the topological means described in previous sections, we will refrain from performing a calculation here and instead settle for description of the quantum information carried by the volume $Q$. 

We begin with what is na\'ively expected and cemented by our analysis, namely that we cannot compute the volume $Q$ purely from the CFT without knowledge of the entirety of entanglement entropies contained in the CFT subregion connected to $Q$. That is, we require all $S(\alpha_g)$, where $(\theta_g,\alpha_g)$ lies in the green regions in figure \ref{fig: BTZoneSideVolume}. Contrary to that neither the entanglemet entropies corresponding to the regions subtended by the red geodesics contribute nor those entanglement entropies of the CFT subregion behind the event horizon.

Nevertheless, we cannot discard the information contained in these regions fully. We have seen above that these regions contribute, for instance, via violet geodesics, which connect the CFT subregion of interest to the remainder of $\p\ads$. Note however that the second endpoint of violet geodesics is mapped to the CFT subregion when performing the quotient to reach the true wormhole geometry. Violet geodesics therefore represent geodesics which wind around the black hole and provide non-minimal geodesics in the BTZ geometry \cite{ZhangChen}, similar to what is known from entwinement \cite{Balasubramanian:2014sra, GerbershagenEntwinement}. Pink geodesics appear special at first since in figure \ref{fig: BTZoneSideVolume} neither of their endpoints is on the CFT subregion. However, both of their endpoints are mapped to the CFT subregion upon taking the quotient and thus their nature is similar to that of the violet geodesics. 

%%%%%%%%%%%%%%%%%%%%%%%%%%%%%%%%%%%%%%%%%%%%%%%%%%%%%%%%%%%%%%%%%%%%%%%%%%%%%%%%
\subsection{Causal Shadow in $\ads_3$ from Kinematic Space}
\label{sec: kinSpaceCausalShadow}
%%%%%%%%%%%%%%%%%%%%%%%%%%%%%%%%%%%%%%%%%%%%%%%%%%%%%%%%%%%%%%%%%%%%%%%%%%%%%%%%
In this section we illustrate how to compute the volume of the causal shadow for two boundary intervals $A_1\cup A_2=A$ in $\ads_3$ using kinematic space. As we will see, the emerging integrals are quite involved. We turn this into a virtue however, since by drawing on the gravity techniques discussed in previous section, we establish an integral identity -- very much in the spirit of integral geometry. This subsection serves as a stepping stone for the actual case of interest, namely islands in wormhole geometries, which are discussed in the next subsection.

%-------------------------------------------------------
\begin{figure}[t]
\begin{center}
\includegraphics[scale=0.45]{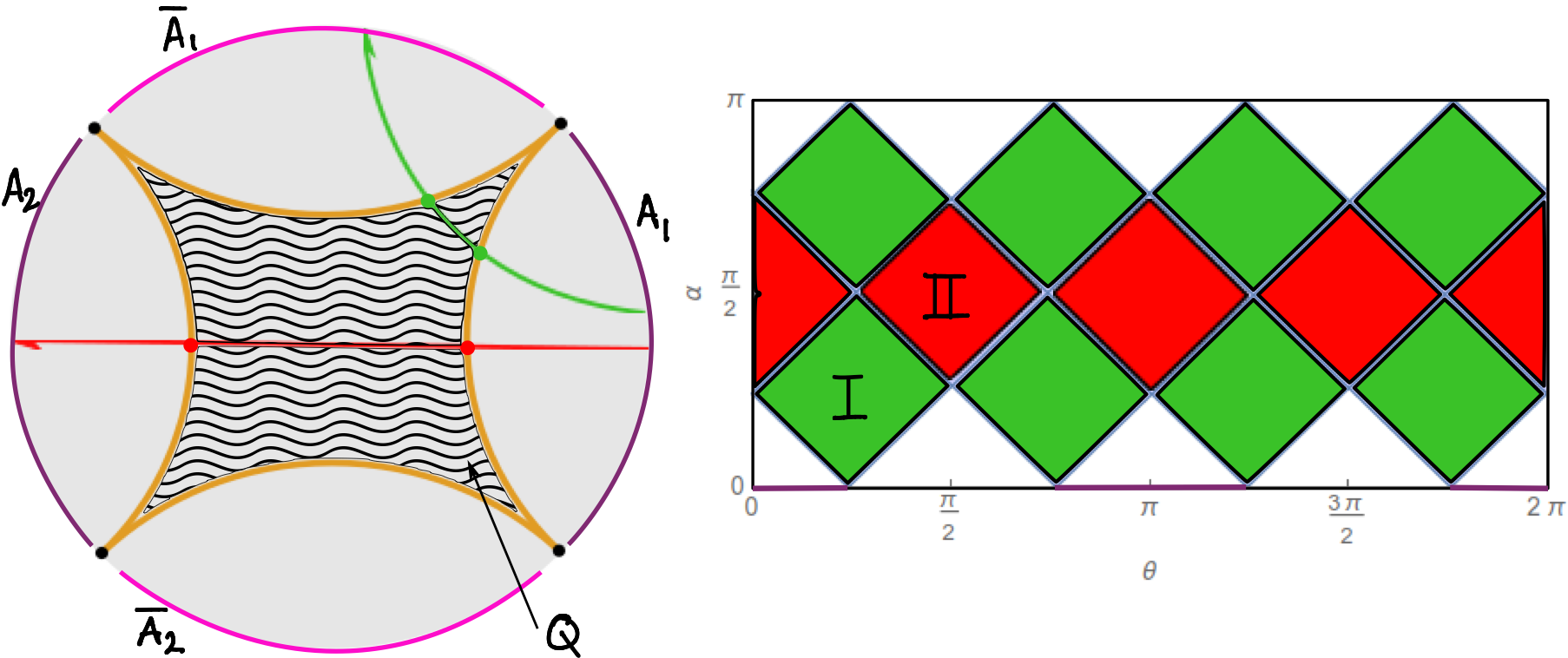} 
\end{center}
\caption{The region $Q$ is the causal shadow emerging for two boundary intervals (purple) $A_1\cup A_2$. The region $\cK_Q\subset\cK$ is the colored region in $\cK$ ( right panel). There are two inequivalent types of geodesics passing through $Q$,  indicated by the green and red squares in $\cK$. All squares of one color contribute equally. Hence we restrict our integration domain to one square of each color, labelled $I$ and $II$, respectively.}
\label{fig: CausalShadowAdS}
\end{figure}
%-------------------------------------------------------
Given that the volume of the causal shadow is determined through topological data, we know that, independently of the boundary interval configuration we choose, we obtain the same result. Therefore, we restrict to the case where the CFT boundary intervals are of equal size, $A_1=A_2$ and placed at opposite sides of global $\ads_3$. The causal shadow, we call it $Q$, appears at the phase transition where the boundary intervals each fill in a quarter circle, as shown in figure \ref{fig: CausalShadowAdS}. Observe that the geodesics which have both of their endpoints located at the same boundary interval, do not contribute to the causal shadow. In other words, the entanglement of the subregions of one boundary interval does not play a role in the causal shadow. The only geodesics that contribute to the causal shadow are those that reach across the boundary intervals; these include the complement of the CFT subregions, $\ol{A}=\ol{A}_1\cup\ol{A}_2$. From the CFT point of view, this means that we need access to \textit{both} CFTs if we want to make sense of the causal shadow as a CFT object through kinematic space. This is in line with considerations of quantum error correcting codes \cite{Almheiri:2014lwa}, where it was noted that a single CFT subregion cannot probe deep into the bulk.

For $Q$ being the causal shadow, $\cK_Q$ splits into twelve squares, see right panel of figure \ref{fig: CausalShadowAdS}. The eight squares adjacent to the boundary (green) correspond to geodesics which leave one boundary region in $A$ and enter $\ol{A}$ or vice versa. All eight such squares yield the same contribution, so we restrict to the square labelled $I$. Geodesics in $I$ are oriented so that they start in $A_1$ and end in $\ol{A}_1$. The squares in the center (red) correspond to those geodesics which start in one half of $A$ $(\ol{A})$ and reach out to the other half in $A$ $(\ol{A})$. All four such squares yield the same contribution, so we restrict to the region labelled $II$. Geodesics in $II$ are oriented so that they start in $A_1$ and end in $A_2$. We are thus in a position to write down the kinematic space integral which computes the volume of the causal shadow $Q$,
\begin{align}\label{CausalShadowVolume}
 V(Q)&=\frac{2G_N}{\pi}\left(8\int_{I}\lambda_I\omega+4\int_{II}\lambda_{II}\omega\right)\\
 &=\frac{2G_N}{\pi}\left(8\int_{-\pi/4}^{\pi/4}du\int_{\pi/4}^{3\pi/4}dv\,\lambda_I(u,v)\omega(u,v)
    +4\int_{-\pi/4}^{3\pi/4}du\int_{5\pi/4}^{3\pi/4}dv\,\lambda_{II}(u,v)\omega(u,v)\right)\notag
\end{align}
We choose lightcone coordinates, see eqn. \eqref{KinSpaceCoordinates}, since they are adapted to the regions of integration $I,\, II$. In these coordinates, $\omega(u,v)=1/(2\sin^2(\frac{v-u}{2}))$. In order to specify the integrand we compute the chords $\lambda_{I/II}$. This is an illustration of the procedure explained around \eqref{pointCurveConstraint}.

First we must fix the point curves \eqref{pointCurve} corresponding to the points where a given geodesic $\gamma=(u,v)\in I$ $(II)$ enters $Q$, $\pin\in\ads_3$, and where it exits $Q$, $\pout\in\ads_3$. In order to fix $(\rho_0,\chi_0)$ for each point curve, we use that this point curve runs through two points of $\cK_Q$. For instance, pick a $\gamma=(u,v)\in I$. The point curve $\pin$ runs through $\gamma=(u,v)$ itself and it also runs through the geodesic subtending $A_1$, $\Gamma_1=(\theta,\alpha)=(0,\pi/4)$. This fixes the point curves $\betain$ of $\pin$ in terms of $u,\,v$,
\begin{align}
 \betain(\eta)&=\arccos\left(\tanh(\rhoin)\,\cos\left(\chiin-\eta\right)\right)\\
 \tan\chiin(u,v)&=\frac{\sqrt{2}\cos\left(\frac{v-u}{2}\right)-\cos\left(\frac{v+u}{2}\right)}{\sin(\frac{v+u}{2})}\\
 \tanh\rhoin(u,v)&=\frac{1}{\sqrt{2}\cos\chiin}
\end{align}
where we have chosen a new coordinate name, $(\theta,\alpha)\rightarrow(\eta,\beta)$ in order to avoid confusion with the parametrization of the integrals over the regions $I/II$ in \eqref{CausalShadowVolume} (in conjuction with \eqref{KinSpaceCoordinates}). Note that any $\gamma=(u,v)\in II$ also runs through $\Gamma_1=(0,\pi/4)$, and thus their point curves are parametrized through $\betain$ as well. 

The exit points however are distinct for geodesics in $I$ and $II$, and so we adjoin an extra label, $\poutI$ and $\poutII$. The point curve for $\poutI$ runs through $\Gamma_2=(\theta,\alpha)=(\pi/2,\pi/4)$ and the point curve for $\poutII$ runs through $\Gamma_3=(\theta,\alpha)=(\pi,\pi/4)$. This fixes their point curves to be
\begin{align}
 \betaoutI(\eta)&=\arccos\left(\tanh(\rhooutI)\,\cos\left(\chioutI-\eta\right)\right)\\
 \cot\chioutI&=\frac{\sqrt{2}\cos\left(\frac{v-u}{2}\right)-\sin\left(\frac{v+u}{2}\right)}{\cos(\frac{v+u}{2})}\\
 \tanh\rhooutI&=\frac{1}{\sqrt{2}\sin\chioutI}
\end{align}
and
\begin{align}
 \betaoutII(\eta)&=\arccos\left(\tanh(\rhooutII)\,\cos\left(\chioutII-\eta\right)\right)\\
 \cot\chioutII&=-\frac{\sqrt{2}\cos\left(\frac{v-u}{2}\right)+\cos\left(\frac{v+u}{2}\right)}{\sin(\frac{v+u}{2})}\\
 \tanh\rhooutII&=-\frac{1}{\sqrt{2}\cos\chioutII}
\end{align}
Note that $\rhooutII>0$ as it should, because $\chioutII\in[3\pi/4,5\pi/4]$.
The sought after chords are then found through \eqref{eq: geodesic distance},
\begin{subequations}\label{chordsIandII}
\begin{align}
 \frac{\lambda_I}{\adsL}=
 \,\text{arctanh}\left(\frac{\sin\left(\frac{v+u}{2}-\chiin\right)}{\sqrt{2\cos^2\chiin-\cos^2\left(\frac{v+u}{2}-\chiin\right)}}\right)
 -
 \,\,\text{arctanh}\left(\frac{\sin\left(\frac{v+u}{2}-\chioutI\right)}{\sqrt{2\sin^2\chioutI-\cos^2\left(\frac{v+u}{2}-\chioutI\right)}}\right)\\
 \frac{\lambda_{II}}{\adsL}=
 \,\text{arctanh}\left(\frac{\sin\left(\frac{v+u}{2}-\chiin\right)}{\sqrt{2\cos^2\chiin-\cos^2\left(\frac{v+u}{2}-\chiin\right)}}\right)
 -
 \,\,\text{arctanh}\left(\frac{\sin\left(\frac{v+u}{2}-\chioutII\right)}{\sqrt{2\cos^2\chioutII-\cos^2\left(\frac{v+u}{2}-\chioutII\right)}}\right)
\end{align}
\end{subequations}
While there is little hope to evaluate the integral \eqref{CausalShadowVolume}, we can turn our logic around and still capitalize on this integral, very much in the spirit of integral geometry. We know from the gravity side that the volume of the causal shadow evaluates to $V(Q)=2\pi\adsL^2$. Therefore, the appropriate way of reading eqn. \eqref{CausalShadowVolume} is as an integral identity on integrals of this type. This is the main result of this section.
\begin{corollary}
 The $\omega$-weighted integral of the functions \eqref{chordsIandII} over regions $I$ and $II$ in $\cK$ is a constant,
 \begin{align}\label{CausalShadowIntegral}
 2\pi\adsL^2=\frac{8G_N}{\pi}\left(2\int_{I}\lambda_I\omega+\int_{II}\lambda_{II}\omega\right)\, .
\end{align}
It measures the volume of the causal shadow for two CFT subregions whose combined size exceeds $\pi$. It is independent of the relative placement of the CFT subregions and depends only on the topology of the causal shadow.
\end{corollary}
\begin{observation}
 The entanglement entropy of the subregions of $A_1$ and $A_2$ does not enter in the volume of the causal shadow \eqref{CausalShadowIntegral}. The contributions come exclusively from geodesics reaching to other boundary regions. This cements that, should we compute the volume of the causal shadow from the CFT, we require non-local information between the sectors $\cH_{A_1}$ and $\cH_{A_2}$ of Hilbert space.
\end{observation}
The attentive reader might wonder if the chords of green geodesics or the chords of geodesics between $\ol{A_1}$ and $\ol{A_2}$ are included in this reasoning. Indeed they are, since we assume that we have access to both $\cH_{A_1}$ and $\cH_{A_2}$. This means that we know the boundary points of both CFT regions on $\p\ads$, which in turn implies we know all geodesics which constitute the boundary of the causal shadow. As we have seen, this information is enough to construct the chords of all geodesics $\gamma\in\cK_Q$. 
%%%%%%%%%%%%%%%%%%%%%%%%%%%%%%%%%%%%%%%%%%%%%%%%%%%%%%%%%%%%%%%%%%%%%%%%%%%%%%%%
\subsection{Kinematic Space and Islands}
\label{sec: kinSpaceIslands}
%%%%%%%%%%%%%%%%%%%%%%%%%%%%%%%%%%%%%%%%%%%%%%%%%%%%%%%%%%%%%%%%%%%%%%%%%%%%%%%%
We are finally in a position to discuss islands using kinematic space. In this subsection we explain which kind of integral arises in the computation of islands for $\n$-exit wormholes. We will derive again an integral identity. This time however we omit details on the computation of the chords since they are derived in the same way as before with appropriate choices for the geodesics which are identified in going to the wormhole geometry.

%-------------------------------------------------------
\begin{figure}[t]
\begin{center}
\includegraphics[scale=0.45]{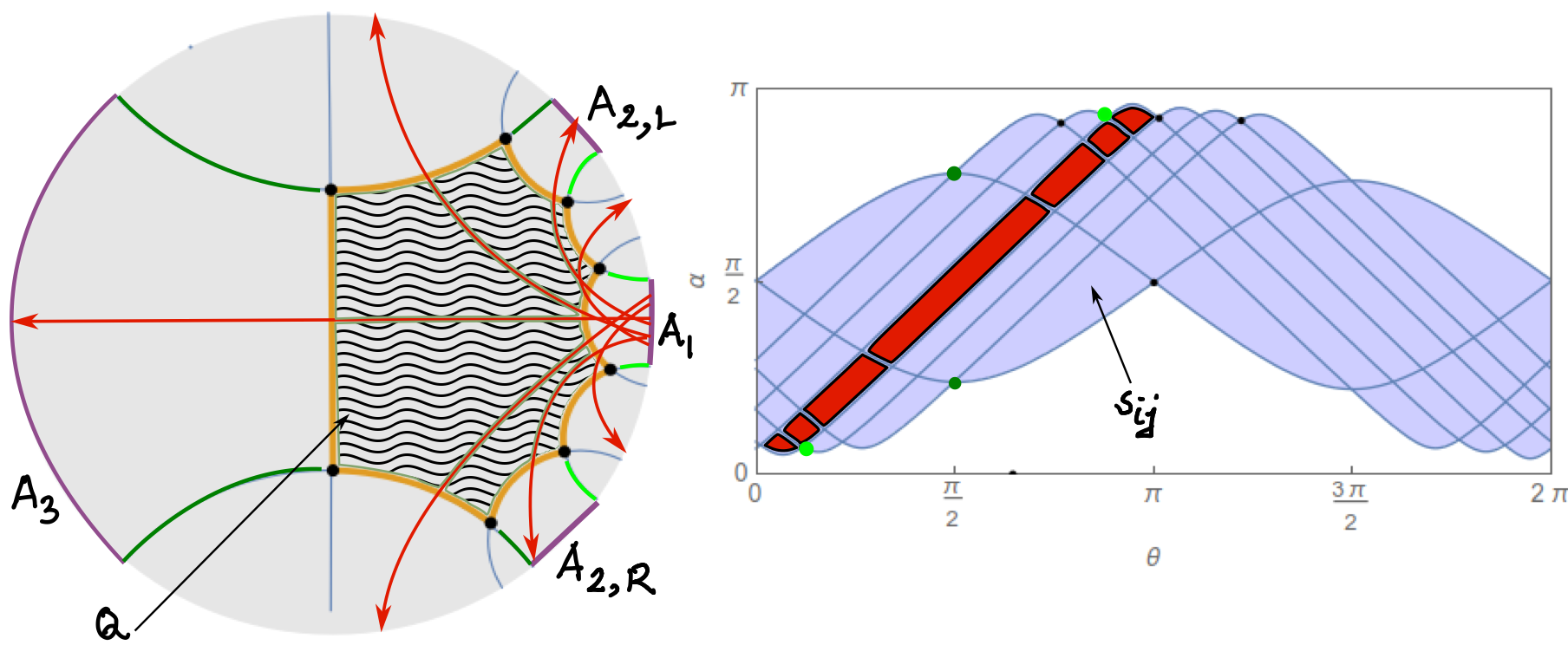} 
\end{center}
\caption{The region $Q$ is the island arising for a three-exit (purple) wormhole. The geodesics which are identified by hyperbolic elements of the Fuchsian group are colored pairwise in light and dark green. Black dots in $\cK$ correspond to event horizons. The colored region in $\cK$ is $\cK_Q$ and its segments are the sectors $\s_{ij}$ described in the text, of which we have $\s=56$. We have higlighted sectors $\s_{ij}$ for a fixed $i$ and varying $j\neq i$ in red.}
\label{fig: IslandsAdS}
\end{figure}
%-------------------------------------------------------
Given an $\n$-exit wormhole, the arising island -- we call it $Q$ as usual for the region of interest -- has $\e=4(\n-1)$ edges, each edge being a geodesic segment. See figure \ref{fig: IslandsAdS} for the case $\n=3$, where the island is an octagon, $\e=8$ (generally an $\e$-gon). We count the edges via a label $i=1,\dots,\e$. The region $\cK_Q\subset\cK$ splits into $\s=\e(\e-1)$ sectors, which we call $\s_{ij}$, each one corresponding to geodesics entering the island $Q$ through the $\ith$ edge and exiting through the $\jth$ edge; this naturally accounts for the orientation of the geodesics. Obviously $\s_{ii}=\O$, because the edges of $Q$ are geodesic segments and can thus only be intersected by $\gamma\in\cK_Q$ once, never twice. This secures the number of sectors mentioned above. 

We now define

\begin{equation}\label{shortcutIntegral}
 \cI_{ij}=\frac{2G_N}{\pi}\int_{\s_{ij}}\,\lambda_Q\,\omega\, , \qquad i\neq j
\end{equation}

which compiles the contribution to the island's volume $Q$ stemming from sector $\s_{ij}\subset\cK_Q$. Note that $\cI_{ij}$ is not a subvolume of $Q$, in fact by itself it is \textit{not even a volume}, despite the formal resemblance with \eqref{volumeTheorem}. This is due to the fact that $\s_{ij}$ does not capture all geodesics running through a particular subregion of $\ads_3$. Observe further that $\cI_{ij}=\cI_{ji}$ since orientation reversed geodesics contribute to $Q$ in the same manner as its mirror. This leaves us with $\e(\e-1)/2$ integrals of type \eqref{shortcutIntegral} at the price of introducing a factor 2 in the computation of $Q$. The chord $\lambda_Q$ is determined as in the previous section. One can place all geodesics which give rise to the $\n$-exit wormhole via quotienting at convenient values to simplify the expressions. This does not influence the result, since we know from previous sections that the volume is topological. 

In conjunction with the results \eqref{jump} from above for the island's volume we thus find here, as the main result of this section, an integral identity.
\begin{corollary}
Let $Q$ be a hyperbolic $\e$-gon on a constant time slice of $\ads_3$. It is associated with $\cK_Q=\bigcup_{i\neq j}\s_{ij}$, which splits into sectors $\s_{ij}$. The sum of $\omega$-weighted integrals $\cI_{ij}$ over all sectors $\s_{ij}$ is a constant,
\begin{align}\label{IslandIntegral}
 2\pi(\n-2)\adsL^2=2\sum_{i=1}^\e\sum_{j>i}^\e\cI_{ij}\, ,\qquad \e=4(\n-1)\,.
\end{align}
It measures the volume of the island arising at the Page transition triggered at page time $\n$. It is independent of the relative placement of the CFT subregions and depends only on the topology of the island.
\end{corollary}
\begin{observation}
 The entanglement entropy of subregions of the wormhole exits $A_k$ with $k=1,\dots,\n$, does not enter in the volume of the island \eqref{IslandIntegral}. The contributions come exclusively from geodesics reaching between sectors. This cements that, should we compute the volume of an island from the CFT, we require non-local information between the sectors $\cH_{A_i}$ of Hilbert space.
\end{observation}
As with the example of the previous subsection the chords of geodesics between parts of $\ol{A}=\ol{\cup_k A_k}$ are included in this reasoning since, given the knowledge of all boundary intervals $A_k$, we can reconstruct the boundary of the island, $\p Q$. This then grants access to the chords of all geodesics $\gamma\in\cK_Q$. This concludes our exposition of kinematic space.

%%%%%%%%%%%%%%%%%%%%%%%
\section{Causal shadow \& complexity in tensor networks }\label{tensornets}
Tensor networks have played a major role over the last few years towards understanding the holographic entanglement in parallel in conjunction with gravity. The idea is to prepare a maximally entangled state within a geometry starting from the ground state. It has been able to provide us with explicit realizations of ideas such as sub-region duality, bulk reconstruction and minimal surfaces that match the gravity proposals quite well. The role played by \cite{Vidal:2007hda,Evenbly:2008pza,Swingle:2012wq,Czech:2015xna,Pastawski:2015qua,Franco-Rubio:2017tkt} have been instrumental in giving shape to this program. Since entanglement is generated eventually by introducing tensors with appropriate properties, this program provides the foundations for the encoding of notions of complexity when particular tensors are thought of as introducing gates in preparing a state. The number of legs in that regards can be therefore thought of as a cost that each of the gates carry.

Our inspiration here derives from \cite{Abt:2017pmf}, where the jump in sub-region complexity was succesfully reproduced by introducing tensor networks and counting the number of legs. Hence, in this section, we turn towards understanding our study from the perspective of tensor networks. The notions of tensor networks in the context of the multi-boundary wormholes has been studied previously in \cite{Peach:2017npp}. We take it up from there and discuss how it naturally complements the lessons we learn in our work. Of course, there are some limitations to the program concerning the discretization of hyperbolic space through tessellations instead of a continuous treatment that kinematic space provides. Therefore, the aim of this section will be to understand the area, volume and causal shadows qualitatively from the point of view of discretized tensor networks. 
\par Standard protocol for the implementation of tensor networks on $\ads_3/\cft_2$ is to tessellate $\mathbb{H}^2$, i.e; the time-slice of AdS$_{3}$, with discrete polygons. With respect to each of the edges of the polygons, reflection symmetry is preserved and it naturally gives rise to an embedded group structure in the hyperbolic space, known as the Coxeter group. 
%We give a more mathematically complete description of the Coxeter group in \ref{Coxeter Anindya}. 
Apart from this particular symmetry that we require, there are several choices one can make while discretizing hyperbolic space through the polygons as discussed in \cite{Peach:2017npp}. This picture fits our previous discussion naturally as the hyperbolic polygons also play a crucial role in understanding the area enclosed by the causal shadows which we discussed in detail above.  The choice of tensor dictates the tessellation since the tensors are introduced in the center of the polygons of the lattice dual to the tessellation lattice. A crucial point is that all of these choices are not equally good in realizing a given model of multi-boundary wormhole. For example, in \cite{Peach:2017npp}, the authors have shown that due to error of discretization, there are some choices of tessellations for which even after taking the quotient, one might not obtain a causal shadow. On the other hand, there are also degeneracies in the representations of the minimal bulk geodesics in some of the choices. Having discussed some of the limitations, more of which can be found in \cite{Peach:2017npp, Bhattacharyya:2016hbx}, we choose the particular set of tessellations for which the limitations are minimized. By minimizing, we mean that one has to work with tessellations for which the geodesic degeneracies are not present and the causal shadow is indeed present. 
\par Let us now briefly discuss the tiling of hyperbolic plane and Coxeter group and explain how the minimal bulk geodesics are represented in terms of the tensors of the dual lattice in this formalism. First we discuss mathematical nuances of the Coxeter group in \ref{Coxeter Anindya} and then in \ref{tensors final}, we describe how we can use our previous discussion to understand the causal shadow regions from the tensor network point of view for three-boundary and $n$-boundary wormholes respectively.
           
 %%%%%%%%%%%%%%%%%%%%%%%          

\subsection{A primer on Tessellation And Coxeter Groups:} 
\label{Coxeter Anindya}         
In this subsection we describe the very basics of tessellation of planes and their relation with Coxeter groups. The formal definition of tessellations is as follows.
\begin{definition}
A tessellation of the Euclidean or the hyperbolic plane is a subdivision of the plane into polygonal tiles $P_i$ such that the tiles have the following properties:
\begin{enumerate}
    \item given any two tiles $P_i$ and $P_j$, there exists an isometry $\gamma$ such that $\gamma(P_i)=P_j$.
    \item If $P_i$ and $P_j$ are not the same tile, then only one of the following holds:
    \begin{itemize}
        \item $P_i\cap P_{j}=\O$
        \item $P_i \cap P_j=p$ where $p$ is a single point in the plane and $p$ is a vertex of both $P_i$ and $P_j$.
        \item $P_i \cap P_j=\e$ where $\e$ is a common edge of both $P_i$ and $P_j$. 
    
    \end{itemize}
    \item For any point $p$ on the plane, there exist at least one tile $P_i$ such that $p\in P_i$. If there exists exactly one such tile then $p$ is in the interior of the tile. If $p$ is in exactly two tiles then it is on the common edge of the two polygons. If $p$ is in more that two tiles then p is a common vertex of all tiles containing it. 
\end{enumerate}
\end{definition}
We are mostly interested in tessellations of the hyperbolic plane $\HH$. If $P$ is a right angled hyperbolic polygon and $\TT$ is a Fuchsian group generated by isometries which identifies pairs of edges of $P$ then the collection of images of $P$, $\{\gamma(P)|\gamma\in \TT\}$, tessellates the hyperbolic plane. Given a tessellation of $\HH$, we call it $T$, we can associate a group with $T$, called the \textit{Characteristic group} of $T$ and we will denote it by $\TT_T$. It is defined as follows
\begin{equation}
 \{\gamma \in  Isom(\HH)|\forall P_{i}\exists P_{j}\text{ s.t. }\gamma(P_i)=P_j\}\, .
\end{equation}
In other words $\TT_T$ contains all the isometries which take all the tiles of $T$ to distinct tiles. Note that $P_i$ can be same as $P_j$, that means if the polygonal tiles themselves have symmetries then those will be included in $\TT_T$. 
\par Next we define the Coxeter groups \cite{thomas2018geometric} and describe their relations with hyperbolic tessellations \cite{davis2008geometry}. 
\begin{definition}\label{Coxeter Groups}
 Suppose $S=\{s_i\}_{i\in I}$ is a finite set and $M=(m_{ij})_{i,j\in I}$ be a matrix such that 
\begin{itemize}
    \item $m_{ii}=1,\forall\, i\in I$
    \item $m_{ij}=m_{ji},\forall\, i,j\in I$
    \item $m_{ij}\in\{1,2,3,...\}\cup \{\infty\},\forall\, i\neq j$
\end{itemize}
Then M is called a Coxeter matrix and the associated Coxeter group G is defined by the presentation 
$$G=\langle S\,|\,(s_{i}s_{j})^{m_{ij}}=1,\,\forall i,j\in I\rangle$$

The pair $(G,S)$ is called the Coxeter system.
\end{definition}

\begin{figure}[t]
	\centering
	\includegraphics[scale=0.35]{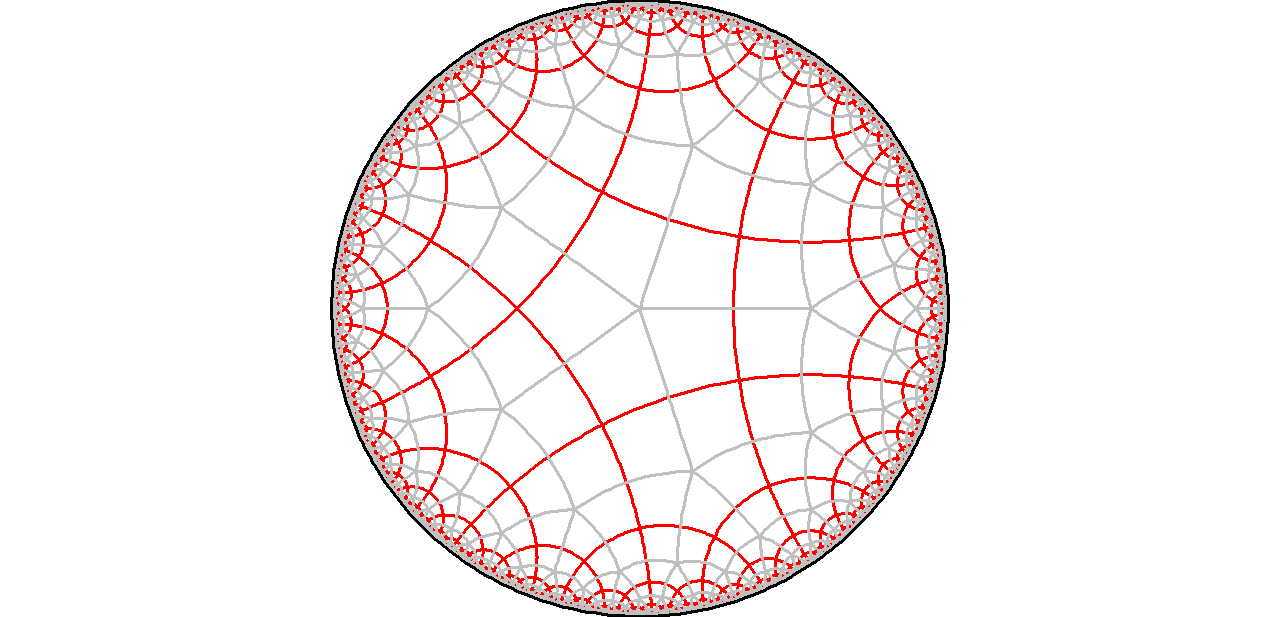}
	\caption{(2,3,5) Tessellation of $\mathbb{HH}$.}
	\label{fig: tessellation1}
\end{figure}

A very useful class of examples of Coxeter groups is the reflection groups, which is described in the following theorem:
\begin{theorem}[\cite{davis2008geometry}]
Let $P$ be a simple convex polygon in $\HH$ with faces $\{e_i\}_{i\in I}$. Suppose $\forall i\neq j$, if $e_i\cap e_j\neq \O$, the angle between $e_i$ and $e_j$ is $\pi/m_{ij}$ for some $m_{ij}=2,3,4,...$. Set $m_{ii}=1$ and $m_{ij}=\infty$ if $e_i\cap e_j=\O$. Let $s_i$ be the isometric reflection w.r.t. the infinite geodesic supported by $e_i$. Then the group $G=\langle s_i\,|\,i\in I\rangle$ satisfies the following properties: 
\begin{itemize}
    \item $G=\langle s_{i}|s_{i}^{2}=1, \,(s_{i}s_{j})^{m_{ij}}=1\rangle$
    \item $G$ is a discrete subgroup of $Isom(\HH)$
    \item $P$ is a fundamental domain of the $G$-action and $P$ tessellate $\HH$.
\end{itemize}
\end{theorem}
The group described in the theorem is an example of a \textit{reflection group} and clearly it is a Coxeter group. There are infinitely many polygons on $\HH$ which tessellate the plane via the Coxeter group generated by reflections. If $(p,q,r)$ is a triple satisfying  $\frac{1}{p}+\frac{1}{q}+\frac{1}{r}<1$, then there exist a hyperbolic triangle with interior angles $\pi/p,\pi/q$ and $\pi/r$ and the Coxeter group generated by their edge-reflections is called the \textit{(p,q,r)-Triangular group} and the triangle tessellate the hyperbolic plane. Also there exist a right angled regular hyperbolic n-gon for all $n\geq 5$ and they also tessellate $\HH$. If $p$ is a regular $n-$gon with interior angles $2\pi/m$ then $P$ tessellate $\HH$ and $m$ copies of $P$ meets at each vertex. We show a simple $\left(2,3,5\right)$ tessellation of $\mathbb{H}^{2}$ in figure \ref{fig: tessellation1}\footnote{These tessellated figures have been generated using the free software available online made by Dmitry Brant.}.

%%%%%%%%%%%%%%%%%%%%%%%

\subsection{Tessellations and multi-boundary wormholes:}
\label{tensors final}
In the construction of an $n$-boundary Riemann surface, one usually considers quotients of $\mathbb{H}^2$ by subgroups of its isometry group $SL\left(2,\RR\right)$ which identifies a pair of geodesics. A given regular tiling of the hyperbolic plane is preserved under elements of the associated Coxeter group. By combining different reflections from the Coxeter group it is possible to construct hyperbolic elements which identify pairs of geodesics that form edges of the tiles \cite{Peach:2017npp, Bhattacharyya:2016hbx}. Thus such elements are also isometries of the tiling and we can quotient by discrete subgroups $\Gamma$ of the Coxeter group consisting of those elements to obtain a tessellation of the Riemann surface $\mathbb{H}^2/\Gamma$.

\begin{figure}[t]
	\centering
	\includegraphics[scale=0.40]{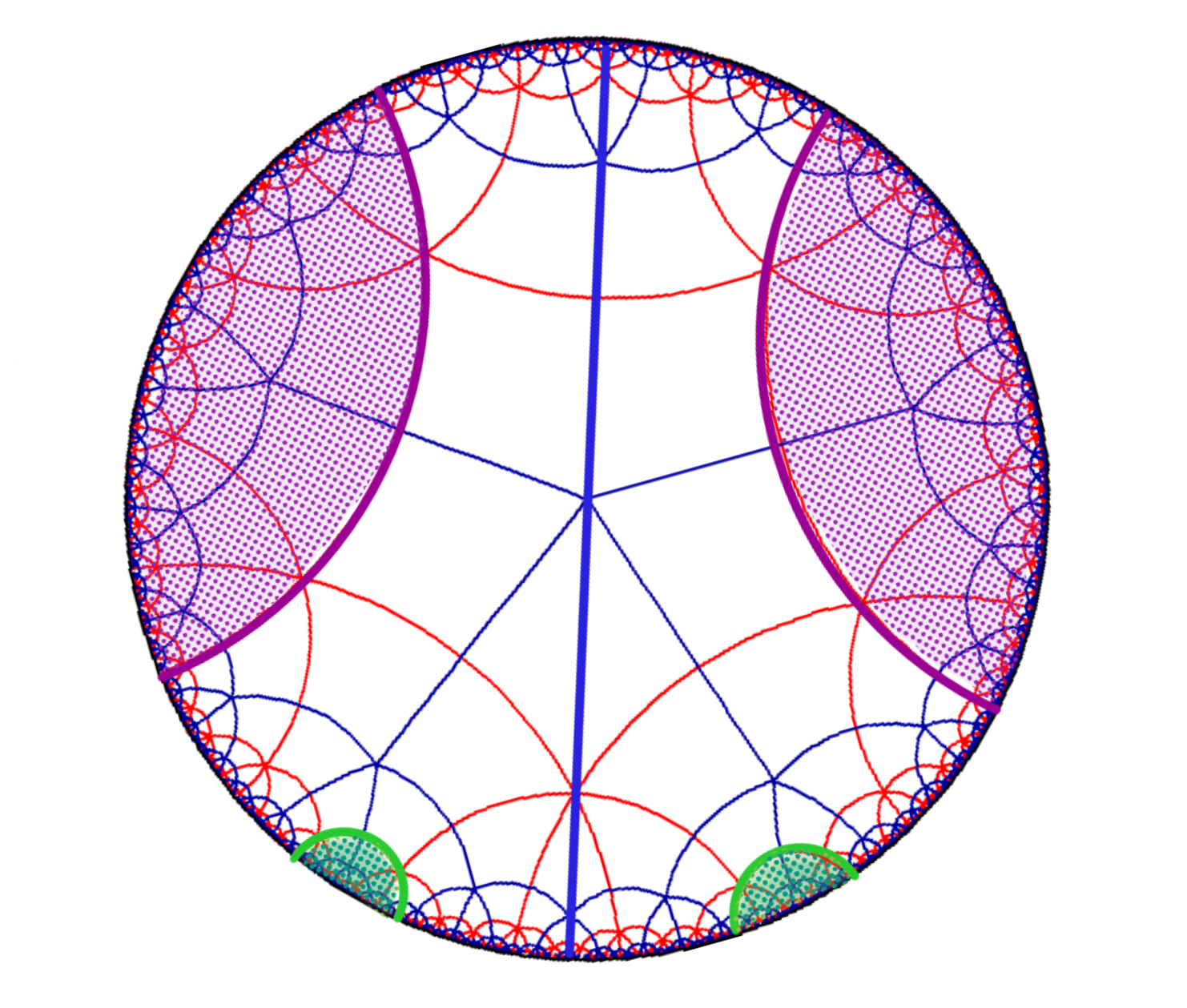}
	\caption{Three-boundary wormhole with (2,5,6)tessellation.}
	\label{fig: tessellation2}
\end{figure}

The authors of \cite{Peach:2017npp} have illustrated multiple possibilities to obtain a tiling of the $3$-boundary wormhole in this way. One of them is shown in figure \ref{fig: tessellation2}, if we let $r_A,\,r_B~\text{and}~r_C$ the reflections about the geodesics coloured blue, purple and green respectively then starting from a $\left(2, 5, 6\right)$ tiling of $\HH$, quotienting by $\Gamma$ generated by $r_Ar_B$ and $r_Br_C$ generates the tiling for the $3$-boundary wormhole with the unshaded region being the fundamental domain of identification.
\par Earlier the minimal closed geodesics between a pair of identified semicircles were identified with the horizon associated with each asymptotic boundary. Their analogues in a tessellation are the minimal closed paths along edges of the tiling homologous to each boundary \cite{Peach:2017npp} (see figure \ref{fig: tessellation2}), they may be degenerate depending on the discretization. Once a tiling has been chosen, the associated tensor network could be constructed by considering the tiling as a graph and placing a tensor on each vertex of its dual graph. Tensor networks help realize a discrete version of the Ryu-Takayanagi formula. 

\begin{lemma}
If we consider only two boundary regions $A$ and $A^c$ then a minimal path $\gamma_A$ as defined above divides the bulk network into two parts with boundaries $\gamma_A \cup A$ and $\gamma_A \cup A^c$. If we denote by $\abs{\gamma_A}$ the number of tensor legs this minimal path cuts through then according to \cite{Pastawski:2015qua} the tightest bound on the entanglement entropy of $A$ and $A^c$ is provided by
\[S_A \leq \abs{\gamma_A}\ln \chi\,, \]
where, $\chi$ is called the bond dimension of the tensors. The bound is not violated for degenerate minimal paths as all of them have same length.
\end{lemma}

With a similar philosophy we can associate the holographic complexity of $A$ with the number of tensor nodes trapped within the region bounded by $\gamma_A \cup A$; in fact this is the definition utilized in \cite{Abt:2017pmf} to study the holographic complexity from tensor network models of $\AdS_3$. To be precise, they embedded the AdS$_{3}$ metric in a 2D Ising model and computed the number of nodes trapped within the entanglement wedge to represent the volume dual to the minimal surface. The subsequent complexity plots in \cite{Abt:2017pmf} show reasonable behaviour and adds to our expectation that this counting works.

\par In our case, we use the tessellations and the elements of the Coxeter group to primarily take the necessary quotients that systematically keep track of the fundamental domain of the multiboundary wormhole. As mentioned before, the tessellations are to be chosen in such a way that the minimal throat horizons corresponding to different exits encode a causal shadow region between them. Once the causal shadow region is understood, as the authors in \cite{Peach:2017npp} have mentioned, one can attribute the constant volume of this region to a constant number of tensors (known as central tensors). Since, discretized tessellations are used instead of some continuous network, the number of tensor nodes within a particular region gives us an approximated measure of volume. Nevertheless, the takeaway message is relatively simpler and can be framed as follows.

\begin{observation}
 In case of the three ($n$) boundary wormhole, the causal shadow region, tessellated by hyperbolic triangles, is a hyperbolic octagon $\left(4\left(n_{Page}-1\right)\text{-gon}\right)$. Therefore, we need eight $\left(4\left(n_{Page}-1\right)\right)$ hyperbolic triangles to tessellate this region.
\end{observation} 
  Now using this fact, it is straight-forward to state that the number of tensor nodes trapped within the $n$-boundary causal shadow is $(n_{Page}-2)$ times the number of tensor nodes trapped within the causal shadow of three-boundary case, see \eqref{jump}. This depends on the respective appropriate tilings and therefore the dual graphs corresponding to the three- and $n$-boundary models. It is noteworthy that the results and lessons from tensor network are qualitatively similar to the kinematic space identities, as mentioned in equation \eqref{IslandIntegral}.

%%%%%%%%%%%%%%%%%%%%%%%%%%%%%%
\section{Discussion and Conclusion:}\label{conclusion}

We have computed the subregion complexity corresponding to the radiation subsystem in the multi-boundary wormhole models in this paper. We have considered two models in which the islands appear, the three-boundary wormhole and the $n$-boundary wormhole. Although the two models are qualitatively similar and the island region in both the models correspond to the causal shadows, there are some differences as well. Building on the finding that the causal shadows tend to play the role of islands in these models, we have approached this from various points of view to build a more concrete understanding of the situation. Starting from the volume from the Ricci scalar, we have also studied what the volume and the causal shadows imply when looked at from the kinematic space and tensor network perspectives. In course of our work, we have repeatedly tried to bridge the physical ideas and models to the mathematical notions. For the wormhole construction, identifications of geodesics and tessellations, we have added a few lemmas and theorems along the way that strengthen these ideas mathematically. In the following, we list down the main results of our paper and conclusions that these results tend to imply.

\textbf{ Sub-region volumes: } This is the central piece of this paper. We computed the volumes corresponding to a bipartite radiation subsystem for the three-boundary wormhole and $(n-1)$-partite one for the $n$-boundary wormhole. The remaining exit in both cases represents the evaporating black hole. As we have mentioned already the proposals in the literature \cite{Alishahiha:2015rta,Abt:2017pmf}, the volumes dual to the subregions capture the complexity of the corresponding state. 
Therefore, the computation of volumes is aimed at enhancing our understanding of the complexity of the radiation state.
%Therefore, the idea of computing volumes is to get some understanding of the complexity of the radiation state.
Recent findings and especially the implications of 
Python's lunch \cite{Brown:2019rox,Bao:2020hsc}, suggest that even though quantum extremal surfaces enable us to reproduce the Page curve, it is still exponentially hard to compute the restricted complexity of the radiation state. Therefore, while Hawking was mistaken about entropy, his statements truly apply to complexity.
Now, since in these three- and $n$-boundary wormhole models, one can reproduce the Page curve consistently, we performed explicit calculations to investigate if the volumes feature precisely such exponential growth. However it is worth noting that within these multiboundary models, the volumes can not capture the exponential restricted complexity. \footnote{In the Python's lunch geometry, one computes the complexity from one of the two sides of the two sided wormhole. This is known as the restricted complexity (exponentially hard) as opposed to the complexity=volume conjecture where one computes the unrestricted complexity (only polynomially hard) by acting with unitaries on both the sides. The bulge in the geometry was conjectured by the authors in \cite{Brown:2019rox} to argue that the restricted complexity is exponential due to presence of such a bulge region in the wormhole connecting the two sides. According to \cite{Brown:2019rox}, the job of shortening of this bulged wormhole to make it a usual thermofield double state without a bulge is exponentially hard and this is the reason for the difference between
the restricted and the unrestricted complexity. In our models, although we compute the volume from the radiation side, the geometries do not have a bulge between the radiation and the evaporating BH. Therefore, the complexity of shortening of the bulge is not considered in our computations of subregion complexity. Nevertheless, it would be interesting to work with a warped multiboundary wormhole with a bulge, extending the construction in \cite{Bao:2020hsc}, and consider the volume of such a bulge region to check if the subregion volume complexity can capture the ideas of restricted complexity and Python's lunch. It is also worth noting that the authors of \cite{Brown:2019rox} argued about the behaviour of complexity purely from the perspective of tensor networks and not from the perspective of volumes. We thank the anonymous referee and the editor-in-charge for asking to clarify on this point.}  %\ab{However, it is worth mentioning at this point that the Python's lunch is a very particular geometric proposal that has a bulge in the bulk connecting the radiation and the black hole which is put in by hand for the sake of the argument. On the other hand , we do not have such a region in the multi-boundary wormholes. Therefore, one should not have high expectations of finding signatures of Python's lunch from our models \footnote{We would like to thank the anonymous referee for pointing this out to us.}. But it would be in general interesting to study similar volumes for geometries constructed in \cite{Bao:2020hsc} to see if the volume in such a construction can teach us something about the scenario described in \cite{Brown:2019rox}. It is also worth noting that the authors of \cite{Brown:2019rox} argued about the behaviour of complexity from purely the perspective of tensor networks and not from the perspective of volumes.}
%\ab{In the Python's lunch geometry, the radiation is again collapsed in a BH and one computes the complexity from one of the two sides of the two sided wormhole. This is known as the restricted complexity (exponentially hard) as opposed to the complexity=volume conjecture where one computes the unrestricted complexity (only polynomially hard) by acting with unitaries on both the sides. The bulge in the geometry was conjectured by the authors in \cite{Brown:2019rox} to argue that the restricted complexity is exponential due to presence of such a bulge region in the wormhole connecting the two sides. According to \cite{Brown:2019rox}, the job of shortening of this bulged wormhole to make it a usual thermofield double state without a bulge is exponentially hard and it is the reason of the difference between restricted and unrestricted complexity. In our models, although we compute the volume from the radiation side, the geometries do not have a bulge between the radiation and the evaporating BH. Therefore, the complexity of shortening of the bulge is not considered in our computations of subregion complexity. Nevertheless, it would be interesting to work with a warped multiboundary wormhole with a bulge extending the construction in \cite{Bao:2020hsc} and consider the volume of such a bulge region to check if the subregion volume complexity can capture the ideas of restricted complexity and Python's lunch.} 

\par What we find is rather surprising. For both models under study, we find two kinds of plots that the volumes dual to the radiation subsystem follow. One is a constantly decaying one whereas the other one is of Gaussian nature. In both the cases, at the Page time, a constant volume is added to the otherwise UV divergent volume due to the change of the minimal surface. The universality goes deeper since the overall plots are very similar inspite of the fact that the nature of the Page curves, especially the Page time is quite different in the two models. In case of the three-boundary model, this volume is simply $2\pi$ whereas for the $n$-boundary analog, it depends on the Page time, here $n_{Page}$. Therefore, the only difference between the nature of the plots is the jump at Page time being independent or dependent of the Page time. It would be interesting to see if this addition of constant volume at Page time is a consequence of three dimensional AdS or not. But since the construction of multi-boundary wormholes is only well known for AdS$_{3}$, it is hard to check this for general spacetime dimensions. 

\par Now let us come back to the nature of the two kinds of plots (figures \ref{fig: 3bdycomp} and \ref{fig: nbdycomp}). In both cases, we find that although the minimal lengths increase steadily before the Page time, there is no guarantee that the volumes also increase. For example, for the three-boundary model the HRT length increases until it reaches the Page time, but the Gaussian plot of volume already starts decreasing \textit{before} the Page time. For the constantly decaying plots, this is even more evident since the volume keeps decreasing irrespective of the nature of the plot that the HRTs follow. Again, the only effect that the Page transition leaves on the volume is a constant jump. This jump is due to the addition of the causal shadow region and the UV divergent part remains unchanged due to a homology constraint of the boundary spatial lengths. There is nevertheless something universal about the nature of these plots since in both models, we end up with very similar graphs with substantially different considerations only distinguished by the quantity of the constant volume that is added at the Page time. Interestingly, none of our plots feature exponential growth. This begs the question whether these volumes represent the complexity of the radiation or not. We do not want to make any strong comment regarding that. But what our results show is how the volumes dual to the radiation subsystems evolve with time within the scope of these models. Now, for coming to a conclusion on how exact these models are, one indeed needs to build a better understanding of the actual evaporating black hole rather than a multi-boundary wormhole model. It would be interesting to check if similar calculations can be done in an actual evaporating black hole situation instead of our simplified models. If the results in those cases also mimic what we find, only then can we say that these multi-boundary wormholes can model the evaporating black holes accurately. Otherwise the conclusion is simply that although within the purview of these models, one can reproduce the Page curves by studying classical HRT surfaces, they are not capable of capturing more complex phenomenona like the complexity of the radiation. It might also be interesting to investigate the nature of these volumes if one works with the eternal BH construction using the multiboundary wormhole geometries. There, we can possibly expect a continuing growth of the volume since the BH exit does not shrink (transparent boundary conditions).

\begin{figure}[t]
	\centering
	\includegraphics[scale=0.20]{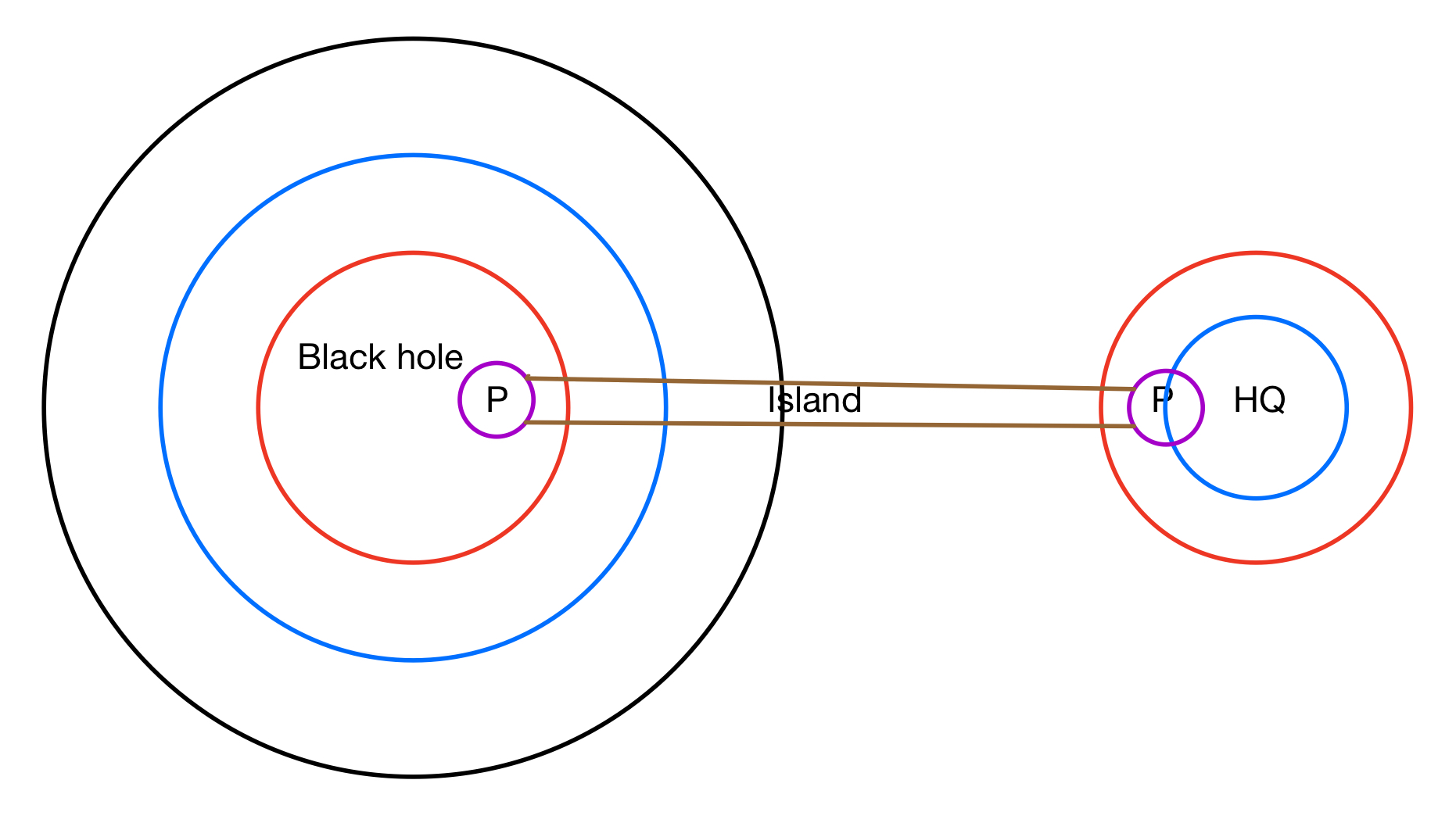}
	\caption{A representative figure of islands and purification: Phase I: In the beginning phase, the black hole is the black sphere and there is no Hawking quanta. Phase II: Evaporating black hole is blue sphere (LHS) and HQ is blue sphere (RHS). Phase III: (Page time) Red spheres in LHS and RHS are BH and HQ respectively. P (purple spheres) are the purified partner modes in both sides. The information of purification is carried by the island region connecting LHS and RHS.}
	\label{fig: purification}
\end{figure}

 \textbf{On complexity of purification:}The volume of the causal shadows for the multi-boundary wormholes have been discussed before briefly in \cite{Fu:2018kcp, Balasubramanian:2018hsu, Caceres:2018blh} in the context of purification complexity. The reappearance of these results in our context strengthens the correspondence between the islands and purification. The correspondence between multi-boundary wormholes and entanglement of purification (EoP) was first advocated in \cite{Bao:2018fso}. In \cite{Bhattacharya:2020ymw}, these similarities were discussed in regards to the multi-boundary wormhole model of islands and multi-partite entanglement of purification. We can therefore argue for a similar but an extended version of this correspondence from the understanding of complexity in this paper. In terms of the results of our paper, the change of complexity ($\Delta C$) due to the island within these models is simply equivalent to tripartite or multipartite complexity of purification (CoP). The way one talks about purification in the context of island is that after the Page time, some of the Hawking modes outside the black hole horizon get purified by their partner modes inside the black hole since the radiation subsystem gets access to those partner modes inside the horizon. This happens due to inclusion of the island regions in the entanglement wedge of the radiation subsystem. In terms of complexity, our results signify that the access to the purifying partner modes also enables the radiation subsystem to access a certain new number of gates which results in the jump at Page time. This jump from no-island to island phase have been also addressed in \cite{Hernandez:2020nem}(Section 4) and has been attributed to the mutual complexity, which matches with the multipartite purification complexity as shown in \cite{Balasubramanian:2018hsu, Caceres:2018blh}. Within the scope of these models, to the best of our understanding, after the Page time the access to this new set of gates characterizes the non-trivial nature of the plots of the subregion complexity (as shown in Figures \ref{fig: 3bdycomp} and \ref{fig: nbdycomp}). We show a representative figure of the above-mentioned event in Figure \ref{fig: purification}. However, the consistent fact apart from the jump in the two candidate curves is that in both the cases the mixed state complexity of the final radiation state becomes zero indicating a final pure state. On the other hand, the initial complexity of the two candidate curves show different features (large value in the decaying one and small or zero value in the other one) indicating that in the very initial phase, the radiation might go through different evolution procedures within these models. 
 
\textbf{ Kinematic Space lessons:} 
Given the prominent role of bulk regions such as causal shadows and islands in our analysis, we elucidated its properties from a complementary angle.
We have shown how to reconstruct volumes of islands in the CFT through use of kinematic space. Our analysis clearly displays which quantum information encodes the volume of islands. For a wormhole geometry with $\n$ exits $A_k$, the correlations responsible for the entanglement of each subregion within a single $A_k$ \textit{never} contribute to the island's volume. The protagonists are always the correlations arising through the entanglement between the exits $A_k$. Other contributions to the island's volume arise from geodesics that are not anchored in $A=\cup_k A_k$. They contribute through their chords piercing the island and we have described how these terms can be computed from the knowledge of $\p A$. Moreover, we have combined the expressions for the volume of causal shadows and islands with our general results from pure gravity analyses to derive integral identities for trigonometric integrals, \eqref{CausalShadowIntegral} and \eqref{IslandIntegral}, in line with the purpose of integral geometry. These identities might be of interest to the mathematical community, and of course any physicist working with trigonometric integrals.

\textbf{ Tensor Networks and Volumes:} Finally, we have discussed how the tensor network approach in multiboundary wormholes can be used to build a parallel understanding of the throat horizon minimal surfaces and the corresponding volumes. While the number of tensor legs cutting a minimal surface quantifies the length of the throat horizons, the total number of tensor legs within any volume encoded by boundary and bulk surfaces quantify the volumes. This is a rough way of quantifying volumes inspired by the study in \cite{Abt:2017pmf}. The limitations of this quantification stem from discretizing hyperbolic space through discrete Coxeter group tessellations. Nevertheless, as argued in \cite{Peach:2017npp}, we also attribute the constant volume of the causal shadow regions in three- and $n$-boundary wormholes to the central tensors of the network. Hence, these also play the role of the islands in our description.

We have worked with various equivalent definitions of area and volumes within the multiboundary wormhole models of island and black hole evaporation. The most quantitative results that we obtain are from the exact volume calculations with the given assumptions of the models in hand. In the other sections, we have partly explained the qualitative lessons and partly turned the qualitative results into quantitative ones through integral identities (kinematic space) and properties of central tensors (tensor network). We have also paid close attention to the mathematical details of these models through a detailed discussion of the construction in section \ref{maths section}. We have discussed briefly the Coxeter group from a mathematical standpoint in section \ref{Coxeter Anindya} before using them in the understanding of wormhole construction using discrete tessellations. 

\par The complexity of the radiation state in the evaporating black hole models have been investigated in detail in this paper. These models apply only within AdS$_{3}$ and it would be really interesting to study higher dimensional situations. The similarities between kinematic space and tensor networks are evident \cite{Czech:2015kbp} and we presume that higher dimensional and time dependent \cite{Czech:2019hdd} understanding of kinematic spaces can teach us something about tensor networks in higher dimensions as well, which in general is a hard numerical problem to address. There are a few more interesting future directions as well. From our results, it is kind of evident that at least within AdS$_{3}$, a constant volume is added to the volume of the radiation subsystem at Page time. It would therefore also be interesting to compute volumes corresponding to eternal black hole models and check whether the volume comes down to zero or keeps growing after the jump at the Page(island-inclusion) time. One might also be interested in computing similar quantities for the doubly holographic braneworld models introduced in \cite{Geng:2020qvw,Chen:2020uac,Chen:2020hmv} and relate them to the ideas of complexity of purification as we have been able to do in our work. The jump in volume presents a phase transition at Page time in the space of states due to the inclusion of islands and our expectation is that this phase transition is universal. This phase transition in terms of complexity of purification signifies the existing new set of gates that the radiation subsystem can access starting from the Page time. Hence, another direction to explore is to look for more signatures of this phase transition in the studies of complexity and track the origin of the new set of accessible gates. This would also help in understanding which modes actually get purified at Page time (as shown in Fig. \ref{fig: purification}).

%\section{Appendix}
%\subsection{ A crash Course on M\"{o}bius transformations:}
%\begin{definition}
%A map  on the complex plane $\CC$, $f:\CC\rightarrow \CC$  is call a M\"{o}bius Transformation if it is of the form $f(z)=\frac{az+b}{cz+d}$ where $a,b,c,d\in \CC$ and $ad-bc\neq 0$.
%\end{definition}
%It is easy to check that any M\"{o}bius transformation defines a self-homeomorphism on $\CC$. Hence the set of all  M\"{o}bius transformations forms a group under conjugation, lets denote the group as M\"{o}b$(\CC)$. 

\acknowledgments
We thank Ignacio Reyes for helpful correspondence. A.B. and S.M. would like to thank Department of Atomic Energy (DAE), Govt. of India for the financial support. A.B. would also like to thank Arpan Bhattacharyya, Aninda Sinha and Shubho Roy for organising of the workshop "Quantum Information in QFT and AdS/CFT", where this work was presented. A.C is thankful to Philip Bowers for fruitful discussions. 
C.N. acknowledges financial support by the Deutsche Forschungsgemeinschaft (DFG, German Research Foundation) under Germany's Excellence Strategy through W\"urzburg‐Dresden Cluster of Excellence on Complexity and Topology in Quantum Matter ‐ ct.qmat (EXC 2147, project‐id 390858490).

%%%%%%%%%%%%%%%%%%%%%%%%%%%%%%%%
\bibliographystyle{JHEP}
%\bibliography{references}
\providecommand{\href}[2]{#2}\begingroup\raggedright\endgroup
%\end{thebibliography}\endgroup

%%%%%%%%%%%%%%%%%%%%%%%%%%%%%%%%
\end{document}